\documentclass[useAMS,usenatbib]{mn2e}
\usepackage{graphicx}
\usepackage{epstopdf}
\usepackage{lineno}
\newcommand{\araa}{Annual Review of Astron and Astrophys}
\newcommand{\pasp} {Publications of the Astronomical Society of the Pacific}
\newcommand{\pasa} {Publications of the Astronomical Society of Australia}
\newcommand{\aap} {Astronomy \& Astrophysics}
\newcommand{\aj}{Astronomical Journal}
\newcommand{\apj}{Astrophysical Journal}
\newcommand{\apjl}{Astrophysical Journal, Letters}
\newcommand{\apjs}{Astrophysical Journal, Supplement}
\newcommand{\mnras}{Monthly Notices of the RAS}

\title[GAMA: The Role of Environment]{Galaxy And Mass Assembly (GAMA): Resolving the role of environment in galaxy evolution}

\author[S. Brough et al.]{S. Brough$^{1}$\thanks{E-mail: sb@aao.gov.au},
S. Croom$^{2}$,
R. Sharp$^{3}$,
A. M. Hopkins$^{1}$, 
E. N. Taylor$^{2,4}$,
I. K. Baldry$^{5}$,
\newauthor M. L. P. Gunawardhana$^{2,1}$,
J. Liske$^6$,
P. Norberg$^7$,
A. S. G. Robotham$^{8,9}$,
\newauthor A. E. Bauer$^{1}$,
J. Bland-Hawthorn$^{2}$,
M. Colless$^{1}$, 
C. Foster$^{10}$,
L. S. Kelvin$^{9,8,11}$,
\newauthor M. A. Lara-Lopez$^1$,
\'A.R. L\'opez-S\'anchez$^{1,12}$,
J. Loveday$^{13}$,
M. Owers$^{1}$,
\newauthor 
K. A. Pimbblet$^{14}$,
M. Prescott$^{15}$\\
$^{1}$Australian Astronomical Observatory, PO Box 915, North Ryde, NSW 1670, Australia\\
$^{2}$Sydney Institute for Astronomy (SIfA), School of Physics, The University of Sydney, NSW 2006, Australia\\
$^{3}$Research School of Astronomy \& Astrophysics, The Australian National University, Cotter Road, Weston Creek, ACT 2611, Australia \\
$^{4}$School of Physics, The University of Melbourne, Parkville, VIC 3010, Australia  \\
$^5$Astrophysics Research Institute, Liverpool John Moores University, Twelve Quays House, Egerton Wharf, Birkenhead CH41 1LD, UK \\
$^6$European Southern Observatory, Karl-Schwarzschild-Str. 2, 85748 Garching, Germany\\ 
$^{7}$Institute for Computational Cosmology, Department of Physics, Durham University, South Road, Durham DH1 3LE, UK \\ 
$^8$ICRAR, University of Western Australia, 35 Stirling Highway, Crawley, WA 6009, Australia\\
$^9$SUPA, School of Physics and Astronomy, North Haugh, St Andrews, Fife, KY169SS, UK \\
$^{10}$European Southern Observatory, Alonso de Cordova 3107, Vitacura, Santiago, Chile \\
$^{11}$Institut f\"{u}r Astro- und Teilchenphysik, Universit\"{a}t Innsbruck, Technikerstra$\rm{\beta}$e 25, 6020 Innsbruck, Austria \\
$^{12}$Department of Physics and Astronomy, Macquarie University, NSW 2109, Australia\\
$^{13}$Astronomy Centre, University of Sussex, Falmer, Brighton BN1 9QH, UK\\
$^{14}$School of Physics, Monash University, Clayton, VIC 3800, Australia \\
$^{15}$Department of Physics, University of the Western Cape, Private Bag X17, Bellville 7535, South Africa \\
}
\begin{document}
\linenumbers
\date{}

\pagerange{\pageref{firstpage}--\pageref{lastpage}} \pubyear{2012}

\maketitle

\label{firstpage}

\begin{abstract}
We present observations of 18 galaxies from the Galaxy And Mass Assembly (GAMA) survey made with the SPIRAL optical integral field unit (IFU) on the Anglo-Australian Telescope.  The galaxies are selected to have a narrow range in stellar mass ($6\times10^9 M_{\odot}<M_{*}<2\times10^{10} M_{\odot}$) in order to focus on the effects of environment.  Local galaxy environments are measured quantitatively using 5th nearest neighbour surface densities.  We find that the total star formation rates (SFR) measured from the IFU data are consistent with total SFRs measured from aperture correcting either GAMA or Sloan Digital Sky Survey single-fibre observations. The mean differences are ${\rm SFR_{GAMA}}/{\rm SFR_{IFU}} =  1.26 \pm 0.23, \sigma=0.90$ and for the Sloan Digital Sky Survey we similarly find ${\rm SFR_{Brinchmann}}/\rm{SFR_{IFU}}= 1.34 \pm0.17, \sigma=0.67$.  Examining the relationships with environment, we find off-centre and clumpy H$\alpha$ emission is not significantly dependent on environment, being present in 2/7 ($29^{+20}_{-11}$ per cent) galaxies in high-density environments ($>0.77$ Mpc$^{-2}$), and 5/11 ($45^{+15}_{-13}$ per cent) galaxies in low-density environments ($<0.77$ Mpc$^{-2}$).  We find a weak but not significant relationship of the total star formation rates of star-forming galaxies with environment.  Due to the size of our sample and the scatter observed we do not draw a definitive conclusion about a possible SFR
dependence on environment.  Examining the spatial distribution of the H$\alpha$ emission, we find no evidence for a change in shape or amplitude of the radial profile of star-forming galaxies with environment.  If these observations are borne out in larger samples this would infer that any environment-driven star-formation suppression must either act very rapidly (the `infall-and-quench' model) or that galaxies must evolve in a density-dependent manner (an `in-situ evolution' model).
\end{abstract}


\section{Introduction}

The galaxy population we see today has some very distinctive features.  One of the most fundamental is the separation of galaxies into a bimodal distribution according to colour (e.g. \citealt{strateva01,baldry06}).  The colour largely relates to the age of the stars, with galaxies on the tight red sequence being mostly passive systems containing old stars. In contrast, the galaxies in the blue cloud generally show a younger stellar population (e.g. Taylor et al. in prep).  However, it is still unclear what drives this separation.

Recent research has focussed on how blue star-forming galaxies can have their star formation quenched, moving them onto the red sequence.  Red-sequence galaxies are preferentially found in denser environments (e.g. \citealt{blanton05, cooper08, thomas10, smith12}) and star formation is also clearly suppressed in those high density environments (e.g. \citealt{lewis02,gomez03,kauffmann04}).  This immediately suggests environmental factors play an important role.  

There is uncertainty, however, in how the change in star-forming properties as a function of environment manifests itself. \cite{balogh04} found that, once luminosity is taken into account, the observed environmental difference is only due to the {\it fraction} of blue galaxies changing in each environment, rather than due to any change in the properties of the galaxy population. Star formation rates measured from single-fibre observations of the H$\alpha$ emission line give similar conclusions: \cite{peng10} observed that the relationship between star formation rate and stellar mass was the same in the highest and lowest density environments. Recent results from the Galaxy And Mass Assembly (GAMA; \citealt{driver11}) survey also show that the fraction of star-forming galaxies falls with increasing environmental density \citep{wijesinghe12,robotham13}, but the star formation rate of the star-forming galaxies depends solely on their stellar mass, showing no change with their environment \citep{wijesinghe12}.  These observations would imply that any mechanism that transforms galaxies in dense environments must be rapid or have happened a long time ago.

In contrast, research examining the strength of the $4000\rm{\AA}$ break and the Balmer absorption lines \citep{vonderlinden10} and ultraviolet imaging from the \emph{GALaxy Evolution EXplorer} (GALEX) space telescope \citep{rasmussen12} suggests that both the star-forming fraction \emph{and} the star formation rate in star-forming galaxies changes as a function of environment, allowing for a longer timescale for any transformation.  

The different conclusions drawn by these observations may have a number of causes, including the different ways that star formation and environment are measured, varying definition for star-forming galaxies and the inability of single-fibre observations to specify where that star formation is happening.  This last point is crucial given that the proposed mechanisms for any modulation of star formation with environment can have very different spatial effects:

{\it Ram-pressure stripping} \citep{gunn72,nichols11}, which can expel the gas from the disk, and {\it strangulation} \citep{larson80}, which results when the gas is removed from the halo, should both preferentially remove gas in the outer parts of galaxies (e.g. \citealt{bekki09,kapferer09}).  These processes may be efficient at removing halo gas, which is observed in galaxy clusters (e.g. \citealt{sun07,randall08}).  Ram-pressure stripping may also act in small and/or compact groups \citep{mccarthy08,rasmussen08} or on the outskirts of clusters (e.g. \citealt{merluzzi12}).   The timescale of $>2$ Gyrs for strangulation \citep{mccarthy08}, however, seems to contradict the short timeframe implied by observations. Although \cite{prescott11} find this to be the likely mechanism for the quenching of star formation in satellites hosted by isolated galaxies. Direct galaxy--galaxy interactions may also play a critical role in either triggering star formation (e.g. \citealt{moss93,ellison08,patton13}) or suppressing it, as seen in the less-massive galaxies of pairs when the pair mass ratio is large \citep{robotham13}.  

Feedback from star formation in low-mass galaxies provides an internal mechanism for transformation.  This provides a solution to the mismatch of the theoretical dark matter halo mass function and the observed stellar mass function (e.g. \citealt{baldry08}) by heating and/or expelling gas in halos.  Extreme outbursts of star formation may be triggered by mergers or interactions (e.g. \citealt{hopkins09}) even with very low luminosity galaxies or tidal debris (e.g. \citealt{angel10,cluver13}), and frequently seen in isolated compact groups (e.g. \citealt{angel04,iraklis10,scudder12}).  This makes a link between internal and environmental effects.  There is observational evidence of feedback from star formation (e.g. \citealt{veilleux05,strickland09,angel12}).  

At present it is still not clear which of these processes dominate in which situations.  In one of the first attempts to study spatially-resolved star formation in a very large sample as a function of a broad range of environment, \cite{welikala08, welikala09} used galaxy colours to demonstrate that star formation is suppressed in the central parts of galaxies in high-density environments, apparently ruling out ram-pressure stripping as a significant influence in the general galaxy population. 

While galaxy colours are a coarse measure of the integrated star formation history of a galaxy, the well-understood H$\alpha$ emission line at 6563\AA\   probes near-instantaneous star formation ($< 10$~Myr, e.g. \citealt{kennicutt98}). Spatially-resolved H$\alpha$ measurements have only been made for samples of local galaxies either in the field or nearby clusters (e.g. \citealt{moss93,vogt04,koopmann04a,meurer06,fumagalli08,rose10,sanchez12}), often with narrow-band imaging, but have not been possible for a data set that covers a wide range in environment.

We present here observations of the spatially-resolved H$\alpha$ emission of galaxies over a wide range of environment from optical integral field unit (IFU) observations of galaxies selected from the GAMA survey.  The primary goal of this paper is to measure the radial distribution of star formation and examine how that varies as a function of environment.  We know that the star formation rates of galaxies are strongly dependent on their stellar mass, but their dependence on environment is less clear.  We therefore use GAMA to select a carefully controlled sample of galaxies with a narrow range of stellar masses ($M_{*}\sim10^{10}M_{\odot}$) in a range of environments.  GAMA is highly spectroscopically complete (97 per cent; \citealt{driver11}), even in the densest regions.  This is achieved by returning to each target area an average of $10$ times, as described in \cite{robotham10}.  This enables accurate environment measurements including 5th nearest neighbour surface densities (e.g. \citealt{wijesinghe12}) and friends-of-friends group determination \citep{robotham11}.

We describe the selection of our sample in Section \ref{sect:sample} and the observations in Section \ref{sect:observations}.  In Section \ref{sect:emissionlines} we describe the method for measuring the emission line properties and then present the total star formation rates and H$\alpha$ surface brightness profiles of the sample in Sections~\ref{Sect:SFR} and \ref{sect:profiles}. We discuss our findings in Section \ref{sect:discussion} before summarising our conclusions in Section~\ref{sect:conclusions}.  Throughout this paper we assume a Hubble constant of $H_0=70$ km s$^{-1}$ Mpc$^{-1}$ and $\Omega_M=0.3$, $\Omega_\Lambda=0.7$.  Equivalent widths for features in emission are quoted as positive numbers. 

\section{Sample}
\label{sect:sample}
We selected our sample from the Galaxy And Mass Assembly (GAMA; \citealt{driver11}\footnote{http://www.gama-survey.org}) survey which combines single-fibre spectroscopy \citep{hopkins13} with a diverse set of supporting imaging data.  We specifically selected galaxies from the first phase of the GAMA survey, referred to as GAMA I.  There are $\sim170,000$ galaxies in the GAMA I sample down to $r=19.4$ mag in two regions, each of 48 sq deg, and $r=19.8$ mag in a third region, also of 48 sq deg.  While the majority of the GAMA spectra have been obtained from the Anglo-Australian Telescope (AAT), the spectra and redshifts for brighter galaxies in these regions, like those targeted here, are obtained from the Sloan Digital Sky Survey (SDSS; \citealt{york00}).

In order to focus specifically on the effects of environment rather than stellar mass we targeted galaxies with stellar masses $6\times10^9M_{\odot}<M_{*}<2\times10^{10}M_{\odot}$.   The stellar mass measurements are from spectral energy distribution fits to optical broad-band photometry \citep{taylor11} and have random uncertainties of $\sim0.3$ dex. Given the size of the uncertainties, no narrower window in stellar mass would be appropriate.  We corrected the redshifts for the effects of peculiar velocity using the \cite{tonry00} multi-attractor flow model ($z_{\rm{TONRY}}$; \citealt{baldry12}) and limited the sample to low redshifts, $0.02<z_{\rm{TONRY}}<0.06$, so that targets are close enough that we can spatially resolve them.  This reduces the available sample to 688 galaxies.  We are complete in stellar mass over the redshift range considered.

The nearest neighbour surface density, $\Sigma_5$, is calculated for all galaxies with reliable redshifts (nQ$>2$; \citealt{driver11}) .  The 5th nearest neighbour metric is similar to the $\Sigma_1$ metric used in \cite{brough11}. The surface density is defined using the projected co-moving distance to the 5th nearest neighbour ($d_5$) with $\pm 1000$km s$^{-1}$ within a pseudo-volume limited density-defining population: $\Sigma_5=5/\pi d^2_5$.  The density-defining population has absolute SDSS petrosian magnitudes M$_r<$M$_{r,\rm{limit}}$-Q$z$, k-corrected to $z=0$ following \cite{loveday12}, where M$_{r,\rm{limit}}$= -20.0 mag and Q defines the expected evolution of M$_r$ as a function of redshift (Q=0.87; \citealt{loveday12}).  Densities are then corrected for the survey $r-$band redshift completeness as $\Sigma_5=\Sigma_{5,\rm{raw}}\times1/$completeness.  Galaxies where the nearest survey edge is closer than the 5th nearest neighbour have upper limits calculated and flags assigned.  More details on this and other environment metrics available for GAMA will be provided in Brough et al. (in prep).  

\begin{figure}
\begin{center}	
\resizebox{20pc}{!}{\includegraphics{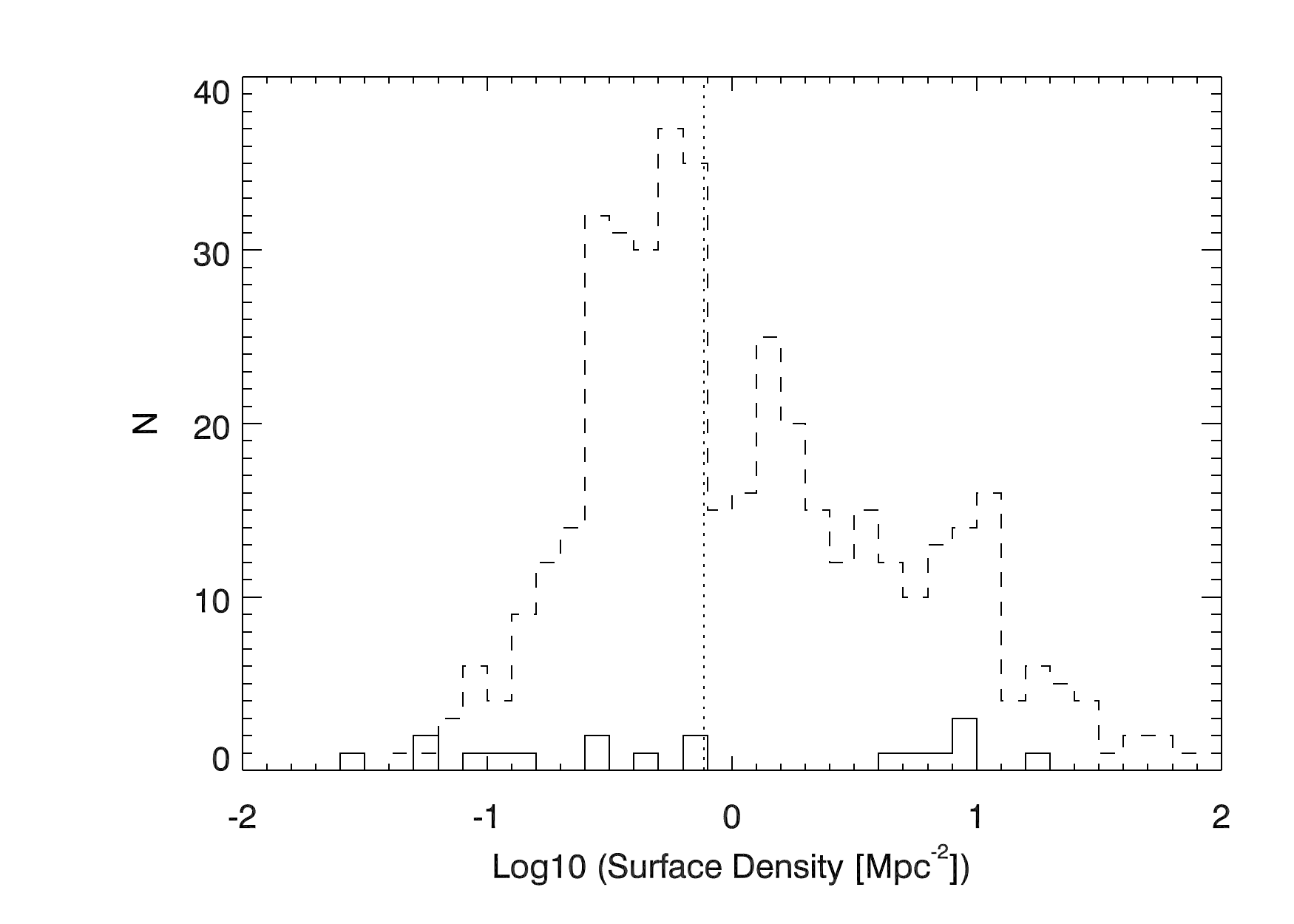} }
\end{center}
\caption{Histogram showing the surface density distributions of the 424 possible targets (dashed line) and the 18 observed galaxies (solid line).  The dotted line indicates the median surface density of $0.77$ Mpc$^{-2}$ that divides the low and high density samples analysed here.  The observed galaxies sample the possible density distribution well and the low and high-density subsamples can be seen to be well separated.}
\label{dens_histo}
\end{figure}

There are 424 galaxies with stellar masses $6\times10^9M_{\odot}<M_{*}<2\times10^{10}M_{\odot}$ and $0.02<z_{\rm{TONRY}}<0.06$ that are not flagged as having been affected by a survey edge.  These 424 galaxies have environmental densities $0.02 < \Sigma_5 (\rm{Mpc}^{-2})<78$, with a median $\Sigma_5=0.77~\rm{Mpc}^{-2}$.  We randomly selected 18 galaxies across 2 density bins around the median density ($<0.77$ Mpc$^{-2}$; 11 galaxies and $>0.77$ Mpc$^{-2}$; 7 galaxies).  The surface density distribution of the 424 possible targets and the 18 selected are illustrated in Figure~\ref{dens_histo}.  The 18 targets selected have apparent SDSS petrosian magnitudes m$_r<17.6$ mag and a mean effective semi-major axis radius, from 2D Sersic surface brightness fits to re-processed SDSS r-band imaging \citep{kelvin12}, of R$_{e,r}=3.4^{\prime\prime}$.  The properties of the 18 observed galaxies from the GAMA survey are described in Table~\ref{sample}.  

We determined the effect the number of the nearest neighbour used has on the sample selected and the results presented here.  We also calculated $\Sigma_ {N=10}$ for the parent sample considered here (galaxies with stellar masses $6\times10^9M_{\odot}<M_{*}<2\times10^{10}M_{\odot}$ and $0.02<z_{\rm{TONRY}}<0.06$). The mean difference $\Sigma_{N=5}-\Sigma_{N=10}=2.0$ Mpc$^{-1} \pm6.8$ Mpc$^{-1}$.  The median $\Sigma_ {N=10}=0.95$ Mpc$^{-1}$, which does not move galaxies between the high and low density bins defined here.  However, using N=10 a large fraction of this sample (12 out of 18) are affected by survey edges, we therefore present all of our results using N=5.

The GAMA groups catalogue \citep{robotham11} is not volume limited so we cannot draw strong conclusions from the group properties of these galaxies.  However, we do note that all galaxies in high-density environments are found in groups and that these generally have higher total dynamical masses ($7\times10^{12}M_{\odot}<M_{dyn}<4\times10^{14}M_{\odot}$) than the 3/11 galaxies in low-density environments that are found in pairs and groups ($7\times10^{11}M_{\odot}<M_{dyn}<7\times10^{12}M_{\odot}$).  For information we also indicate in Table~\ref{sample} whether the galaxies in groups are the central galaxy in their group (C), a satellite (S) or one of a pair of galaxies (P), where the group centre is defined following an iterative centre-of-light analysis \citep{robotham11}.  None of these galaxies are at the centre of a group.

\begin{table*}
 \centering
  \caption{Properties of the observed sample of galaxies from GAMA (described in the text).  The galaxies are divided into the two environmental density bins: high density (top of table) and low density (bottom of table).}
  \label{sample}
  \begin{tabular}{llllllllll}
  \hline
GAMA ID&RA&Dec.&Stellar Mass& z$_{\rm{TONRY}}$ &$\Sigma_5$  &Group Mass&   Central? &     R$_{e,r}$  &Sersic n$_r$\\
&J2000&J2000&Log(M$_\odot$)&&Mpc$^{-2}$ & Log(M$_\odot$)&&$^{\prime\prime}$ & \\
\hline

  136624&  11:43:17.06& -01:38:39.0 & 10.30  & 0.0463   &   16.33    &      13.8  & S &	2.24   & 	2.4    \\
  220328& 12:04:16.65& +01:32:46.5 &  9.83   & 0.0221    &  9.91         &  13.8 	& S	 & 5.20 & 	 	2.5 \\
  618152&  14:18:05.49& +00:13:38.6&  10.03  & 0.0543     & 8.25          & 14.6  & S& 	3.56 	 & 	0.9    \\
  227278&  14:11:19.16& +01:18:34.3&  10.13  & 0.0259     & 8.06       &    12.8   & S&	2.05 	  & 	2.6   \\
  600916&  09:12:06.92& +00:20:12.7& 10.05  & 0.0549     & 7.71     &      12.9  & S&	5.50 	 & 	1.1  \\
  136880&  11:45:20.90& -01:48:46.1 &  9.78    &  0.028     &  6.04    &       13.8   & S&	3.79 	 & 	1.6   \\
  600978&  09:12:45.34& +00:20:24.9 &  9.91   & 0.0549     & 4.89       &     12.9       & S&   	5.08 	 & 	0.8  \\
   \hline
  422359&   08:42:13.17& +02:37:28.6&  10.10  & 0.051     &  0.71    &       -            &    &  	2.47 	 & 	1.6  \\
  106252&   14:21:05.39& +00:51:54.3&  9.83  & 0.0550     &  0.65    &       -            &      &	2.34 	 & 	0.6 \\
  227962&  14:22:01.09& +01:11:50.3&  9.88   & 0.0558     & 0.42       &    12.8 	 & S	& 2.02	  & 	0.9  \\
  92770&     14:30:14.98& +00:37:17.0&   9.87    & 0.0271     & 0.31      &     -             &   &  	3.28   & 	1.2  \\
  418448&   09:04:50.24& +02:30:23.3&  10.02   & 0.0557    &  0.26    &       -          &      & 2.03 	 & 	2.5 \\ 
  375909&   08:45:32.05& +01:17:36.0&  9.96  & 0.0450      & 0.14       &    12.9 	 &  P	&4.23 	 & 	0.8     \\
  536005&   12:00:00.48& -01:01:40.7& 10.09 &  0.0483     & 0.12     &      -            &     & 2.62 	 & 	1.4 \\
  535319&  11:49:15.69& -00:58:36.9&   9.89   & 0.0606     & 0.09        &    11.9  & P	&4.08 	 & 	0.8     \\
  55150&    12:03:01.01& -00:17:28.2&  10.09  & 0.0417     & 0.06        &   -             &     &	4.47  &  	0.7  \\
  371177&  08:41:39.24& +00:58:26.7&   10.03 &  0.0608     & 0.06       &     -          &      &  2.63 	 & 	1.7  \\
  583637&  11:43:17.98& -00:10:53.8&   10.01 &  0.0577     & 0.03       &    -          &     &	 3.28  & 	1.2  \\
 \hline
\end{tabular}
\end{table*}

\section{Observations}
\label{sect:observations}
The data were taken in 2011 April,  2012 February and 2012 May with the SPIRAL IFU. SPIRAL is a $32\times16$ element rectangular microlens array coupled via an optical fibre feed to the dual-beam AAOmega spectrograph \citep{saunders04, sharp06}.  It has a spatial sampling of $0^{\prime\prime}.7$/spaxel with no gaps, giving a field of view of  $22^{\prime\prime}.4 \times 11^{\prime\prime}.2$.  We observed with the low-resolution 580V grating in the blue and the higher resolution 1000R grating in the red.  These settings correspond to wavelength ranges of $3700-5700\rm{\AA}$ and $6200-7300\rm{\AA}$ and spectral resolutions of 1900 and 5000 respectively. Accounting for the sample redshift range, this targets the emission lines H$\beta$ in the blue and H$\alpha$ in the red.  Observations were made during dark time, with an average seeing of $1.5^{\prime\prime}$ (FWHM). Each galaxy was observed for $3\times2400$s with individual observations dithered by 1-2 spaxels in Right Ascension and Declination in order to avoid four isolated dead elements in SPIRAL.  Spectrophotometric standard stars were also observed each night, in order to prepare a sensitivity function.  

Initial data reduction, from raw detector output to dark-subtracted, bias-subtracted, wavelength-calibrated, sky-subtracted, 1D-extracted spectra, was achieved using the \texttt{2dfdr} pipeline \citep{croom04}.  The root mean square dispersion around the wavelength solution is $0.12\rm{\AA}$ in the blue and $0.03\rm{\AA}$ in the red spectra.  The dispersion around the $5577\rm{\AA}$ sky line is $0.09\rm{\AA}$.  Twilight flat-field frames were also observed in order to account for relative fibre-to-fibre transmission variations.  As none of our targets completely fill the SPIRAL field-of-view a sky background spectrum was calculated by taking the median over pixels without galaxy light.  The final data analysis was carried out using custom IDL routines.  The sensitivity function determined from comparing the total observed flux from spectrophotometric standard stars to that predicted as a function of wavelength was applied.  The final flux calibration was done by applying SDSS $g$- and $i$-band fiber magnitudes (measured in the $3^{\prime\prime}$ SDSS fibre) to the blue and red SPIRAL spectra integrated over a $3^{\prime\prime}$ aperture respectively for each galaxy.  The calculated offset was then applied to the individual SPIRAL spectra.  Comparison with the flux-calibrated SDSS spectra indicates a 6 per cent uncertainty in the flux calibration level of the blue spectra (covering the H$\beta$ line) and 10 per cent in the red spectra  (covering the H$\alpha$ line).  Following flux calibration individual frames were aligned and mosaiced using telescope offset information. Frames are scaled based on a comparison of overlap regions in the mosaic, to account for minor variations in transparency and seeing.

\section{Emission-Line Measures}
\label{sect:emissionlines}

In order to examine the radial distribution of star formation, high signal-to-noise spectra were produced by combining spectra within annuli for each galaxy. Each annulus is defined as a radial de-projection of the galaxy, based on position angle and inclination information derived from GAMA analysis of SDSS imaging data \citep{kelvin12}. Prior to stacking, each spectrum from individual spaxels is velocity matched based on a velocity fields derived from emission and absorption fits to a block-averaged (3x3 spaxel) data-cube. The [N\textsc{ii}], H$\alpha$, [N\textsc{ii}] region is fitted with a four component model, three emission lines and H$\alpha$ absorption, assuming common velocities and widths for the emission components and with the ratio of the flux of the nitrogen lines fixed at 3.28. The emission redshift is allowed to float with respect to that of the absorption line. Errors are estimated from a quadrature summation of the statistic values returned for the best-fit model and the parameter distribution from a bootstrap re-sampling of each composite spectrum.

Individual Gaussian fits to the H$\alpha$ and H$\beta$ emission lines in each spaxel of each SPIRAL data-cube were also made for examination purposes. The resulting H$\alpha$ flux and velocity maps as well as SDSS thumbnail images for the same field-of-view are shown for each galaxy in Appendix~\ref{maps}.  The maps are presented in order of environmental density (highest density environment first).  There are some interesting features in the H$\alpha$ flux maps including off-centre (600978, 92770, 371177) and clumpy emission (618152, 535319, 55150, 583637).  \cite{gerssen12} also observe clumpy emission in their VIMOS IFU observations of SDSS galaxies.  Two of the off-centre and clumpy emission features are in galaxies in high-density environments (618152 and 600978; $29^{+20}_{-11}$ per cent), while the remaining 5 are in galaxies in the low-density environments ($45^{+15}_{-13}$ per cent).  We conclude that off-centre and clumpy H$\alpha$ emission does not significantly depend on environment in our data. 

The H$\alpha$ velocity maps (Appendix~\ref{maps}) show that all galaxies with strong H$\alpha$ emission show ordered rotation in that emission line, even when the emission is clumpy or off-centre.  This is consistent with observations at higher redshifts showing that clumpy galaxies are well fit by ordered disk models (e.g. \citealt{wisnioski11}).  

\section{Star Formation Rates}
\label{Sect:SFR}

We first examine the total star formation rates (SFR) of the galaxies.  The total SFR measurements are made by summing the obscuration-corrected H$\alpha$ flux ($f_{H\alpha}$ [ergs s$^{-1}$ cm$^{-2}$]) over the observed extent of the galaxy.  The flux is obscuration corrected using the Balmer decrement (BD; H$\alpha$ flux/H$\beta$ flux) measured in individual spaxels . The H$\alpha$ luminosity is then:
\begin{equation}
\rm{L}_{H\alpha} \rm{(W)}=4\pi d_L^2 ~f_{H\alpha} (BD/2.86)^{2.36}, 
\end{equation}
where $d_L$ is the luminosity distance in centimetres.  The Balmer decrement is a unitless obscuration sensitive parameter and its departure from the Case B recombination value of 2.86 indicates dust attenuation along the line of sight. While dust geometries are complex, this approach implicitly models the dust as a foreground screen averaged over the galaxy \citep{calzetti01}.  The exponent in the dust obscuration correction factor is equal to $k(\lambda_{H\alpha})/[k(\lambda_{H\beta}) - k(\lambda_{H\alpha})$], and $k(\lambda)$ at a given $\lambda$ is determined from the \cite{cardelli89} Galactic dust extinction curve (derived from observations of the UV extinction of stars). This is found to well describe the obscuration of the ionised gas in star-forming galaxies \citep{calzetti01,gunawardhana11}.  The SFRs are then calculated using the relationship given by \cite{kennicutt98} assuming a \cite{salpeter55} initial mass function (IMF), i.e. SFR $= \rm{L}_{H\alpha} \rm{(W)}/1.27\times10^{34}$.  These values are given in Table~\ref{SFRtable}.   

\begin{table*}
 \centering
  \caption{Spectral measurements from IFU observations and the GAMA survey.  Total H$\alpha$ fluxes and obscuration-corrected star formation rates (SFR) are given for the IFU observations. The GAMA measurements (described in the text) are made from SDSS stellar absorption-corrected single-fibre spectra.  The column `Class' details whether a galaxy is classified as an AGN by GAMA and whether it would be classified as Star Forming (SF) or Non-Star Forming (NSF) by Wijesinghe et al. (2012). The galaxies are divided into the two environmental density bins: high density (top of table) and low density (bottom of table). Galaxy 220328 only shows H$\alpha$ in absorption and Galaxies 136880 and 227278 are active galactic nuclei (AGN).}  
  \label{SFRtable}
  \begin{tabular}{lllllll}
  \hline
GAMA ID&f$_{H\alpha,\rm{IFU}}$ & SFR$_{\rm{IFU}}$&H$\beta$ EW$_{\rm{SDSS}}$ &BD$_{\rm{GAMA}}$ & SFR$_{\rm{GAMA}}$ & Class\\
&$10^{-17}$ergs s$^{-1}$ cm$^{-2}$ & $M_{\odot}$yr$^{-1}$& $\rm{\AA}$ &  & $M_{\odot}$yr$^{-1}$\\
\hline
      136624   &   266 &   0.02&      0.10	  & 1.50     &   0.01 &  NSF\\
      220328     & -  &-&-0.08   &  2.42      &    0.01 & NSF\\
      618152      &618 &    0.79&  4.21    &  4.57    &      1.58 &    SF\\
      227278      &85    &-&       0.66   & 3.37    &      -   &  AGN\\
      600916      &276    & 0.35&         1.46  &   5.09         &   0.79  &   NSF\\
      136880      &44    &-&      0.10  &  5.22 &         -  &   AGN \\
      600978      &1608  &    1.92&        6.92  &   3.96  &        2.14 &     SF\\
      \hline
      422359 &     1328      &1.62&         6.10  &   4.32  &        2.13  &    SF\\
      106252   &   165     &0.19&   3.23   &   5.13  &        0.64  &   SF\\
      227962     & 1930    &  2.36&  8.38   &   3.61   &       1.87   &   SF  \\
      92770     &366    &0.07   &  0.69   &  3.07      &     0.04  &  NSF     \\
      418448     & 793   &   1.03&   4.19  &    3.62     &     0.88    & SF    \\
      375909     & 2976   &   4.04&   8.43   &   4.85  &        5.09   &   SF\\
      536005      &2140     & 1.68&   8.76   &   4.57      &    4.11   &    SF\\
      535319      &1425     & 2.04& 6.42  &    3.75    &     1.72   &   SF\\
      55150     & 5630      &5.89&     5.91  &   4.27     &     4.26   &   SF \\
       371177     & 730      &1.37&  3.00  &    3.41    &      0.56   &  SF\\
      583637      &958      &1.42&   2.18   &   4.04   &       0.65   &  NSF\\
 \hline
\end{tabular}
\end{table*}

\subsection{Dependence on Environment}

\begin{figure}
\begin{center}	

\resizebox{18pc}{!}{\includegraphics{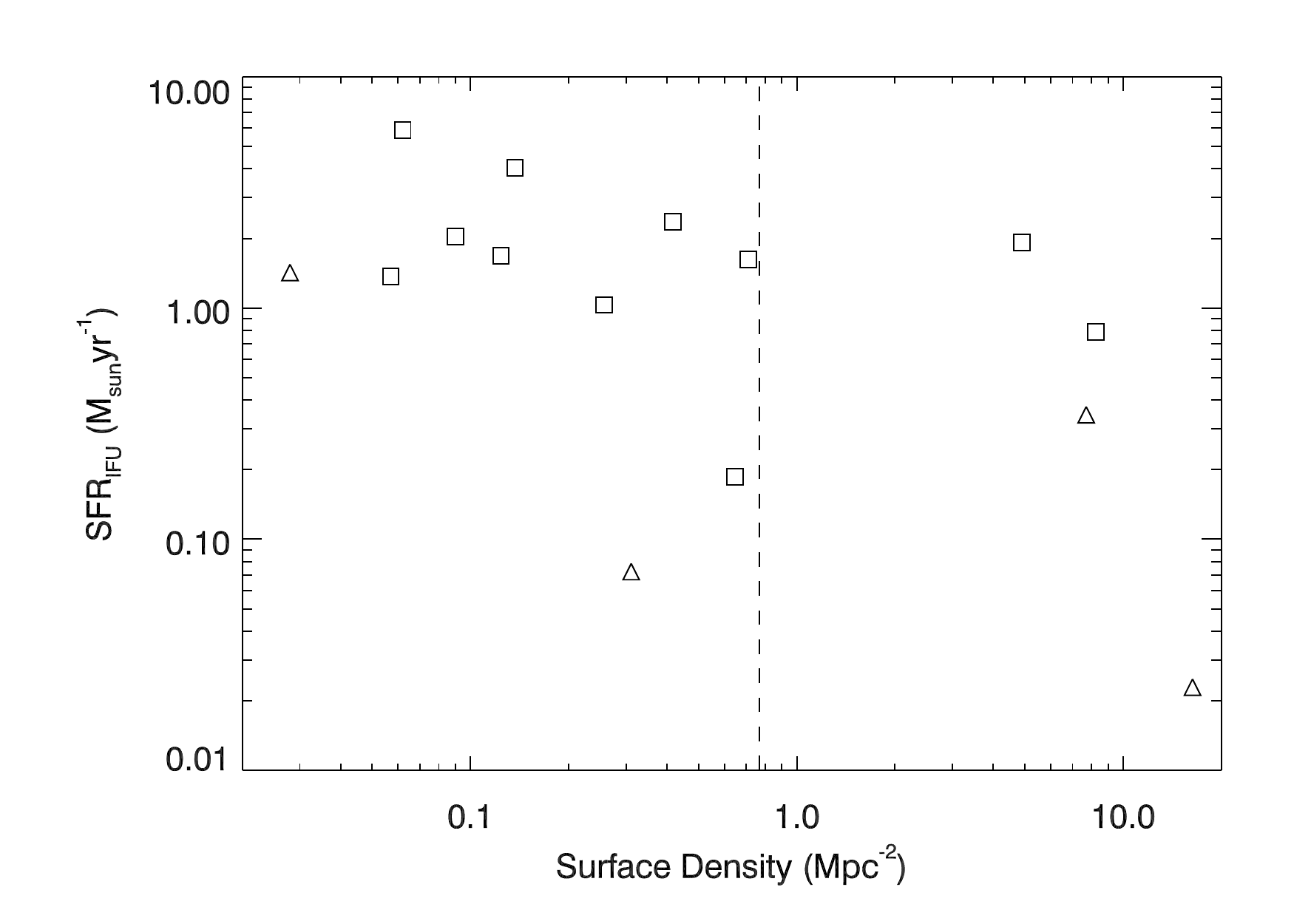} }

\end{center}
\caption{Total star formation rates of the star-forming galaxies as a function of environmental density.  The dashed line indicates the median surface density of 0.77 Mpc$^{-2}$ that divides the low and high-density subsamples analysed here.  There is a weak correlation between log10(SFR$_{\rm{IFU}}$) and log10($\Sigma_5$), significant only at the $1.8\sigma$ level.  We indicate which galaxies would have been classified by Wijesinghe et al. (2012) as star forming (squares) and non-star forming (triangles); some of those classified as non-star forming are still forming stars.  }
\label{SFR}
\end{figure}

We compare the total SFRs of the star-forming galaxies (SFR$_{\rm{IFU}}$) with their environmental density in the top panel of Figure~\ref{SFR}.  We note that the only absorption-dominated galaxy (galaxy 220328) in this sample is found in the highest density environment.  The observed mean SFR$_{\rm{IFU}}$ (low density)$=1.97\pm0.51$ is more than a factor of two higher than that at high density ($=0.77\pm0.42$). To determine the significance of an environmental dependence we apply a Kolmogorov-Smirnov two sample test to the SFRs in the low and high-density environments.  This gives a probability of 19.6 per cent that the two samples are drawn from the same parent population.  A Spearman rank correlation of the relationship between log10($\Sigma_5$) and log10(SFR), shows that the correlation between these parameters is only significant at the $1.8\sigma$ level.

Given the small size of our sample and the scatter observed we examine whether we can detect a significant relationship with environment.  We tested this by adjusting the mean SFRs of the galaxies in the high-density environments while maintaining their standard deviation and re-ran our statistical tests.  A weakly significant correlation between environment and SFR is observed (at a $2.3\sigma$ level) if the mean SFRs in high-density environments are a factor of 5 lower than those in the low-density environments (a factor of 2 lower than observed).  

We observe a weak but not significant relationship of SFR with environment in this sample.  Due to the small sample size and observed scatter we cannot draw a strong conclusion about a universal SFR dependence, or lack of, on environment.

\subsection{Comparison with Wijesinghe et al. (2012)}

We use the total SFR measurements made from our IFU observations to analyse the \cite{wijesinghe12} star-forming galaxy classification.  They define star-forming galaxies as those not classified as active galactic nuclei (AGN) using the \cite{kewley01} definition with non-absorption corrected H$\beta$ equivalent widths (EW) $> 1.5\rm{\AA}$, BD $< 15$ and SFR $> 10^{-3}M_{\odot}$yr$^{-1}$.  We use the single-fibre GAMA measurements to determine whether our sample would meet their star-forming classification.  

GAMA utilises the MPA/JHU emission-line catalogue\footnote{http://www.mpa-garching.mpg.de/SDSS/DR7/} to obtain absorption-corrected line fluxes and equivalent widths (EW) for these bright galaxies.  These line measurements are made from stellar absorption-corrected SDSS spectra.  The GAMA SFRs are calculated as described in \cite{hopkins13} and \cite{gunawardhana13} and are given in Table~\ref{SFRtable}.  In summary, the H$\alpha$ luminosity is calculated from the H$\alpha$ EW, which is aperture corrected and extinction corrected using the Balmer decrement as per the IFU measurements.  The GAMA SFRs are then calculated using the \cite{kennicutt98} relationship.  We note that dust obscuration is not excessive in any of the galaxies in this sample: their mean Balmer decrements are $3.9\pm0.9$ (Table~\ref{SFRtable}) which translates to a dust obscuration factor of $\sim2$.   

\cite{wijesinghe12} do not correct H$\beta$ equivalent widths for stellar absorption for the definition of their star-forming sample, however, they do correct H$\alpha$ equivalent widths by adding a constant correction of $0.7\rm{\AA}$.  We therefore use that addition here, i.e. stellar absorption corrected H$\beta$ EW$>2.2\rm{\AA}$, to determine whether our sample would meet their star-forming classification.  Table~\ref{SFRtable} shows that 2 of the galaxies in this sample are classified as AGN from the single-fibre analysis and 5 do not make the \cite{wijesinghe12} star-forming classification, based on their H$\beta$ EW.  Figure~\ref{SFR} shows that 4 of these `non-star forming' galaxies are still forming stars at some level.  These galaxies are a clear indication of the need to take care when separating galaxies into distinct star-forming and non-star-forming populations.

\subsection{Aperture Corrections}

\begin{figure}
\begin{center}	

\resizebox{18pc}{!}{\includegraphics{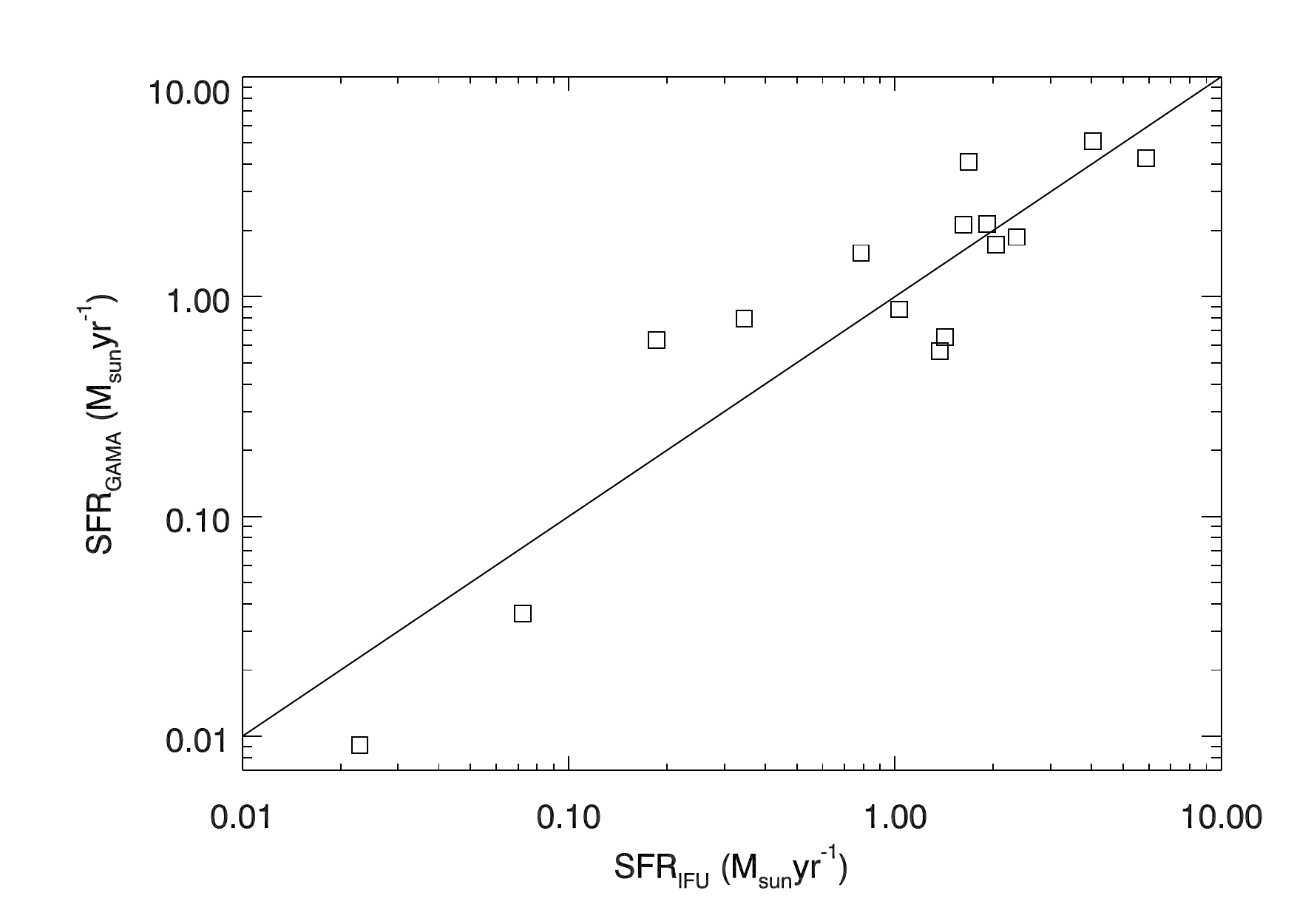} }

\resizebox{18pc}{!}{\includegraphics{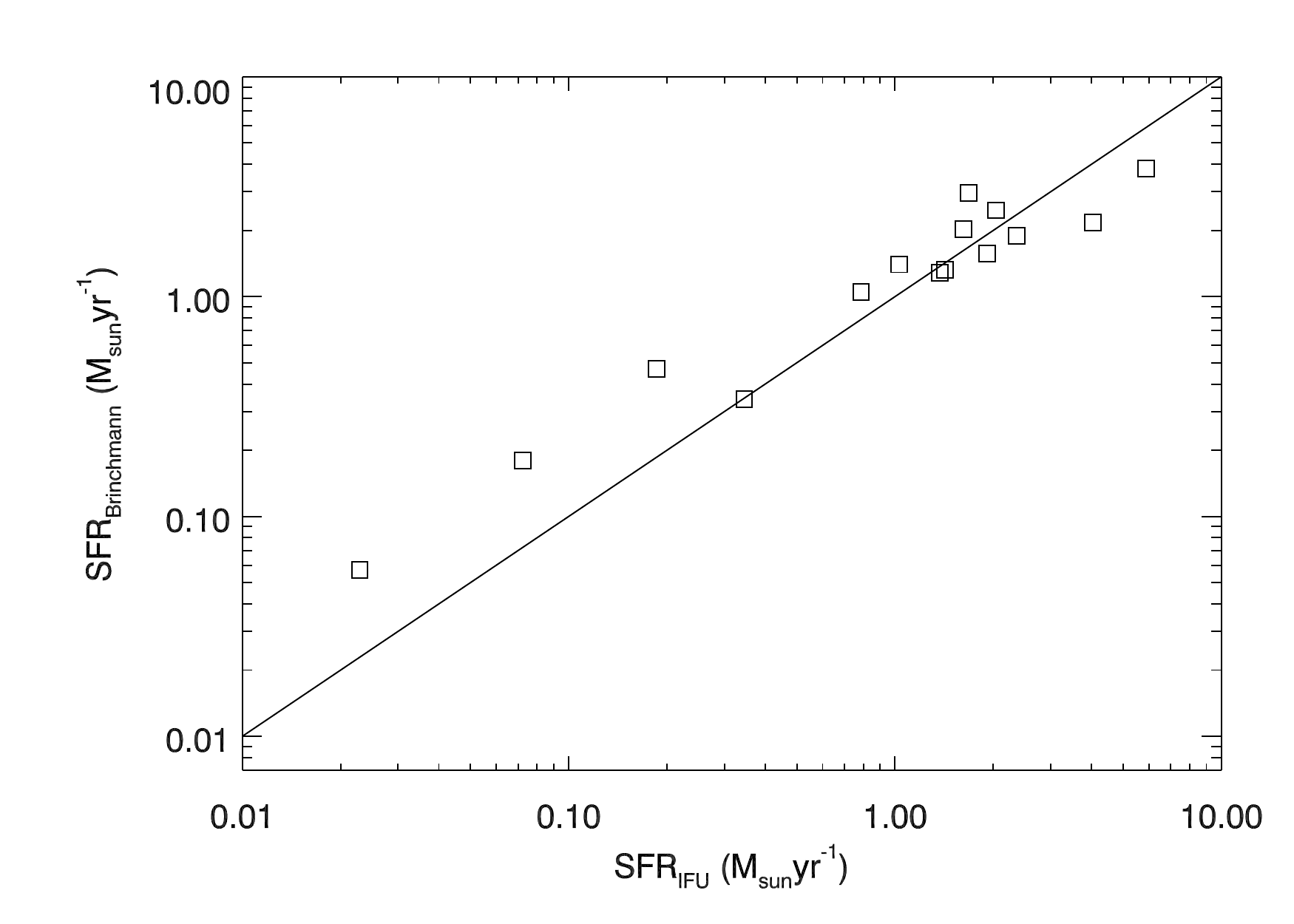} }	
\end{center}
\caption{Comparing total star formation rates from IFU and single-fibre observations of the star-forming galaxies.  The solid lines indicate the 1:1 relationships. The \emph{top panel} shows the comparison between the total IFU star formation rates and the total aperture-corrected GAMA single-fibre star formation rates.  The mean ratio is $\rm{SFR_{GAMA}} / {\rm SFR_{IFU}}=1.26\pm0.23$.  The \emph{lower panel} shows the comparison between the total IFU star formation rates and the total SDSS single-fibre SFR aperture-corrected as per Brinchmann et al. (2004).  The mean ratio $\rm{SFR_{Brinchmann}} / \rm{SFR_{IFU}}=1.34\pm0.17$. The aperture-corrected SFRs are in relatively good agreement with those from the IFU measurements, no matter which aperture correction method is used. However, the uncertainties can still be large for individual systems.} 
\label{aperture}
\end{figure}

Calculating the total SFR of galaxies from single-fibre observations requires a correction for the portion of the galaxy enclosed by the size of the fibre used: an aperture correction.  In the GAMA survey the total SFR is calculated by aperture correcting the H$\alpha$ flux measured within the fibre following the method of \cite{hopkins03}.  This aperture correction relies on the assumption that the line emission scales directly with the stellar continuum, as measured by the $r$-band magnitude.  However, there is obviously some uncertainty in that assumption.  We use the total SFR measurements made from our IFU observations to analyse this correction further.  The IFU SFR are compared to the GAMA SFR in the top panel of Figure~\ref{aperture}.  

We do not include either galaxy 220328 or the two AGN in this analysis due to a lack of observable star formation and AGN contamination respectively.  The mean difference $\rm{SFR_{GAMA}} / \rm{SFR_{IFU}}=1.26\pm0.23, \sigma=0.90$ i.e. SFR$_{\rm{GAMA}}$ is on average 26 per cent higher than SFR$_{\rm{IFU}}$ with a broad dispersion (the standard error on the mean is calculated as $\sigma/ \sqrt N$).  

\cite{gerssen12} analysed the aperture correction applied to calculate total SFR from SDSS spectra by \cite{brinchmann04} using IFU observations of 24 star-forming (H$\alpha {\rm EW}> 20\AA$, $f_H\alpha>448\times10^{-17}$ergs s$^{-1}$ cm$^{-2}$), low-mass ($1\times10^8M_{\odot}<M_{*}<3\times10^{10}M_{\odot}$) SDSS galaxies. They found the \cite{brinchmann04} aperture corrections to underestimate the total SFR by a factor of 2.5 with a dispersion of 1.75, significantly larger than the factor of 1.26 difference and 0.90 dispersion we find between our IFU SFR and the GAMA aperture-corrected SFRs.  We analyse whether the GAMA aperture correction is significantly different to that used by \cite{brinchmann04} by also comparing SFR$_{\rm{IFU}}$ with the most recent total SFR estimated by Brinchmann for the SDSS DR7 data release\footnote{These are the total values given in gal\_totsfr\_dr7\_v5\_2.fits available at: http://www.mpa-garching.mpg.de/SDSS/DR7/sfrs.html} in the bottom panel of Figure~\ref{aperture}.  

\cite{brinchmann04} determine total SFRs using a Bayesian approach to calculate the likelihood of fits of the observed spectrum to \cite{charlot01} models, which incorporate an obscuration model. They note that, to first approximation, the dust corrections are based on the H$\alpha$/H$\beta$ ratio.  The \cite{brinchmann04} total SFR are calculated with a \cite{kroupa01} IMF and we convert to the \cite{salpeter55} IMF used here by multiplying their measurements by 1.5. They aperture correct in an empirical manner using the distribution of the SFR/M* ratio at a given ($g-r, r-i$) colour and the photometry outside the fibre to correct the fibre SFR. This aperture correction is updated for the DR7 data release by calculating the light outside the fibre for each galaxy, and then fitting stochastic models to the photometry.

We find a mean difference $\rm{SFR_{Brinchmann}} / \rm{SFR_{IFU}}=1.34\pm0.17, \sigma=0.67$, i.e. we also find that the Brinchmann correction overestimates the SFR.  This is a much smaller difference than \cite{gerssen12} found.  This also suggests a marginal trend toward higher SFR estimates by aperture correcting using either method, although again with a high dispersion. These results suggest that contrary to the claim by \cite{gerssen12}, the aperture-corrected SFRs for these low-mass galaxies are in relatively good agreement with those estimated from the IFU measurements, no matter which method is used. The large dispersion does however mean that the uncertainties can still be large for individual systems. In addition, this is still only a small sample, and reliable statistics on total SFR estimates compared to those from aperture-corrected measurements will need a much larger sample.  

\section{Radial H$\alpha$ Profiles}
\label{sect:profiles}
We determine whether any dependence of SFR on environment is evident in the spatial distributions of H$\alpha$ emission in these galaxies.  

\begin{figure}
\begin{center}	
\resizebox{20pc}{!}{\includegraphics{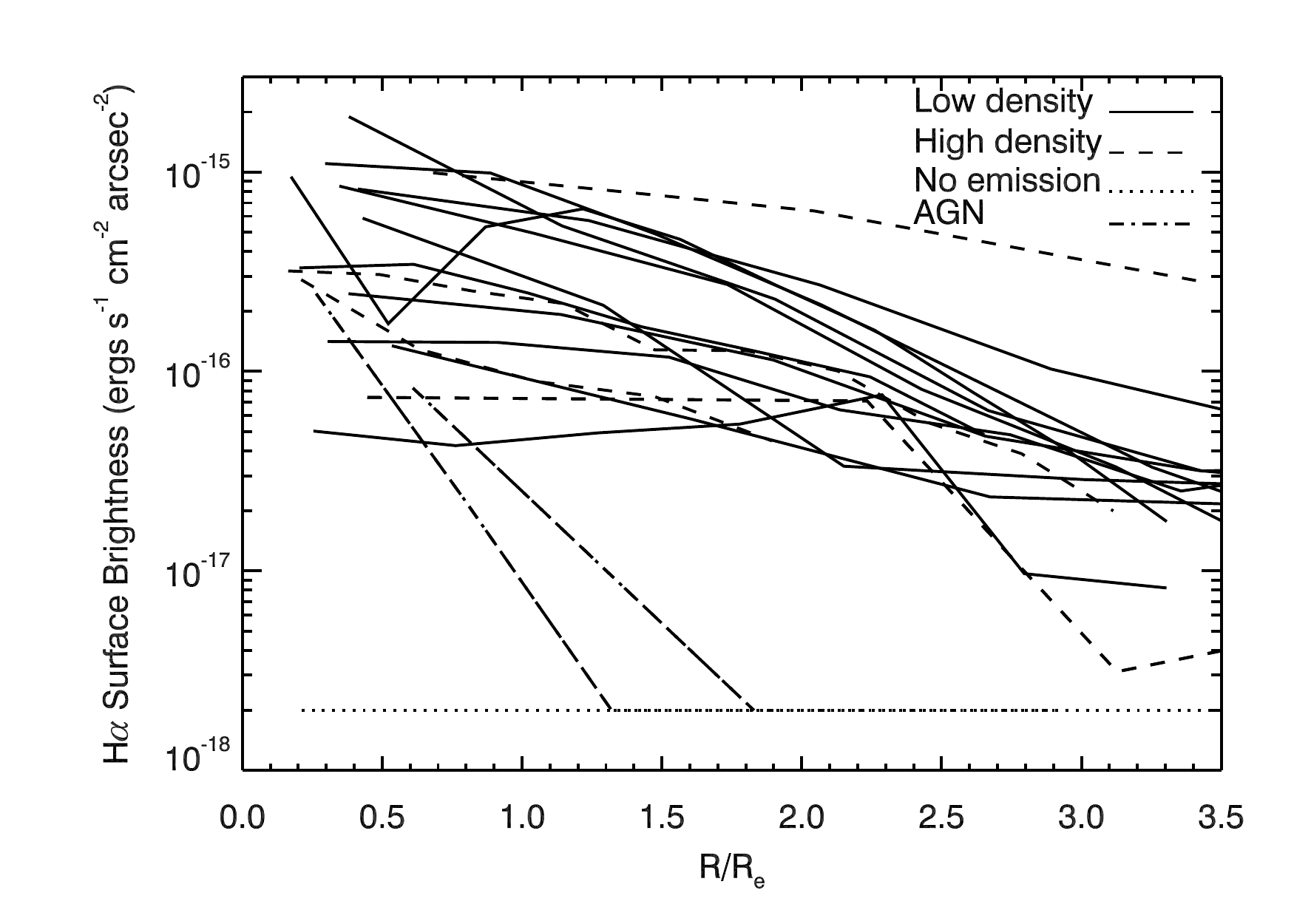} }
\end{center}
\caption{Radial H$\alpha$ surface brightness profiles as a function of the effective radius of the galaxy.  Solid lines show galaxies in low-density environments ($<0.77$ Mpc$^{-2}$) and dashed lines show galaxies in high-density environments ($>0.77$ Mpc$^{-2}$), dotted lines indicate regions that do not show emission above the detection limit of $2\times10^{18}$ ergs s$^{-1}$ cm$^{-2}$ arcsec$^{-2}$ and dot-dashed lines indicate the emission of the two AGN.  The average seeing approximates to R/R$_{\rm{e}}\sim0.2$.  The profiles do not show a dependence on environment. 
}
\label{profiles}
\end{figure}

The H$\alpha$ surface brightness is calculated by dividing the summed flux in each elliptical annulus by the area of the annulus.  These radial profiles are shown in Figure~\ref{profiles}.

Figure~\ref{profiles} shows that the radial H$\alpha$ surface brightness profiles of galaxies in both environments (solid and dashed lines) are very similar to one another, being centrally concentrated with high surface brightnesses over all radii studied.  In contrast, the surface brightness profiles of the galaxies with no emission above the detection limit of $2\times10^{18}$ ergs s$^{-1}$ cm$^{-2}$ arcsec$^{-2}$ (dotted lines) are only present in the highest density environments.  Of these, galaxy 220328 is dominated by absorption and the 2 AGN (136880, 227278) show central emission (dot-dashed lines) and no significant emission beyond that.

\begin{figure}
\begin{center}	
\resizebox{20pc}{!}{\includegraphics{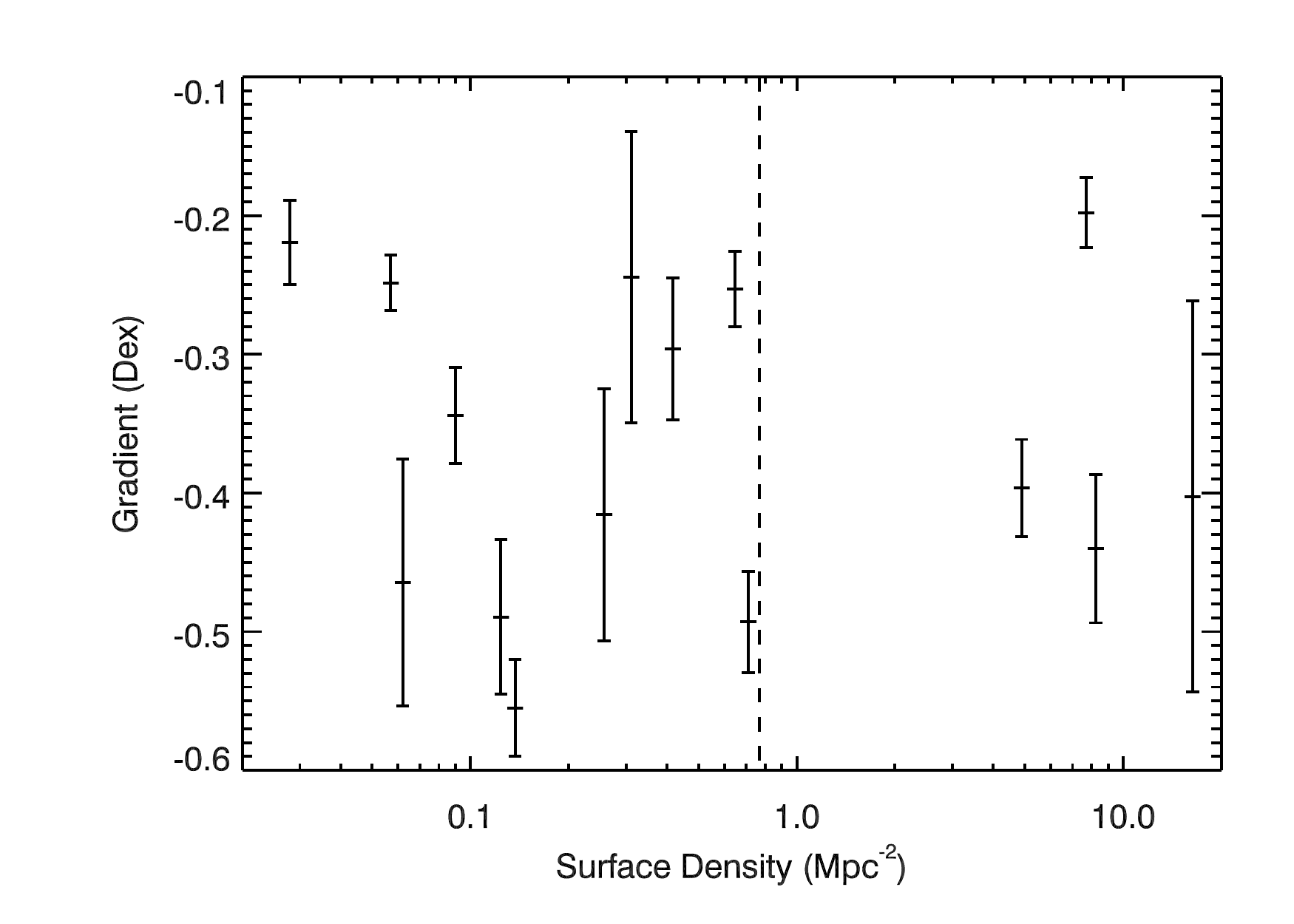} }

\resizebox{20pc}{!}{\includegraphics{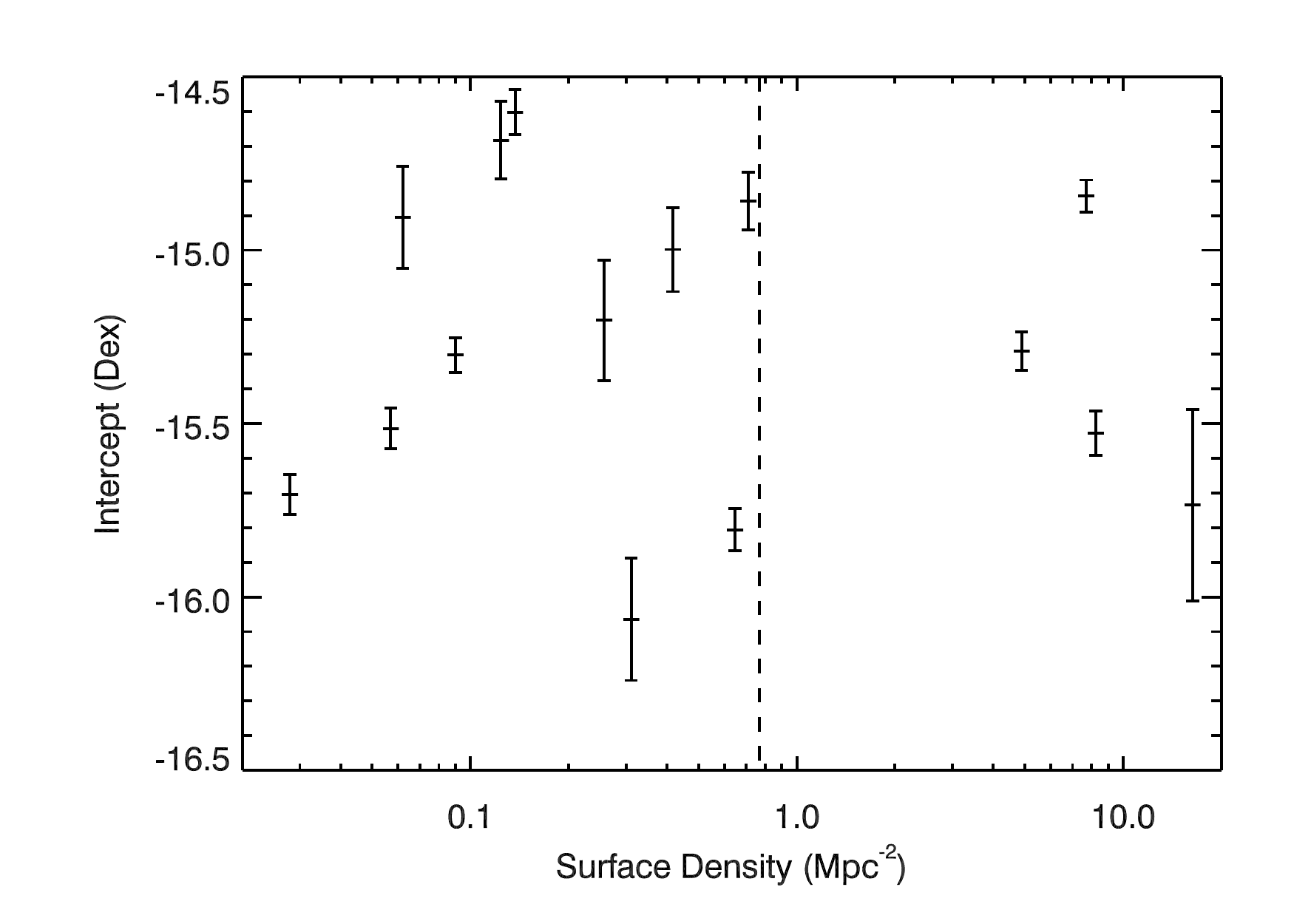} }
\end{center}
\caption{Parameters of straight-line fits to H$\alpha$ surface brightness profiles of the star-forming galaxies as a function of environmental density.  The \emph{top panel} shows the gradient of the fits. The gradients are correlated with environment at a significance of only $0.25\sigma$.  The \emph{lower panel} shows the fitted intercept.  The intercepts are correlated with environment at a significance of $0.4\sigma$.  The error bars show the $1\sigma$ uncertainties on the profile fits. The fits to the surface brightness profiles do not show a dependence on environment.}
\label{profile_fits}
\end{figure}

We further analyse the relationship of the profiles of the star-forming galaxies with their environment with straight-line fits to the H$\alpha$ surface brightness profiles taking into account the uncertainties in the H$\alpha$ flux measurements.  We do not include the 3 galaxies with undetected emission in this analysis.  We show the fitted gradient and intercept values and $1\sigma$ errors of the 15 star-forming galaxies as a function of environment in Figure~\ref{profile_fits}.  

We test the correspondence between these parameters with a Spearman rank correlation, finding that the gradients are correlated with log10($\Sigma_5$) at a significance of only $0.25\sigma$ and the intercepts are correlated at a significance of $0.4\sigma$.  There is no dependence of the radial distribution of the H$\alpha$ emission as a function of environment. 

The galaxies observed here all have very similar stellar masses and we are observing no significant dependence in either H$\alpha$ surface brightness profile shape or amplitude of the star-forming galaxies as a function of environment. 

\section{Discussion}
\label{sect:discussion}

We have presented observations of the spatially-resolved H$\alpha$ emission of galaxies over a wide range of environment with the aim of examining how the radial distribution of star formation varies as a function of environment.  We observe a weak but not significant difference in total star formation rate and no difference of radial profile of the star-forming galaxies' H$\alpha$ emission as a function of local galaxy environment in this sample of 18 galaxies with stellar masses $\sim10^{10}M\odot$.  

Before making general comments on the effect of environment on star formation based on our observations, we show that our sample is unbiased and representative of the broader galaxy population in this narrow stellar mass range.  The uncertainties given below are $1\sigma$ binomial errors \citep{cameron11}.  There are 424 galaxies in GAMA with accurate surface densities (i.e. not affected by survey edges) that have stellar masses, $6\times10^9M_{\odot}<M_{*}<2\times10^{10}M_{\odot}$, and redshifts, $0.02<z_{\rm{TONRY}}<0.06$.  We use an updated version of the \cite{wijesinghe12} classification (those not classified as AGN with stellar absorption-corrected H$\beta$ EW $> 1.0\rm{\AA}$, BD $< 15$ and SFR $> 10^{-3}M_{\odot}$yr$^{-1}$) to quantify star-forming galaxies.  Of the 424 galaxies, there are roughly equal numbers of AGN in each environmental density: 21/212 ($10^{+2}_{-2}$ per cent) in low-density environments and 19/212 ($9^{+2}_{-2}$ per cent) in high-density environments.  There are 147/212 ($69^{+3}_{-3}$ per cent) of galaxies in low-density environments that would make our star-forming galaxy criteria and 93/212 ($44^{+3}_{-3}$ per cent) in high-density environments.

In the sample observed here, in the low-density environments, there are no AGN and 11/11 ($100_{-14}$ per cent) of the galaxies make the updated star-forming galaxy criteria.   In the high-density environment there are 2/7 ($29^{+20}_{-11}$ per cent) AGN and a further 2/7 non-star-forming galaxies, meaning that 3/7 ($43^{+18}_{-15}$ per cent) make our updated star-forming galaxy criteria.  

We can conclude from this that the numbers of galaxies in this sample meeting the star-forming criteria in each environment are within $2.2\sigma$ of those in the broader sample.  This sample therefore follows the distributions of the general population.

The sample studied here is broadly representative of the general population and we find that the total SFRs of the star-forming galaxies do not depend significantly on their local environmental density (Figure~\ref{SFR}). However, due to the small size of the sample and the scatter observed we do not draw a definitive conclusion about a possible SFR dependence on environment.  Examining the spatial information provided by the H$\alpha$ surface brightness profiles, we observe no difference in amplitude or shape of the H$\alpha$ surface brightness profile of star-forming galaxies as a function of environment (Figures~\ref{profiles} and ~\ref{profile_fits}). We also find no evidence for `clumpiness' in H$\alpha$ emission depending on environment (Appendix~\ref{maps}).

The only comparable analysis to date examined the spatial star formation histories inferred from the colours of 44,964 galaxies in SDSS \citep{welikala08}.  They found that the mean star formation rate of each galaxy as a function of radius is dominated by star formation in the central regions of galaxies, and that the trend for suppression in high-density environments is driven by a reduction in that central star formation. They also find that the mean star formation rate in the outskirts is independent of environmental effects.  \cite{welikala08} conclude that the environment itself cannot suppress the star formation as the outer regions should otherwise be most affected and therefore this points to an evolutionary or AGN feedback origin.  We do not observe any significant radial dependence of the surface brightness of H$\alpha$ emission as a function of environment, either centrally or in the outer regions.  However, the suppression observed by \cite{welikala08} in their highest star-forming galaxies (SFR$>1.02M_{\odot}$yr$^{-1}$) is a factor of $<2$, of the order SFR$\sim0.002M_{\odot}$yr$^{-1}$.  This difference is significantly smaller than we can detect with these observations so we cannot rule out the suppression they observe.  \cite{welikala09} considered the density-morphology relation in the same sample, observing the strongest relation in the lowest luminosity galaxies with the highest star formation rates.  However, they conclude that it cannot solely explain the observed suppression of star formation in galaxies in high-density environments.  Table~\ref{sample} shows that 3 out of the 4 galaxies with Sersic $n_r>2$ are in high-density environments, of which 2 are not star forming. The mean $n$(high density)~$=1.7\pm0.3$ and (low density)~$=1.2\pm0.2$, suggesting that there are signs of a difference in morphology as a function of environment in our sample but that it alone does not explain the lack of significant dependence of star formation on environment. 

If these observations are borne out in larger samples, then combined with the known decreasing fraction of star-forming galaxies as a function of increasing environmental density (e.g. \citealt{balogh04,baldry06,bolzonella10, mcgee11, wetzel12, wijesinghe12}) and the small numbers of galaxies in transition between star forming and non-star forming observed in large samples (e.g. \citealt{wijesinghe12, mendel13}), this would suggest that if environment does drive the change in fractional contribution it must either act very rapidly (the `infall-and-quench' model; e.g. \citealt{balogh04,bamford08,nichols11,wetzel13}) or have occurred a long time ago due to density-dependent evolution (an `in-situ evolution' model; \citealt{wijesinghe12}), such that galaxies in transition are rare at this time. In-situ evolution would involve galaxies in dense environments evolving faster than galaxies in low-density environments, building their stellar mass faster and earlier, leading to the observed morphology-density relation, and consistent with the measured SFR-density relations at both low and high redshift.  The in-situ evolution model is similar to `downsizing' \citep{cowie96}, `staged evolution' \citep{noeske07} and the `mass quenching' model of \cite{peng10}, however, galaxies of common mass would evolve differently in different environments in order to give rise to the observed population mix.  Transition redshifts, at which the dependence of the specific star formation rates of galaxies on increasing environmental density transitions from increasing to decreasing, have been observed (e.g. \citealt{elbaz07,greene12}) giving weight to this argument.

We note that there are some caveats to this argument.  Firstly, this is a small sample. A larger sample would increase the robustness of our results.  Our sample also does not reach the densest cluster environments where galaxies are observed to be affected by ram pressure stripping \citep{owers12,merluzzi12} and tidal distortions (e.g. \citealt{moss93, vogt04, bretherton13}).  Our choice of environmental metric may play a role as the SFRs of galaxies at a fixed stellar mass have been observed to increase as a function of increasing cluster-centric radius, rather than environmental density \citep{vonderlinden10,rasmussen12,haines13}.  In addition, \cite{vandenbosch08} argue that the relationship between SFR and environment is driven by evolutionary differences between central and satellite galaxies in a dark-matter halo and we are only studying satellite galaxies here. We will examine the effects of our choice of environmental metric in a forthcoming paper (Brough et al., in prep.).  Dust may also play a role as \cite{koyama13} find dustier galaxies in higher-density environments mask a trend of increasing specific SFR with increasing environmental density.

To separate the two scenarios of `infall-and-quench' and `in-situ evolution' and address these caveats requires observations of a very large sample of galaxies, covering a broad range of stellar mass and environment, in order to place stringent limits on the number density of any transition galaxies.  It will be crucial to detect the very faintest levels of star formation present, as well as its spatial dependence. This will require very high signal-to-noise IFU spectra to enable careful decomposition of emission and absorption contributions \citep{sarzi06}.  The new Sydney Australian Astronomical Observatory (AAO) Multi-object-IFU (SAMI; \citealt{croom12}) instrument with 13 deployable IFUs over a 1 degree field-of-view, and associated survey will enable this crucial next step.

\section{Conclusions}
\label{sect:conclusions}
We present observations of the spatially-resolved star formation as a function of local environment from optical integral field unit (IFU) observations of 18 galaxies with stellar masses $M_*\sim10^{10} M\odot$ selected from the GAMA survey.  Our conclusions can be summarised as:

\begin{itemize}

\item{The total star formation rates measured from the IFU data are consistent with the total aperture-corrected star formation rates measured from both the GAMA and SDSS surveys.  The mean differences are $\rm{SFR_{GAMA}} / \rm{SFR_{IFU}}=1.26\pm0.23, \sigma=0.90$; $\rm{SFR_{Brinchmann}} / \rm{SFR_{IFU}}=1.34\pm0.17, \sigma=0.67$.}

\item{Off-centre and clumpy H$\alpha$ emission does not depend on environment.  It is present in 2/7 ($29^{+20}_{-11}$ per cent) galaxies in high-density environments and 5/11 ($45^{+15}_{-13}$ per cent) galaxies in low-density environments show similar features.}

\item{In this sample, we see weak but not significant evidence of a dependence of total star formation rate on environment, using IFU observations for the first time. }

\item{We observe no clear environmental trend on the amplitude or shape of the radial profile of H$\alpha$ emission. This implies that, for this sample, there is no strong outside-in or inside-out quenching.}

\item{The lack of dependence of the radial profile of H$\alpha$ emission shape or amplitude on environment suggests that if environment drives the known change in fractional contribution of star-forming galaxies in different environments, it must either act very rapidly (the `infall-and-quench' model) or galaxies must evolve in a density-dependent manner (an `in-situ evolution' model), to explain the lack of transition galaxies observed in large samples.}

\end{itemize}

In order to identify more precisely how and when any transition due to environment occurs requires high signal-to-noise, spatially-resolved spectra as well as a very large sample that covers a range in stellar mass, environment and star formation stage, including post-starburst galaxies. The new Sydney Australian Astronomical Observatory (AAO) Multi-object-IFU (SAMI; \citealt{croom12}) instrument will address this with its associated survey.

\section*{Acknowledgments}

We thank the anonymous referee for their comments which greatly improved the paper.  SMC acknowledges the support of the 
Australian Research Council via a Future Fellowship (FT100100457).  GAMA is a joint European-Australasian project based around a spectroscopic campaign using the Anglo-Australian Telescope. The GAMA input catalogue is based on data taken from the Sloan Digital Sky Survey and the UKIRT Infrared Deep Sky Survey. Complementary imaging of the GAMA regions is being obtained by a number of independent survey programs including GALEX MIS, VST KIDS, VISTA VIKING, WISE, Herschel-ATLAS, GMRT and ASKAP providing UV to radio coverage. GAMA is funded by the STFC (UK), the ARC (Australia), the AAO, and the participating institutions. The GAMA website is http://www.gama-survey.org. 


\appendix

\section[]{H$\alpha$ emission-line maps}
\label{maps}
\begin{figure*}
\begin{center}	
\resizebox{35pc}{!}{\includegraphics[angle=90,scale=2.65,trim=-0.8cm 0cm 0cm 0cm,clip]{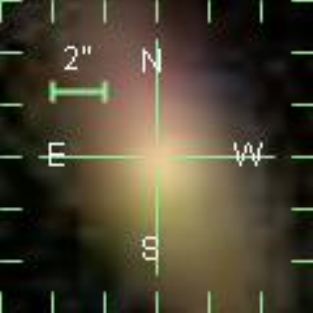} \includegraphics{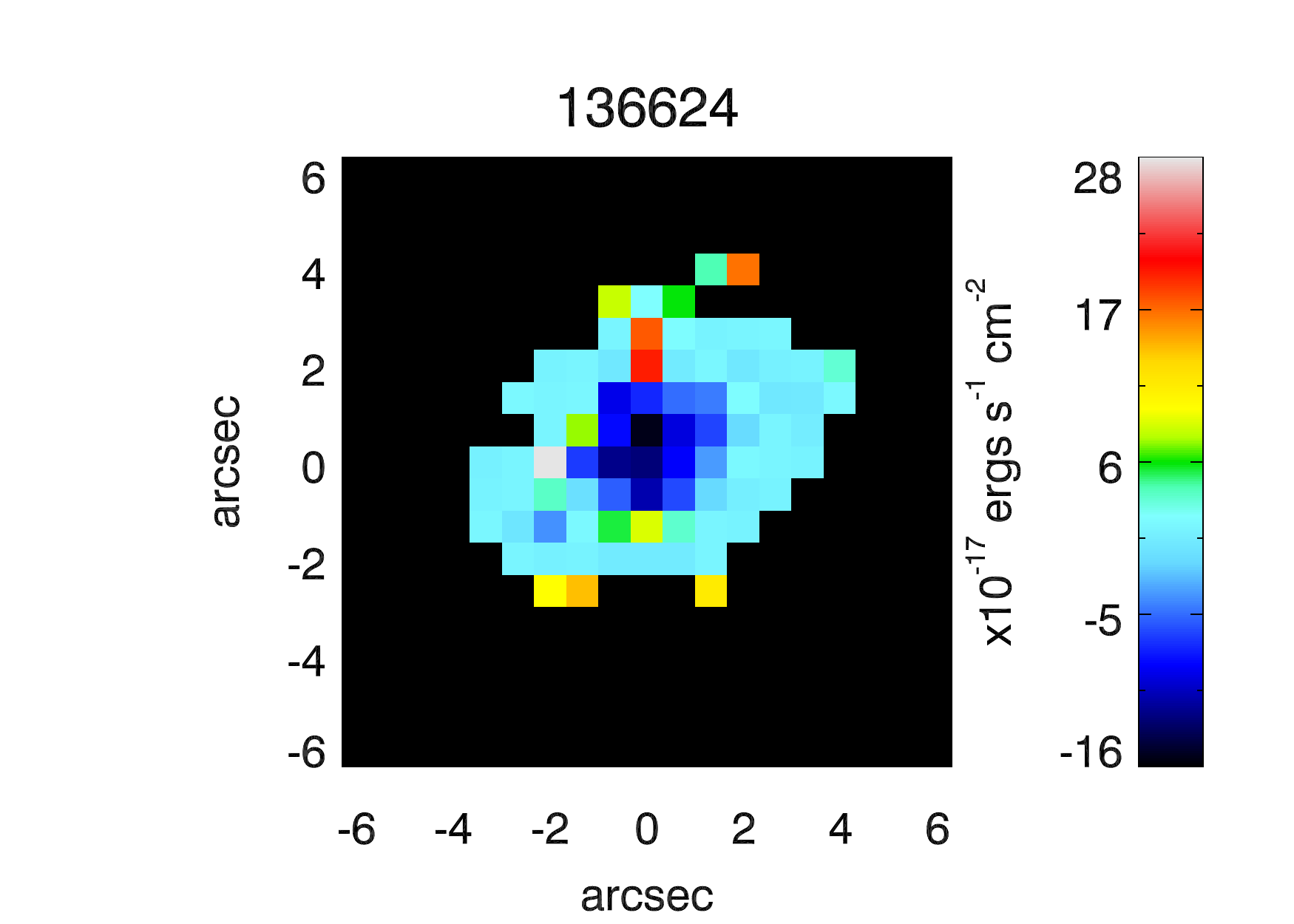} \includegraphics{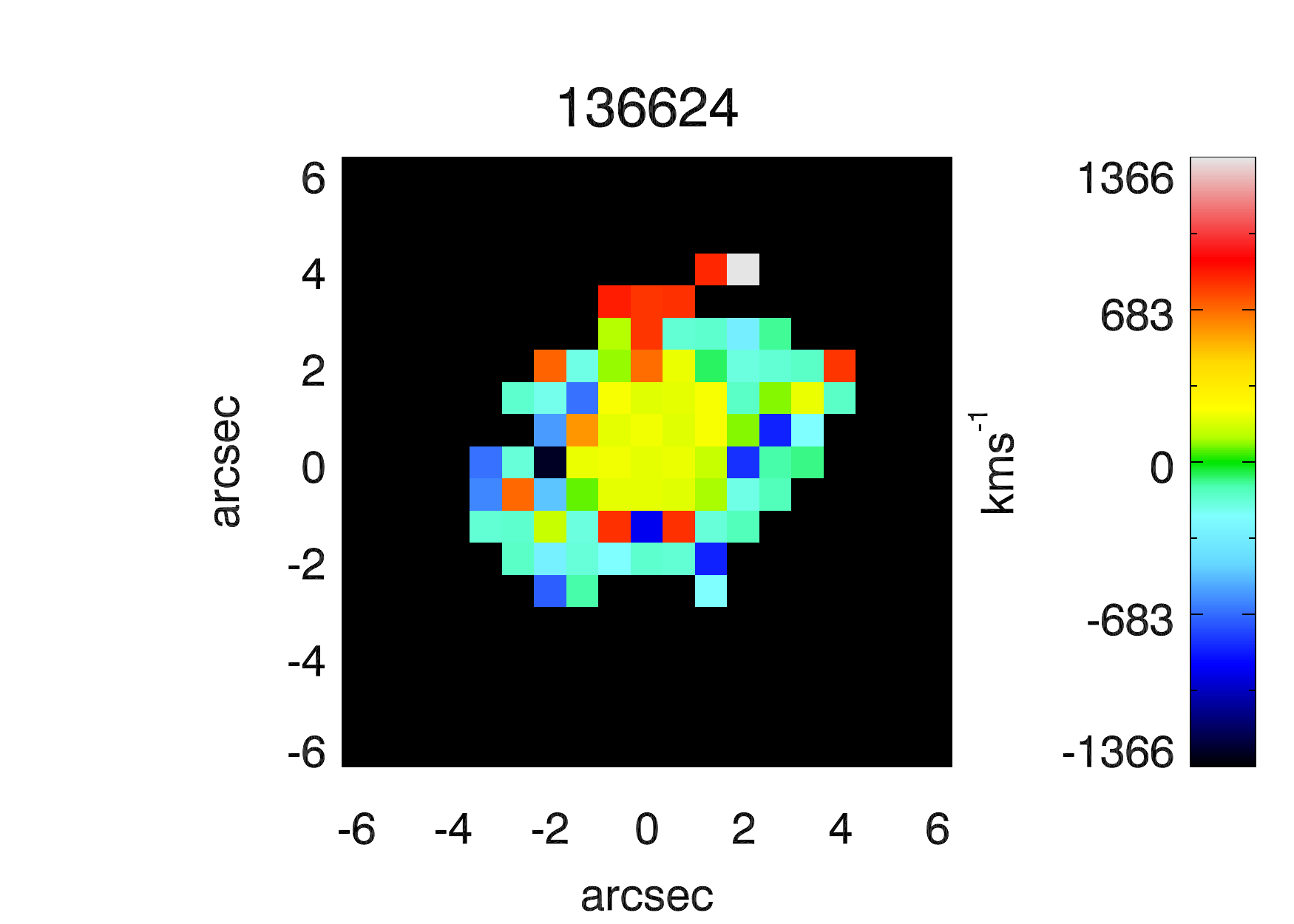} }
\resizebox{35pc}{!}{ \includegraphics[angle=90,scale=2.65,trim=-0.8cm 0cm 0cm 0cm,clip]{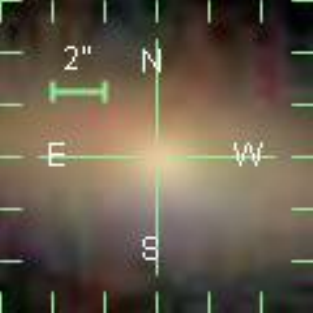} \includegraphics{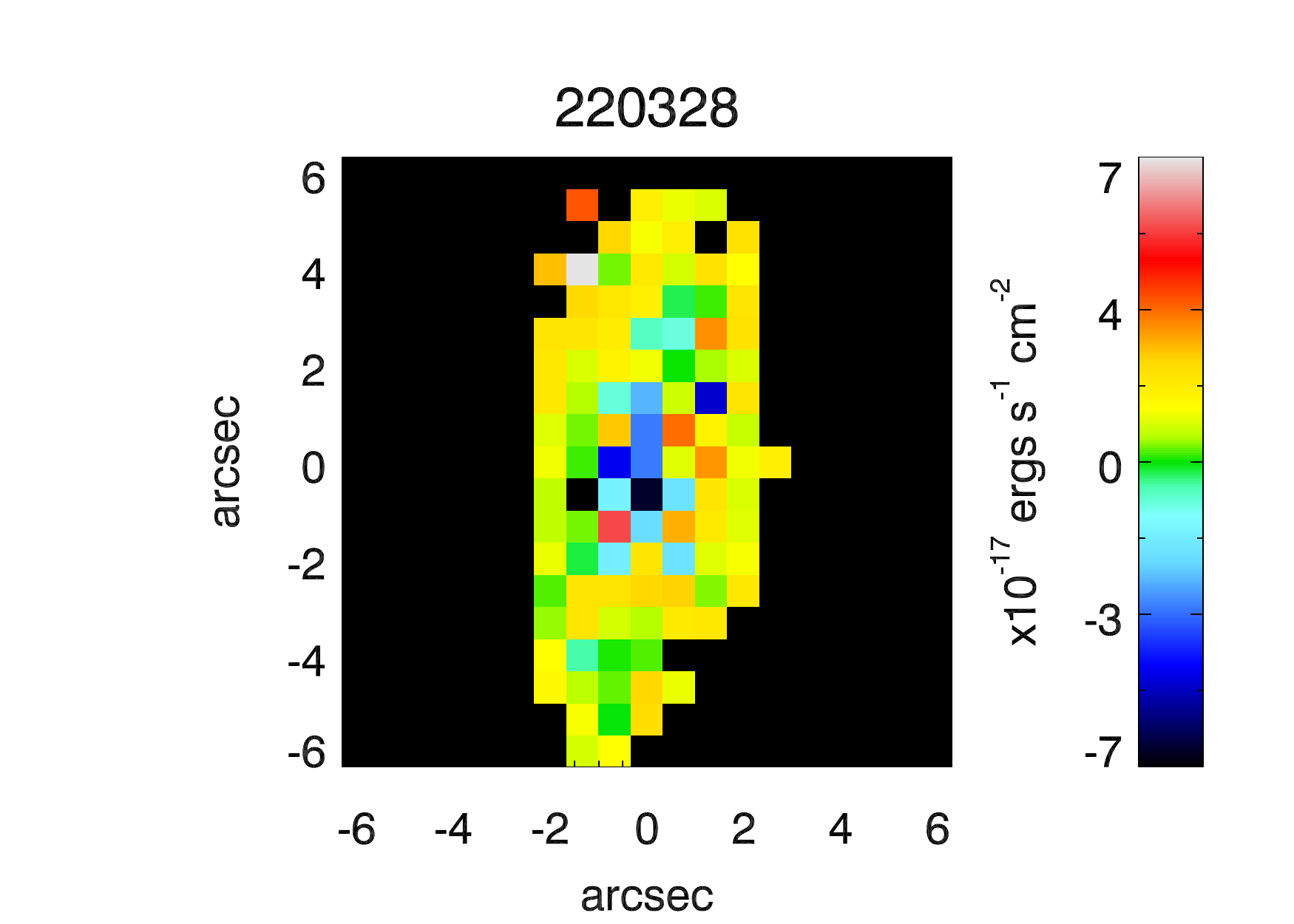} \includegraphics{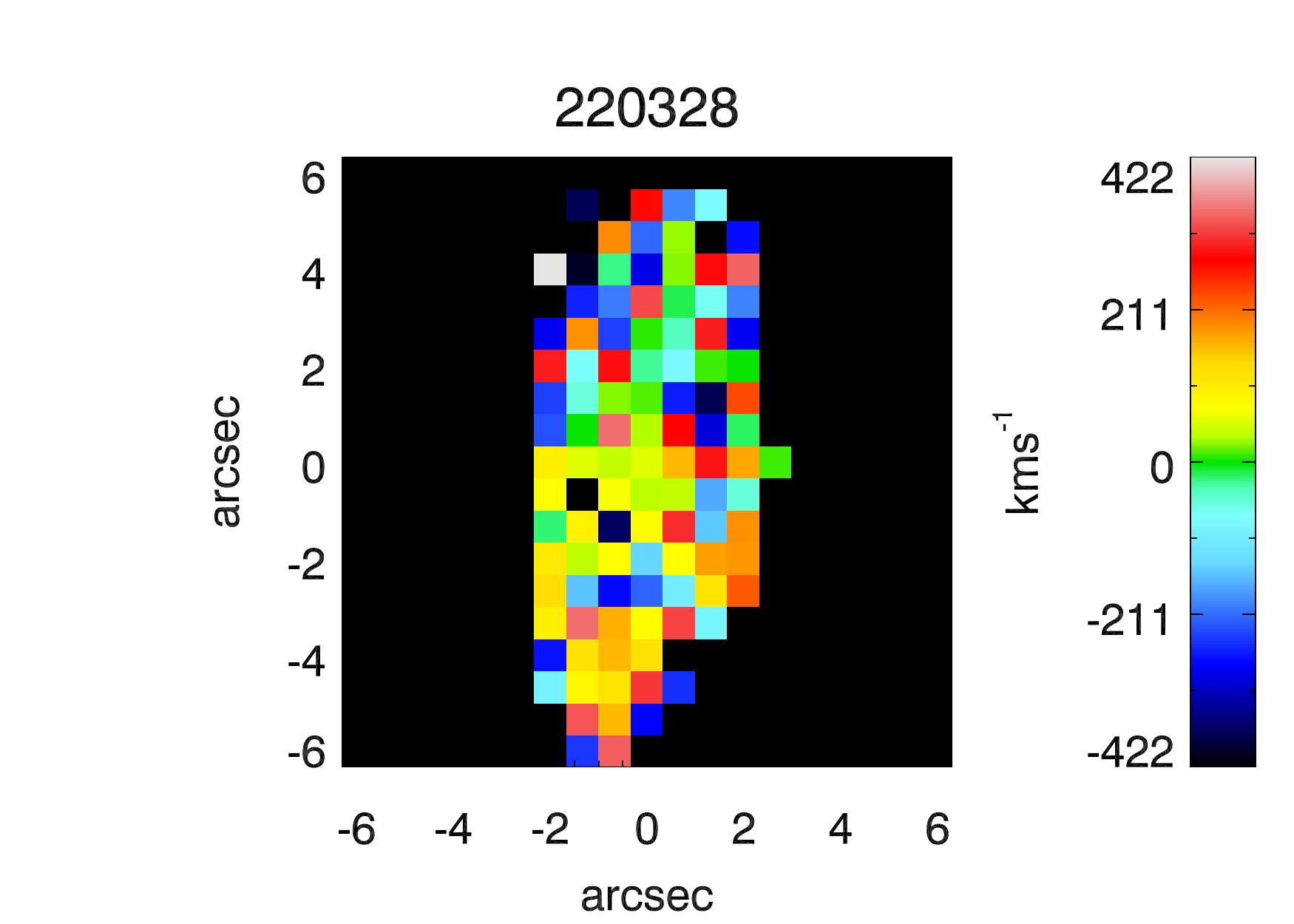} }
\resizebox{35pc}{!}{\includegraphics[angle=90,scale=2.65,trim=-0.8cm 0cm 0cm 0cm,clip]{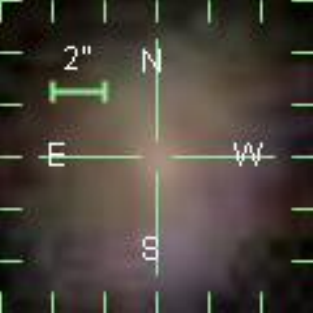}  \includegraphics{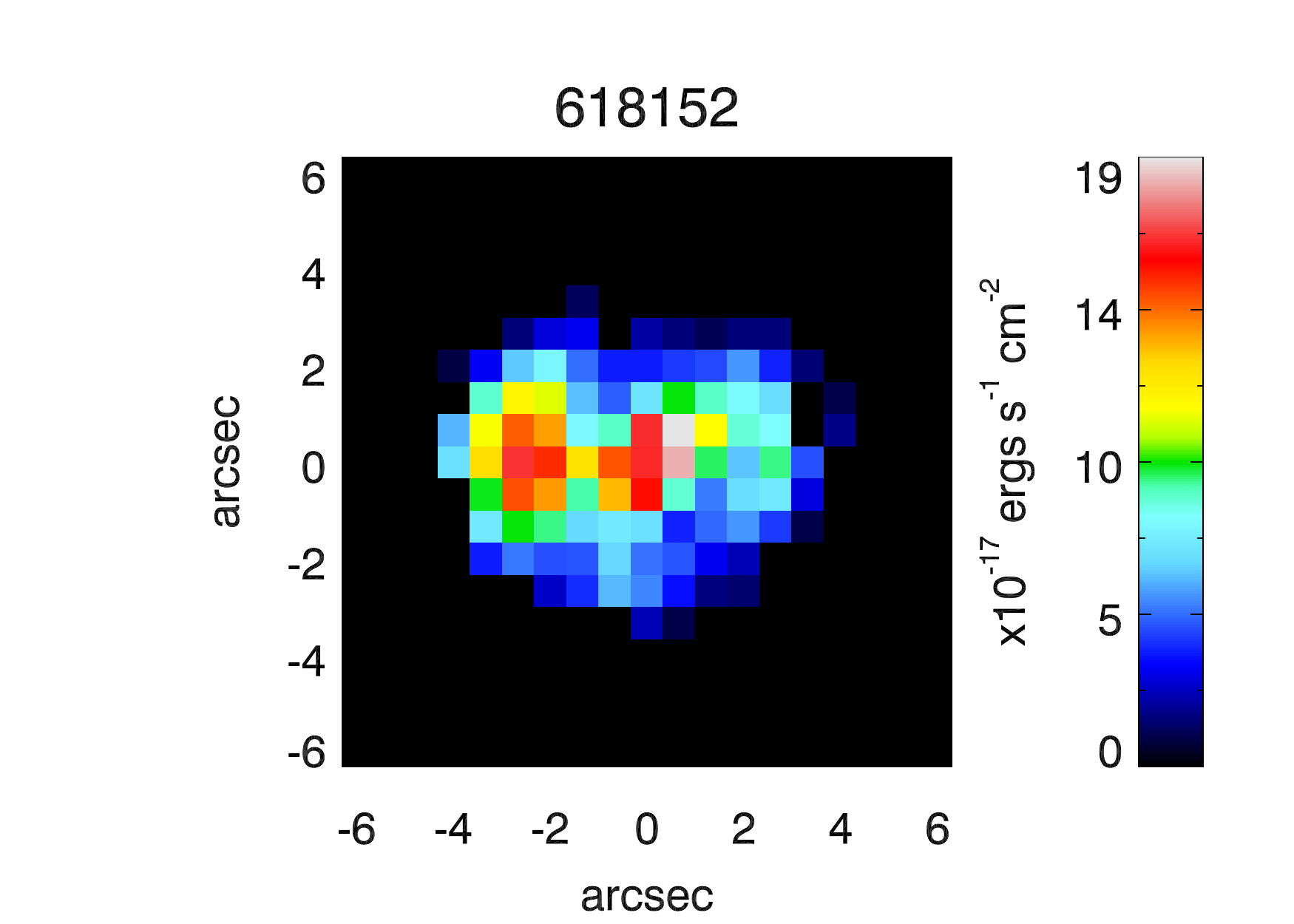} \includegraphics{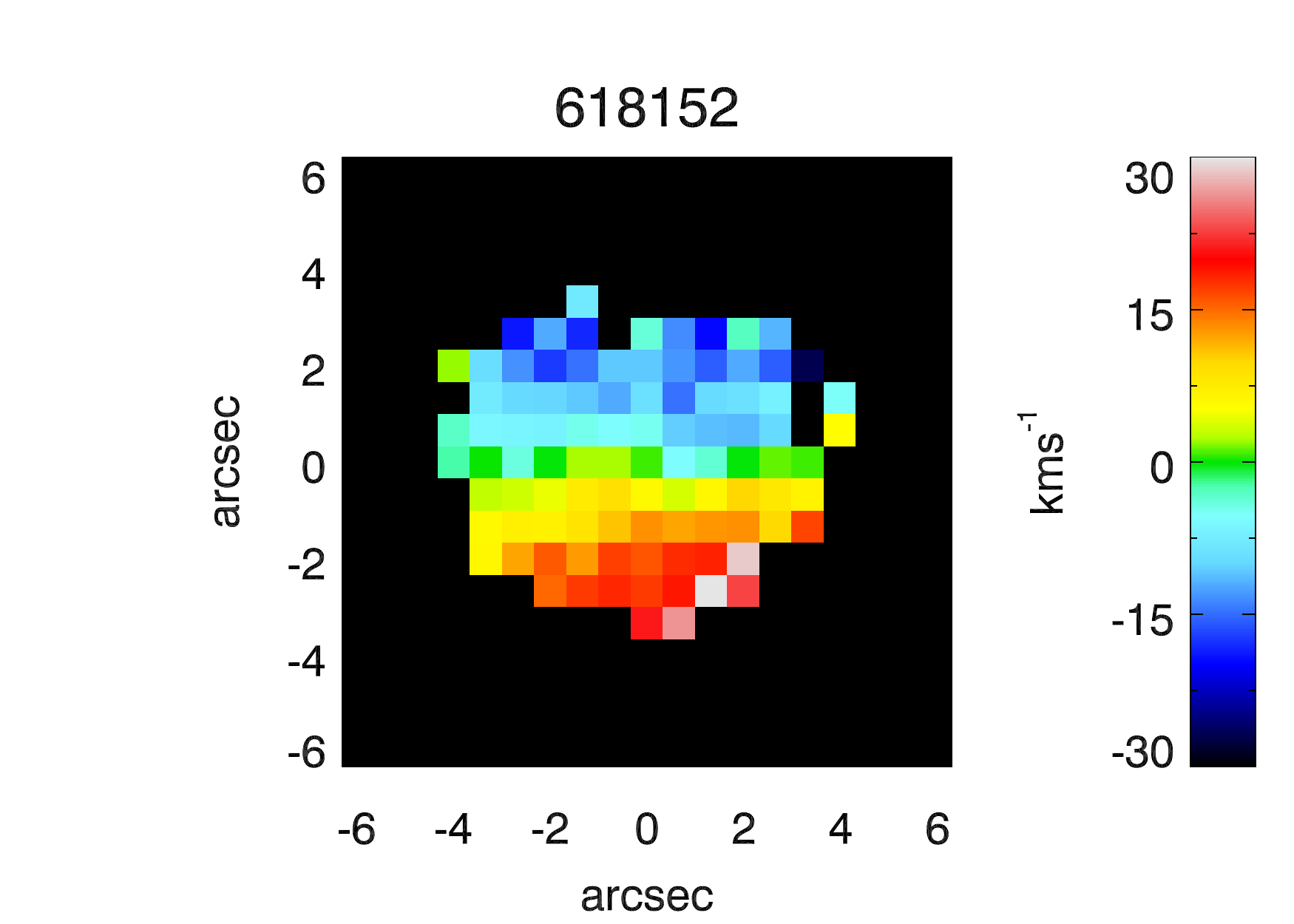}} 
\resizebox{35pc}{!}{\includegraphics[angle=90,scale=2.65,trim=-0.8cm 0cm 0cm 0cm,clip]{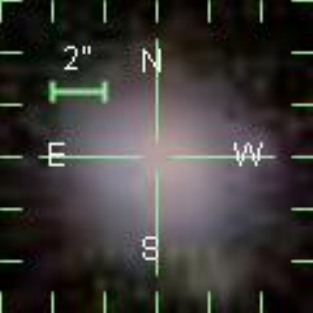} \includegraphics{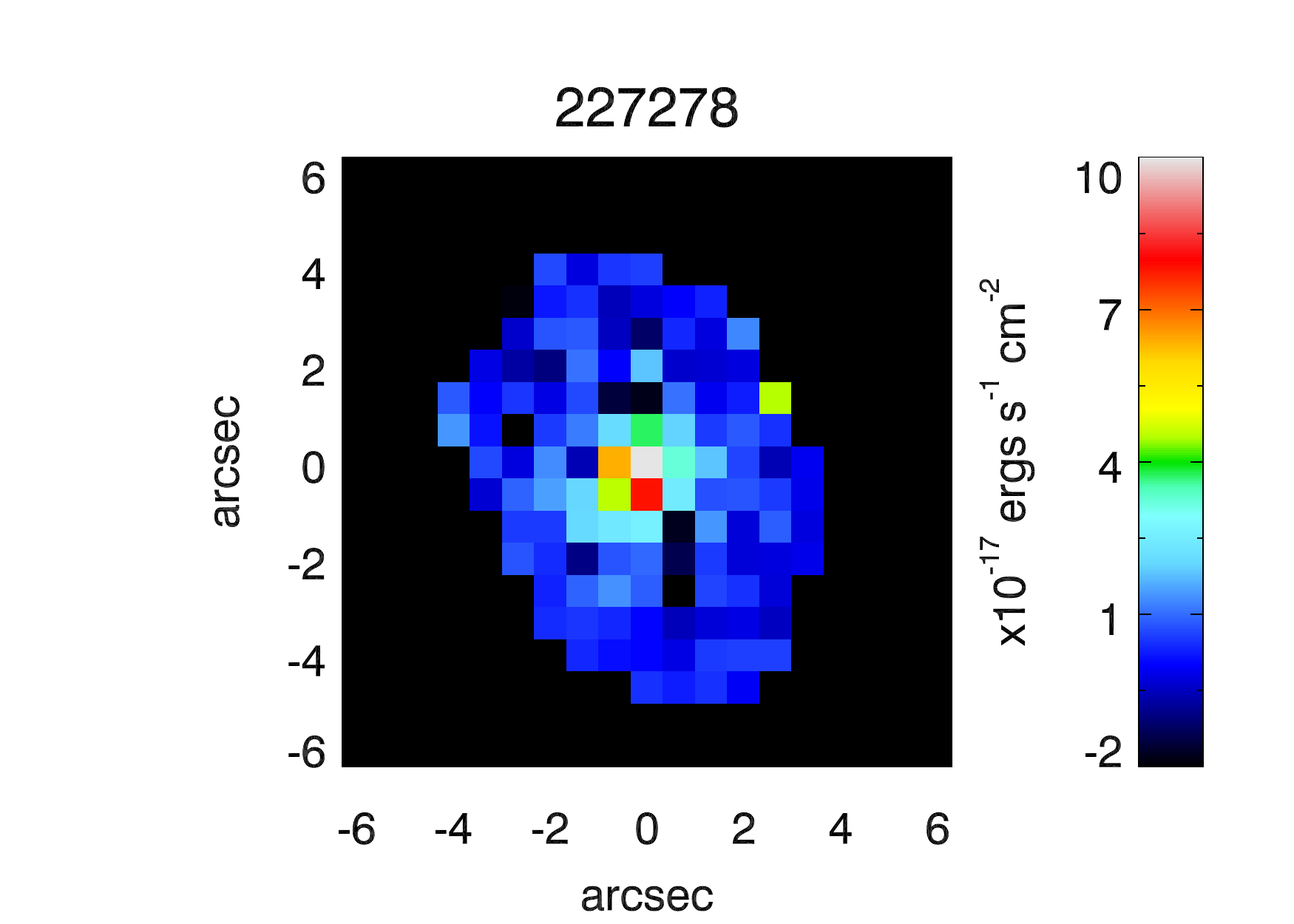} \includegraphics{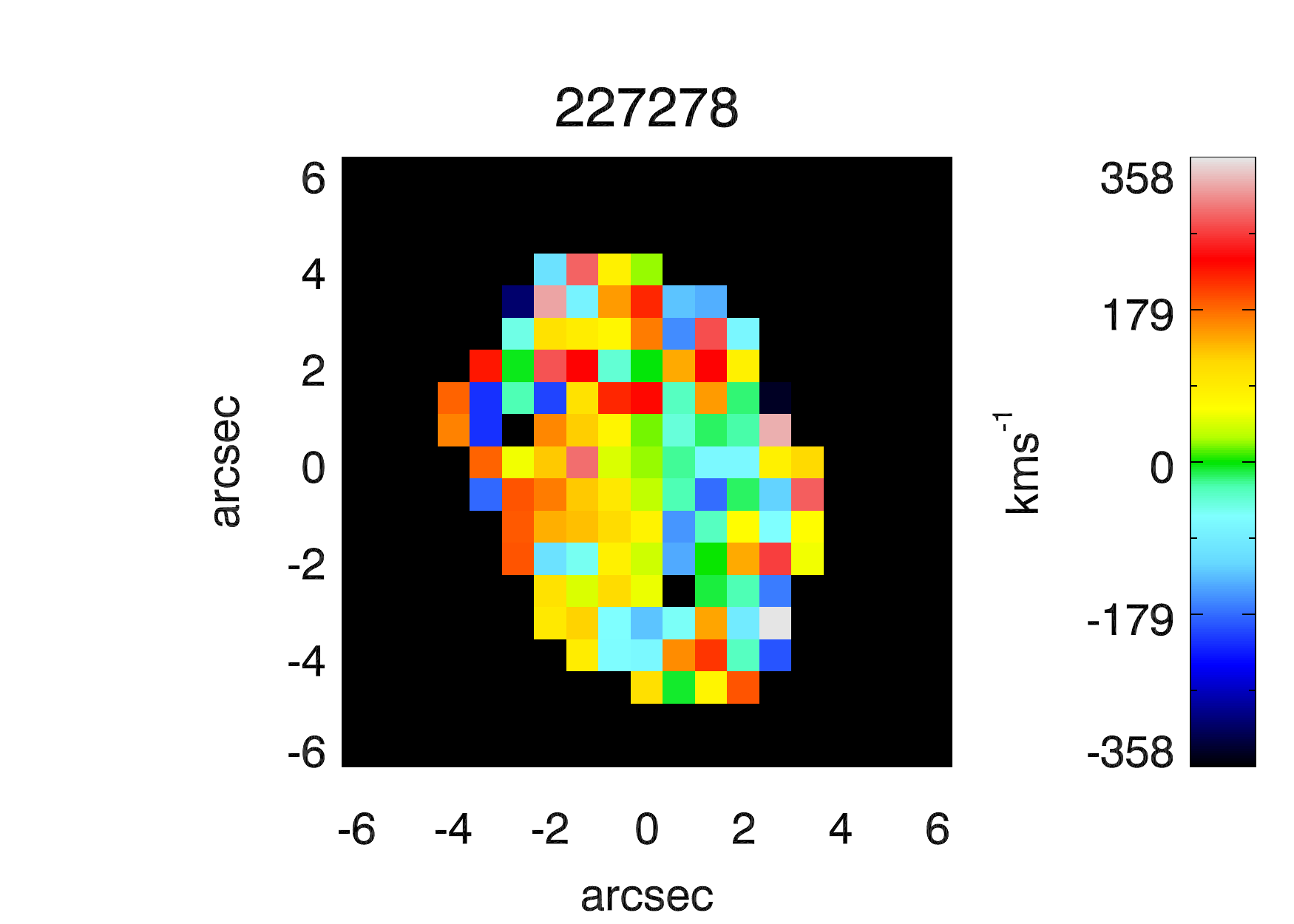}}
\end{center}
\caption{High-density environment galaxy sample.  From left-to-right: SDSS thumbnail image of SPIRAL field-of-view, H$\alpha$ flux map of central region; H$\alpha$ velocity map of central region.  Only spaxels with signal-to-noise ratios $>3$ are shown.}
\label{piccies1}
\end{figure*}

\begin{figure*}
\begin{center}	
\resizebox{35pc}{!}{\includegraphics[angle=90,scale=2.65,trim=-0.8cm 0cm 0cm 0cm,clip]{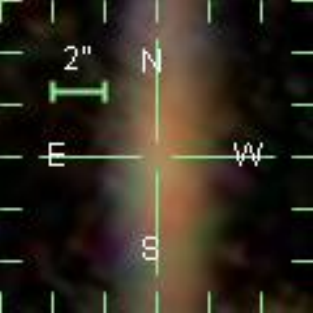} \includegraphics{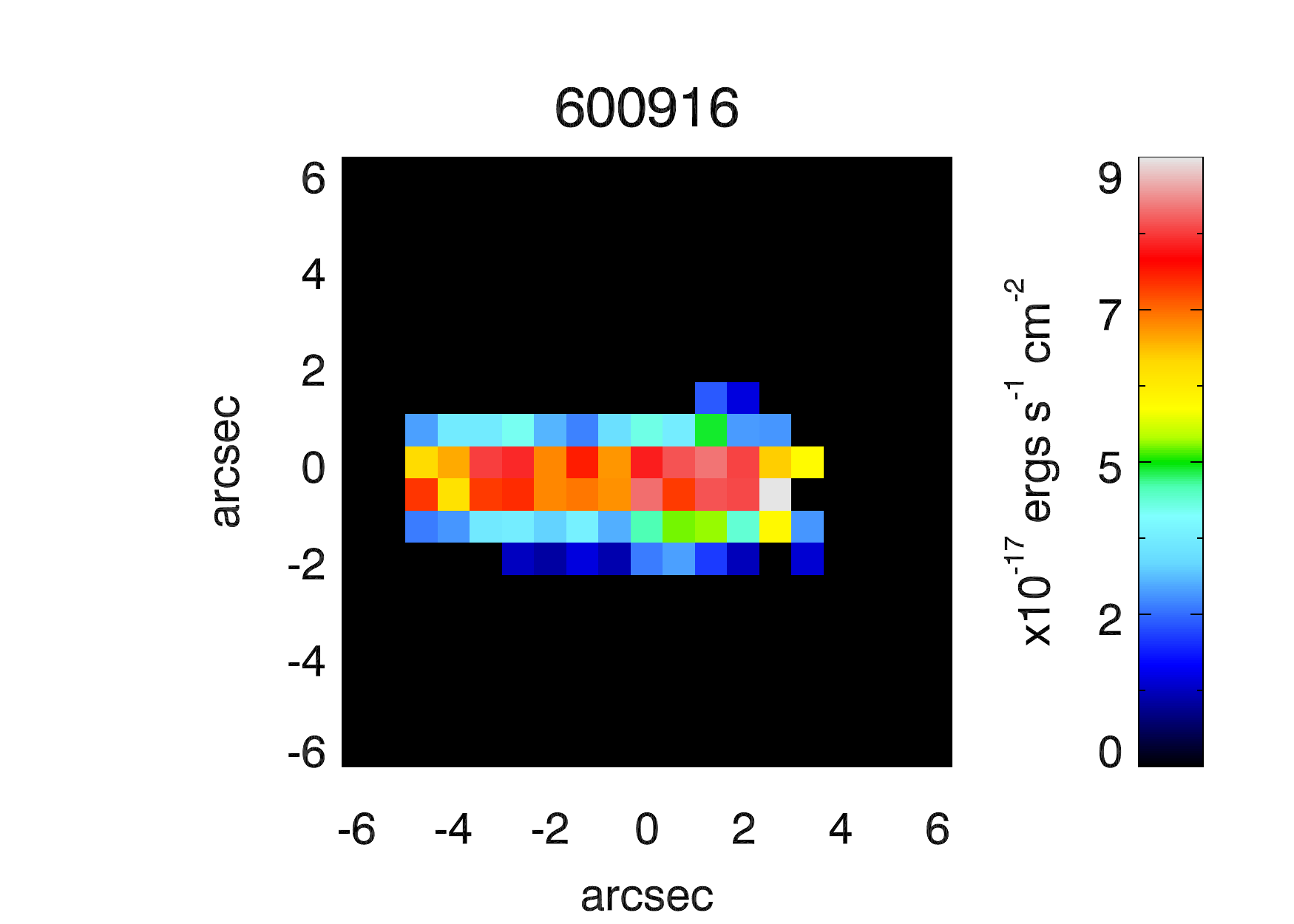} \includegraphics{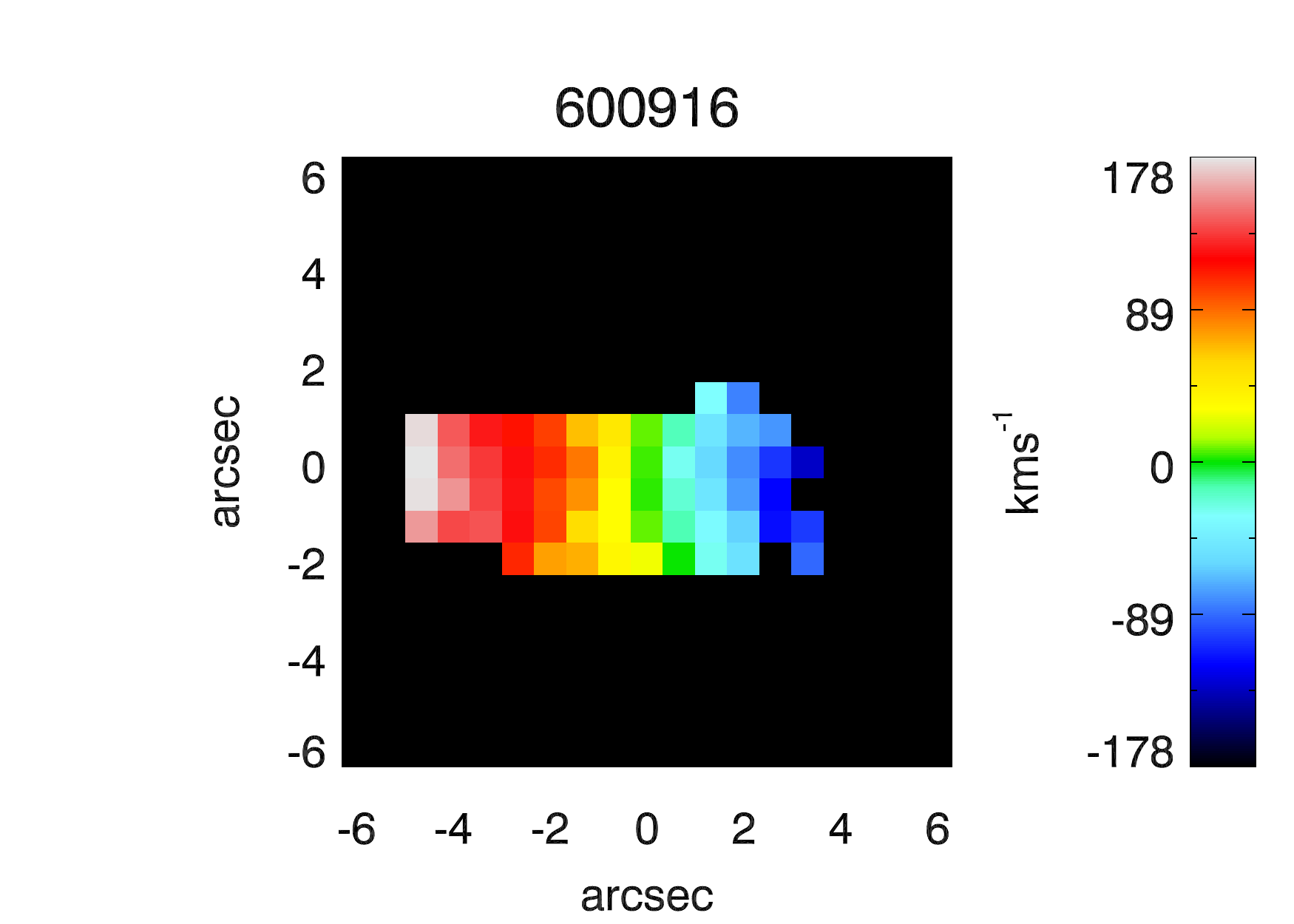}}
\resizebox{35pc}{!}{\includegraphics[angle=90,scale=2.65,trim=-0.8cm 0cm 0cm 0cm,clip]{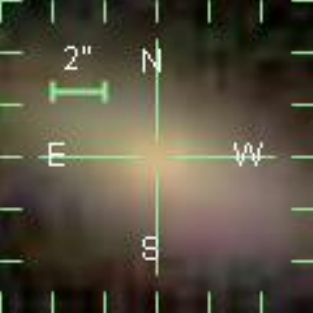} \includegraphics{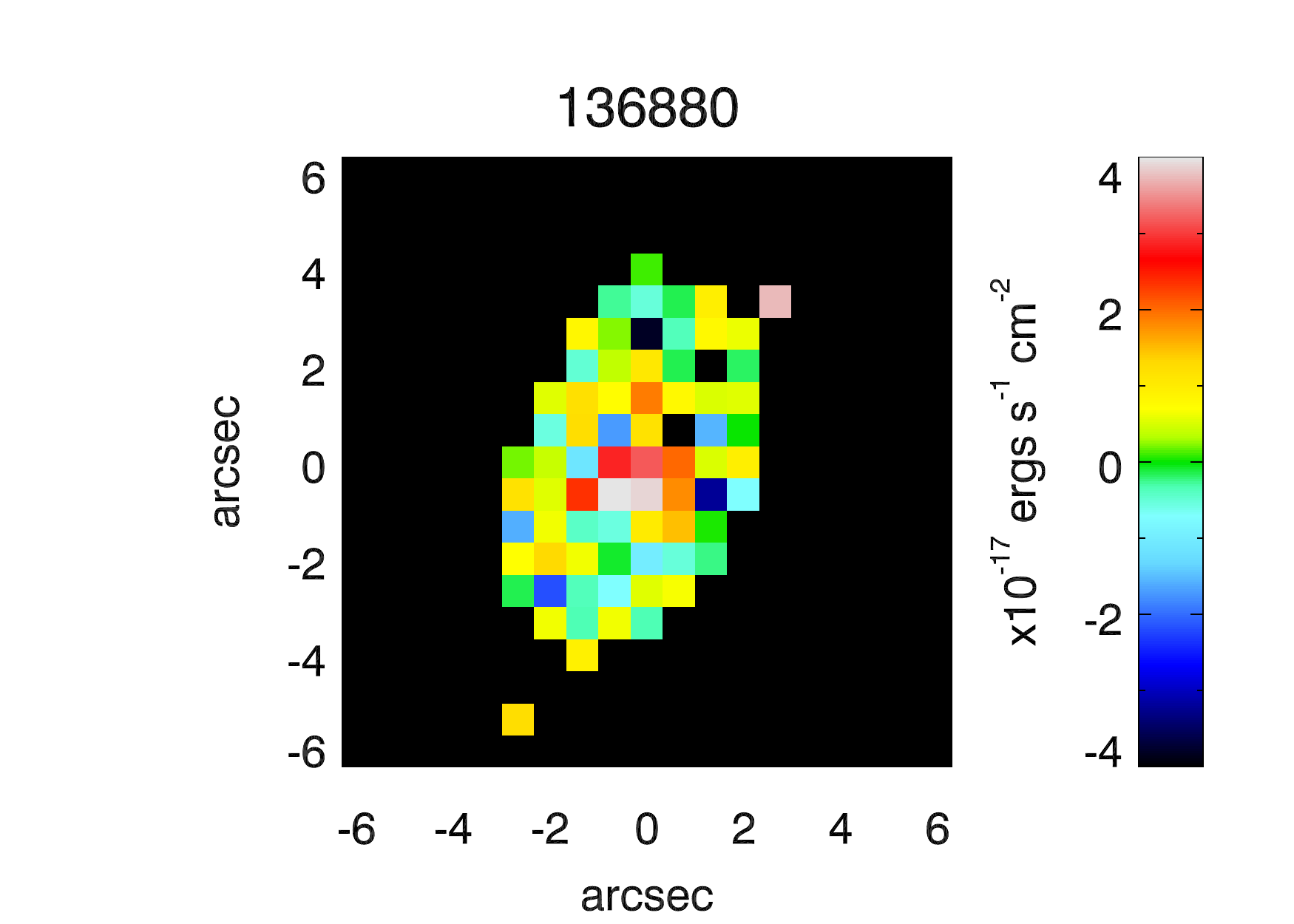} \includegraphics{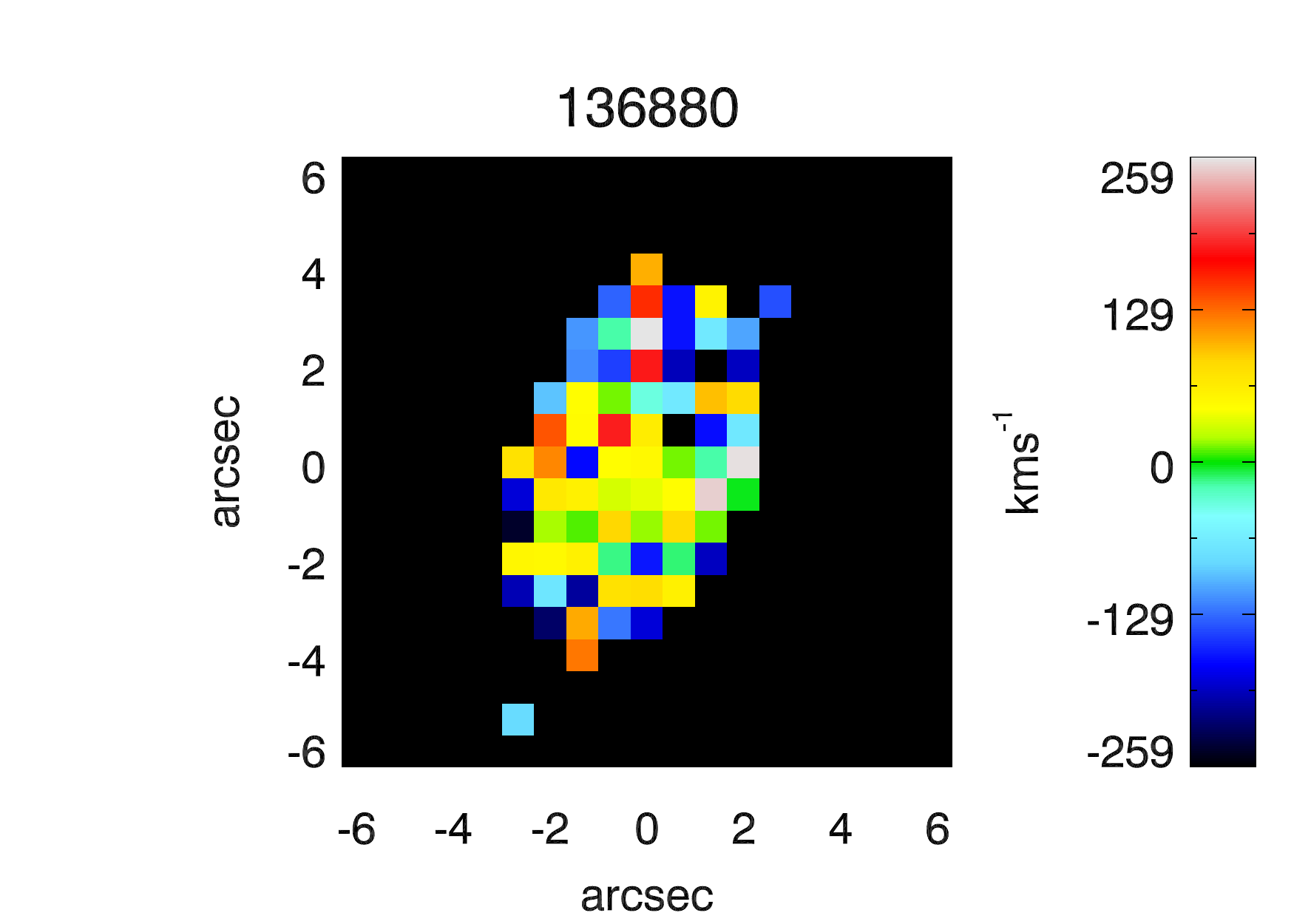} }
\resizebox{35pc}{!}{\includegraphics[angle=90,scale=2.65,trim=-0.8cm 0cm 0cm 0cm,clip]{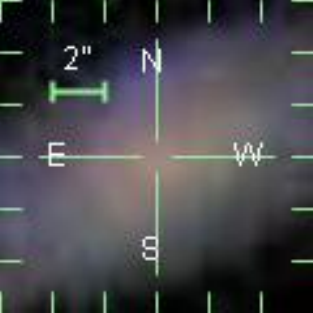} \includegraphics{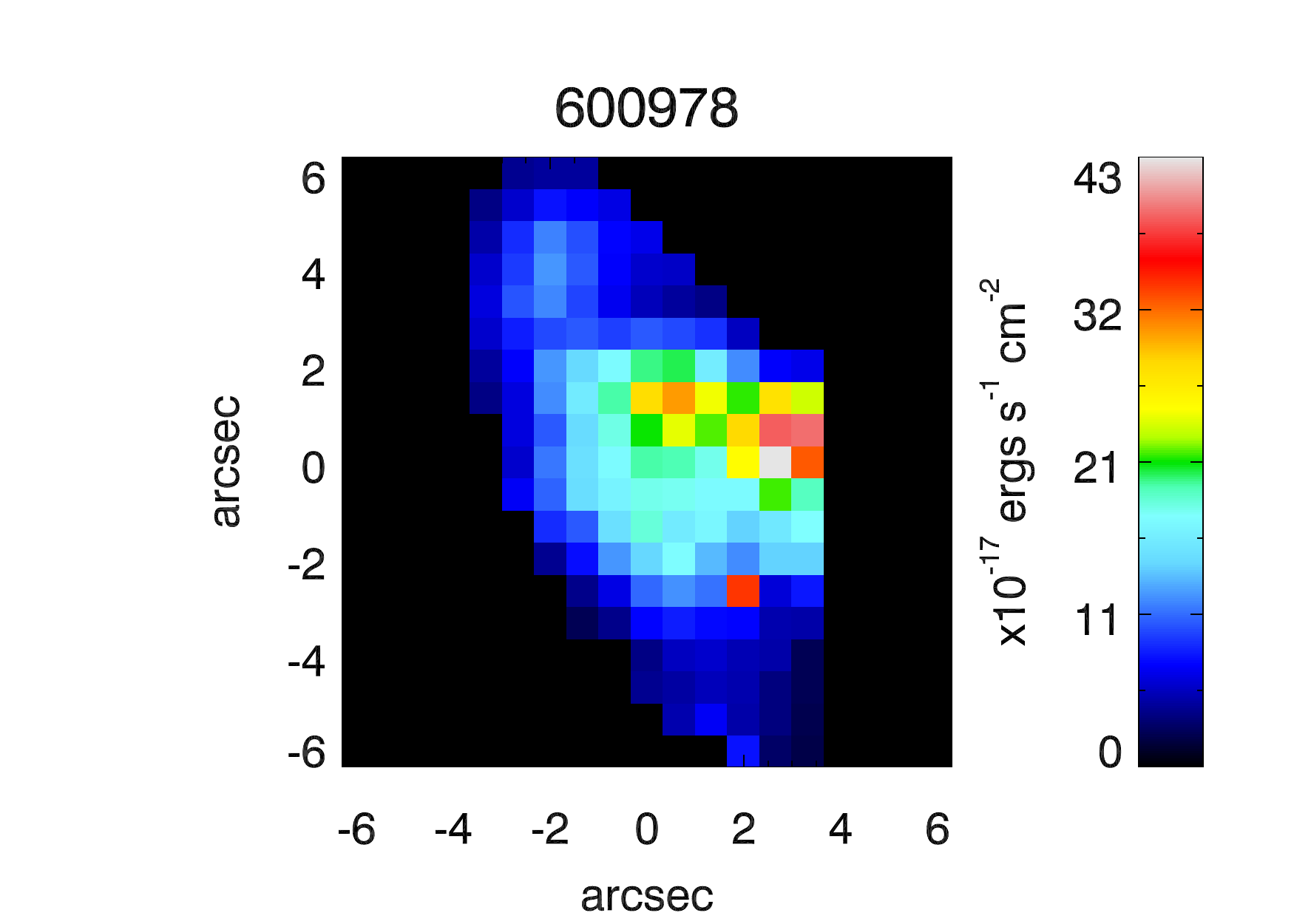}  \includegraphics{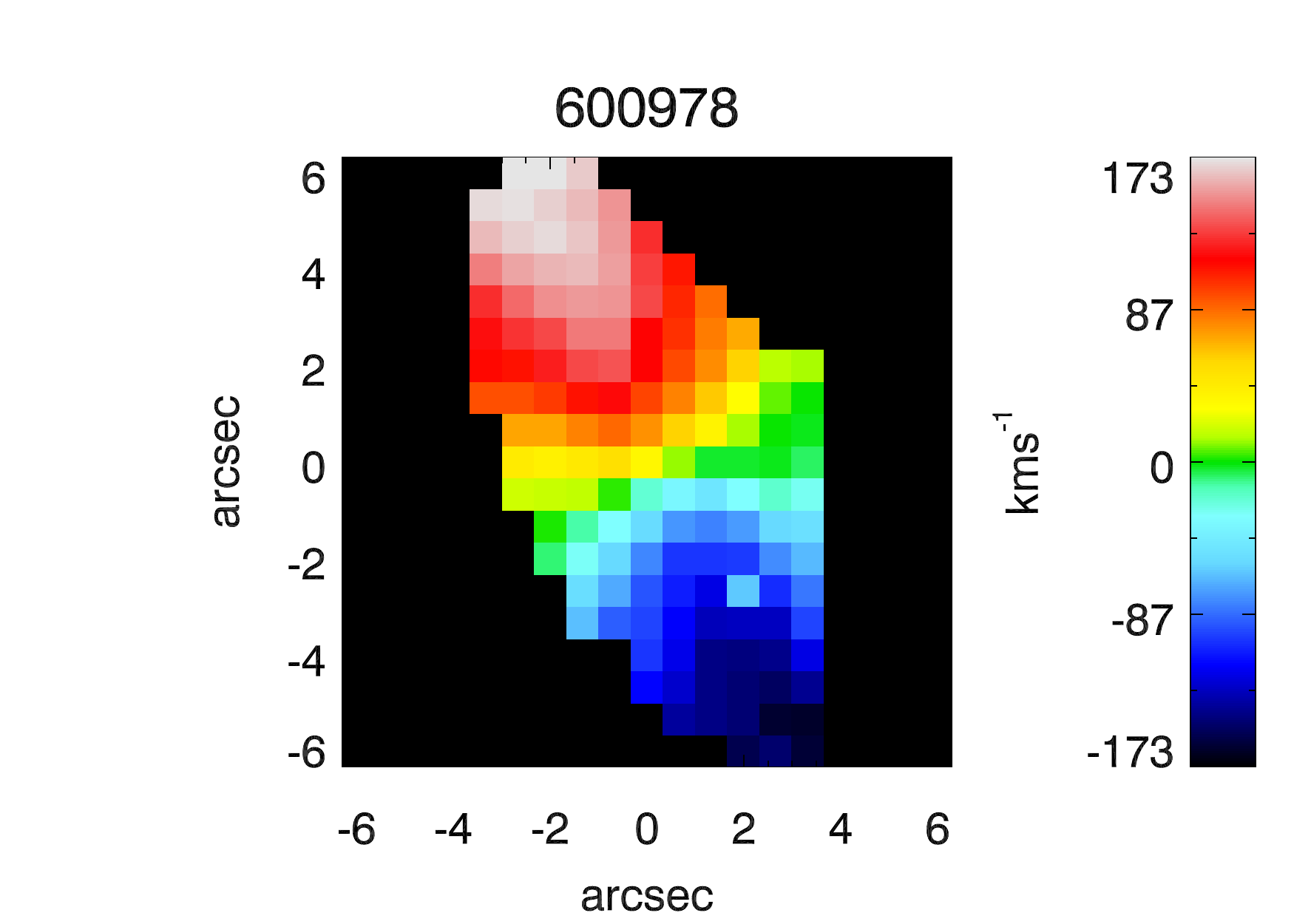}}
\end{center}
\caption{High-density environment galaxy sample - cont. From left-to-right: SDSS thumbnail image of SPIRAL field-of-view, H$\alpha$ flux map of central region; H$\alpha$ velocity map of central region.  Only spaxels with signal-to-noise ratios $>3$ are shown.}
\label{piccies2}
\end{figure*}

\begin{figure*}
\begin{center}
\resizebox{35pc}{!}{\includegraphics[angle=90,scale=2.65,trim=-0.8cm 0cm 0cm 0cm,clip]{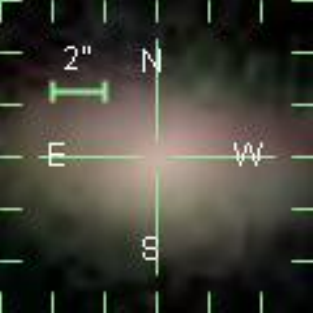} \includegraphics{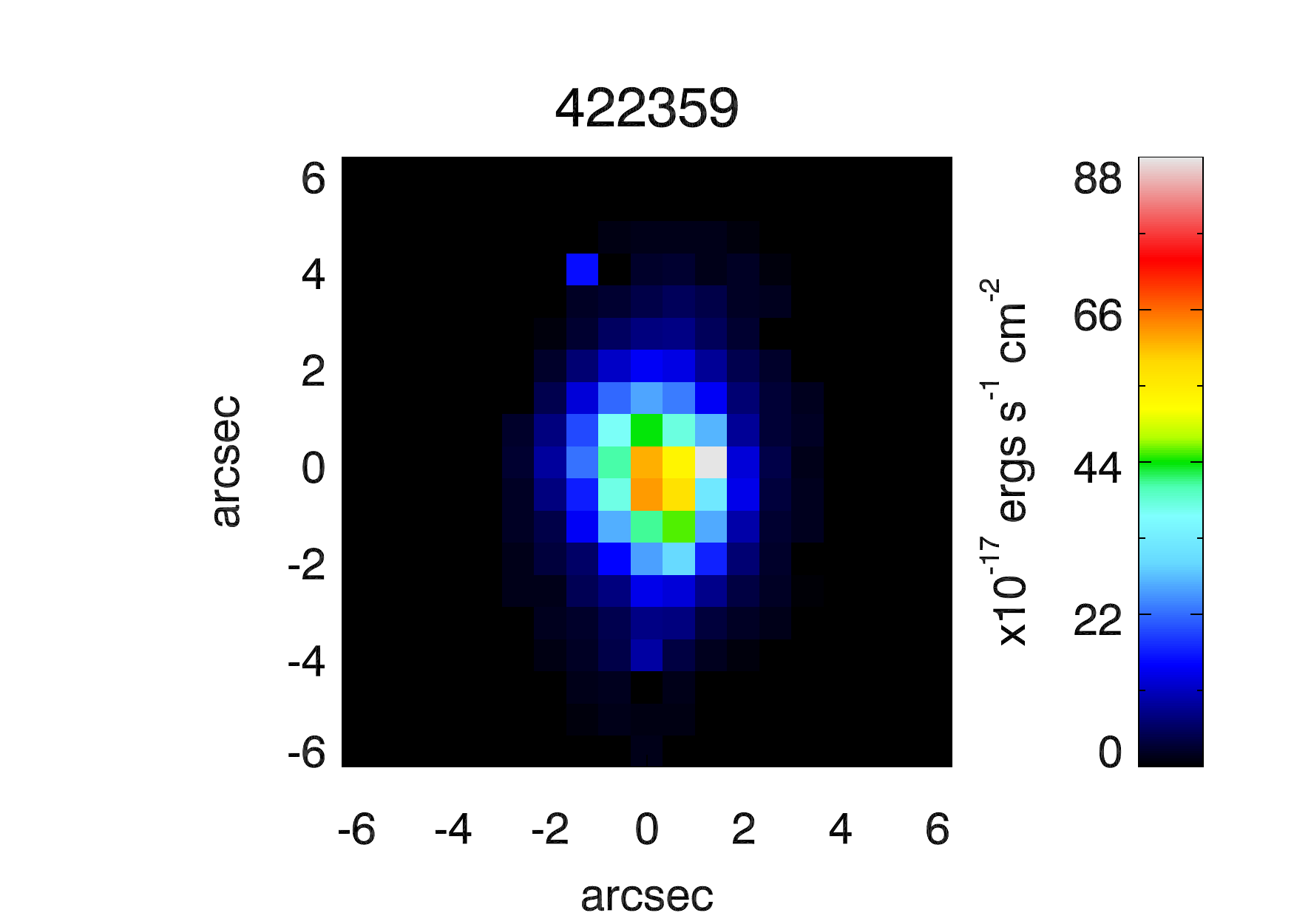} \includegraphics{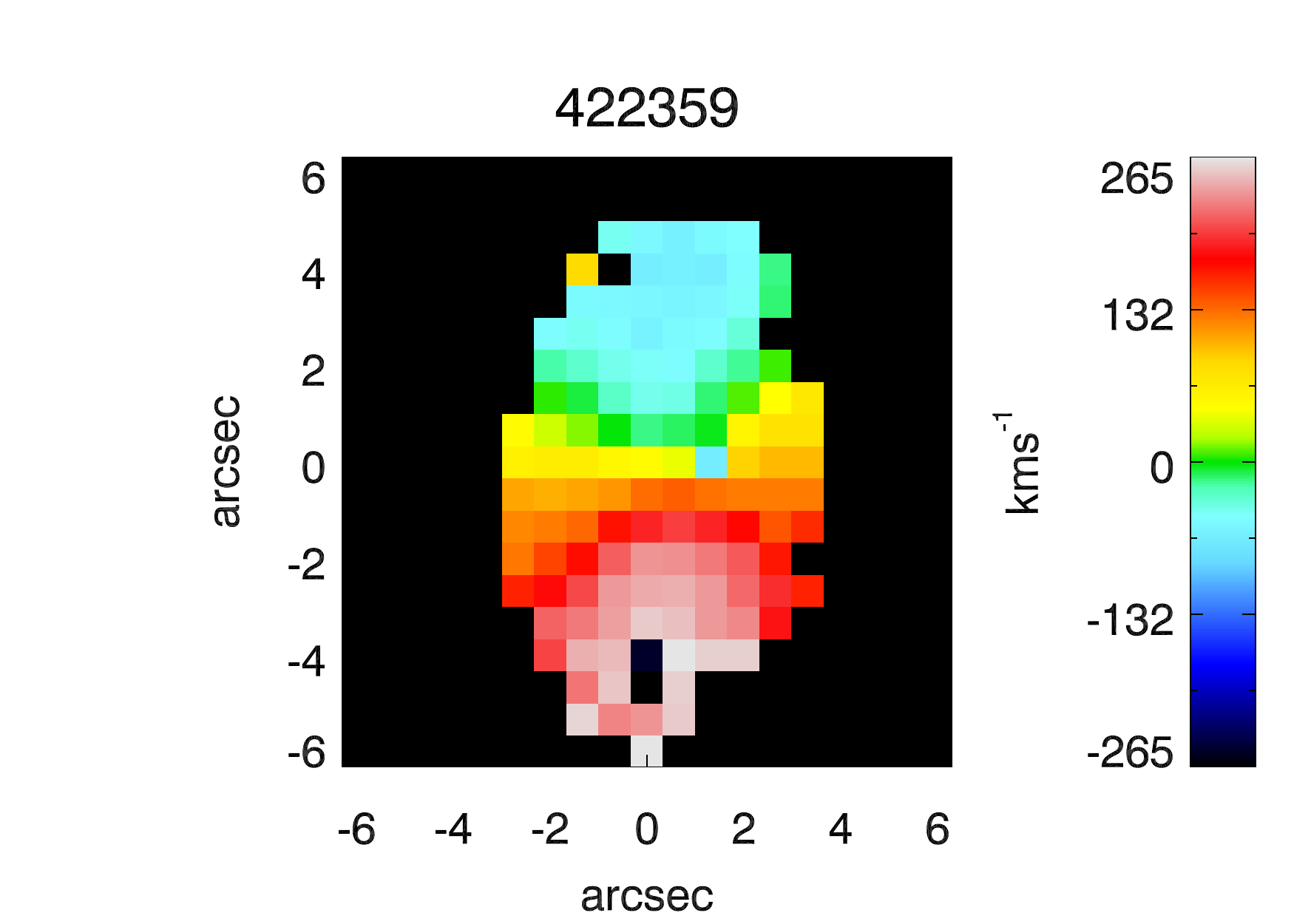}}
\resizebox{35pc}{!}{\includegraphics[angle=90,scale=2.65,trim=-0.8cm 0cm 0cm 0cm,clip]{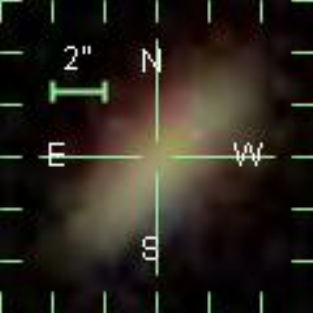} \includegraphics{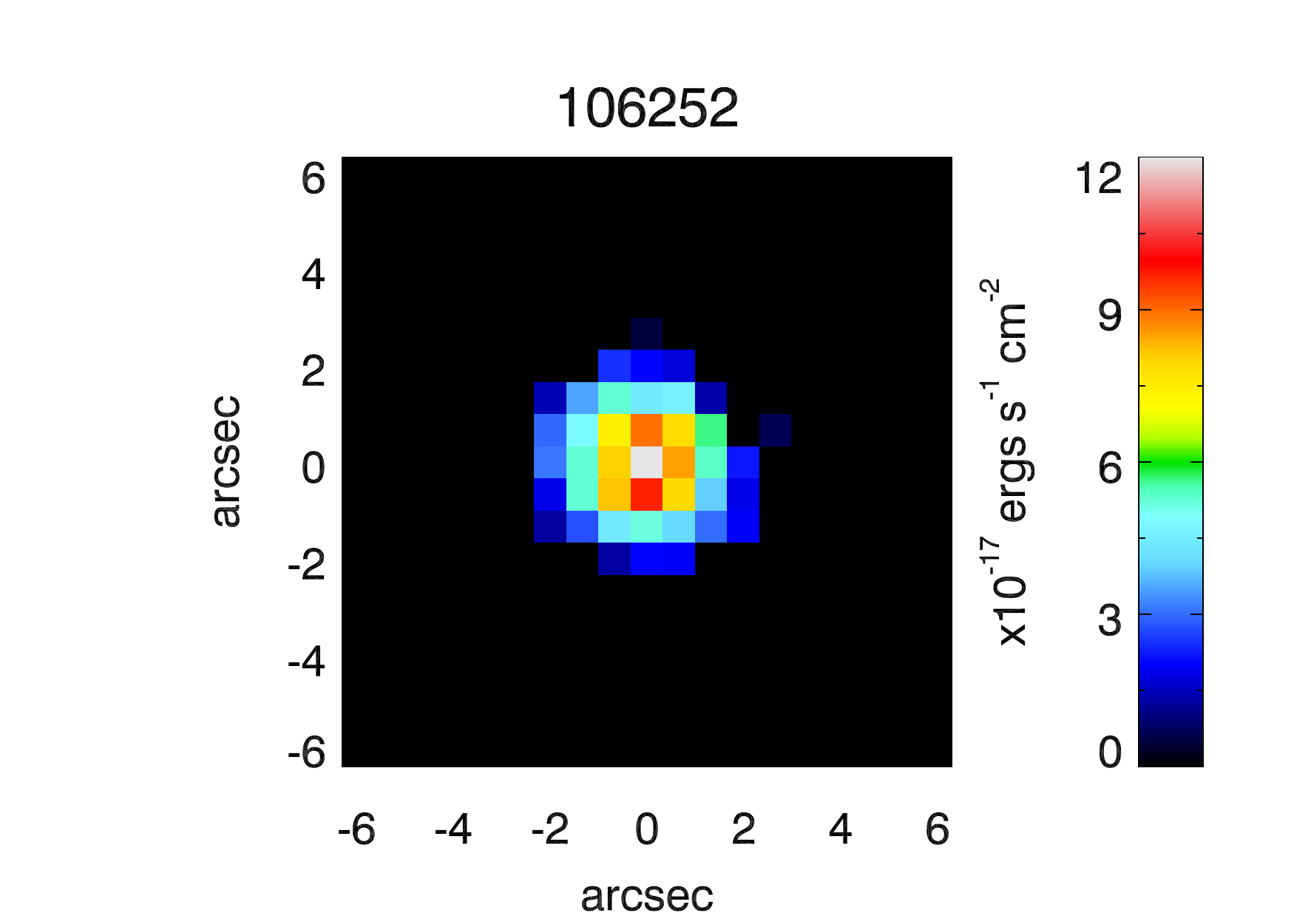} \includegraphics{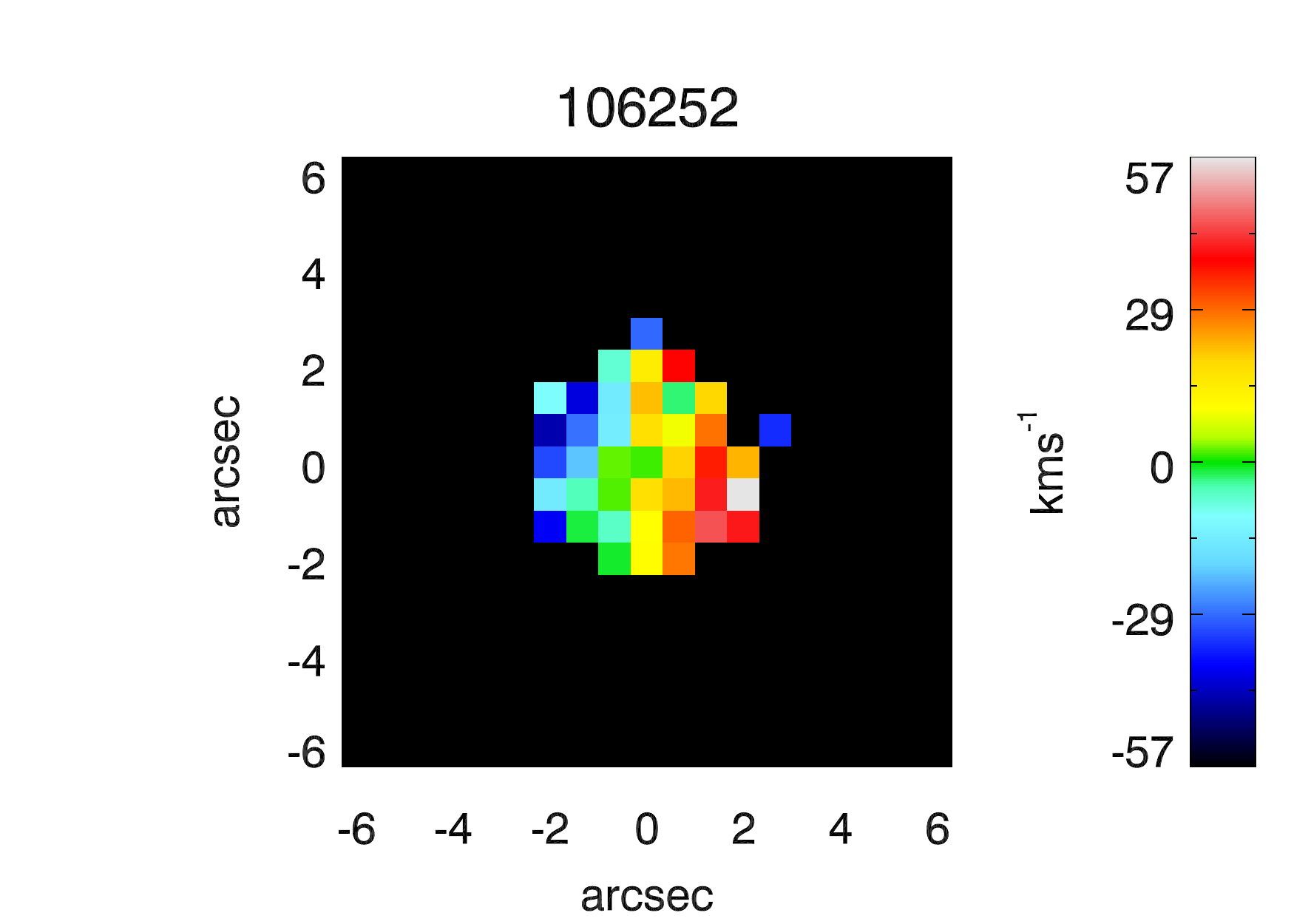}}
\resizebox{35pc}{!}{\includegraphics[angle=90,scale=2.65,trim=-0.8cm 0cm 0cm 0cm,clip]{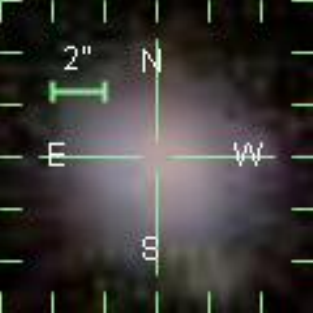} \includegraphics{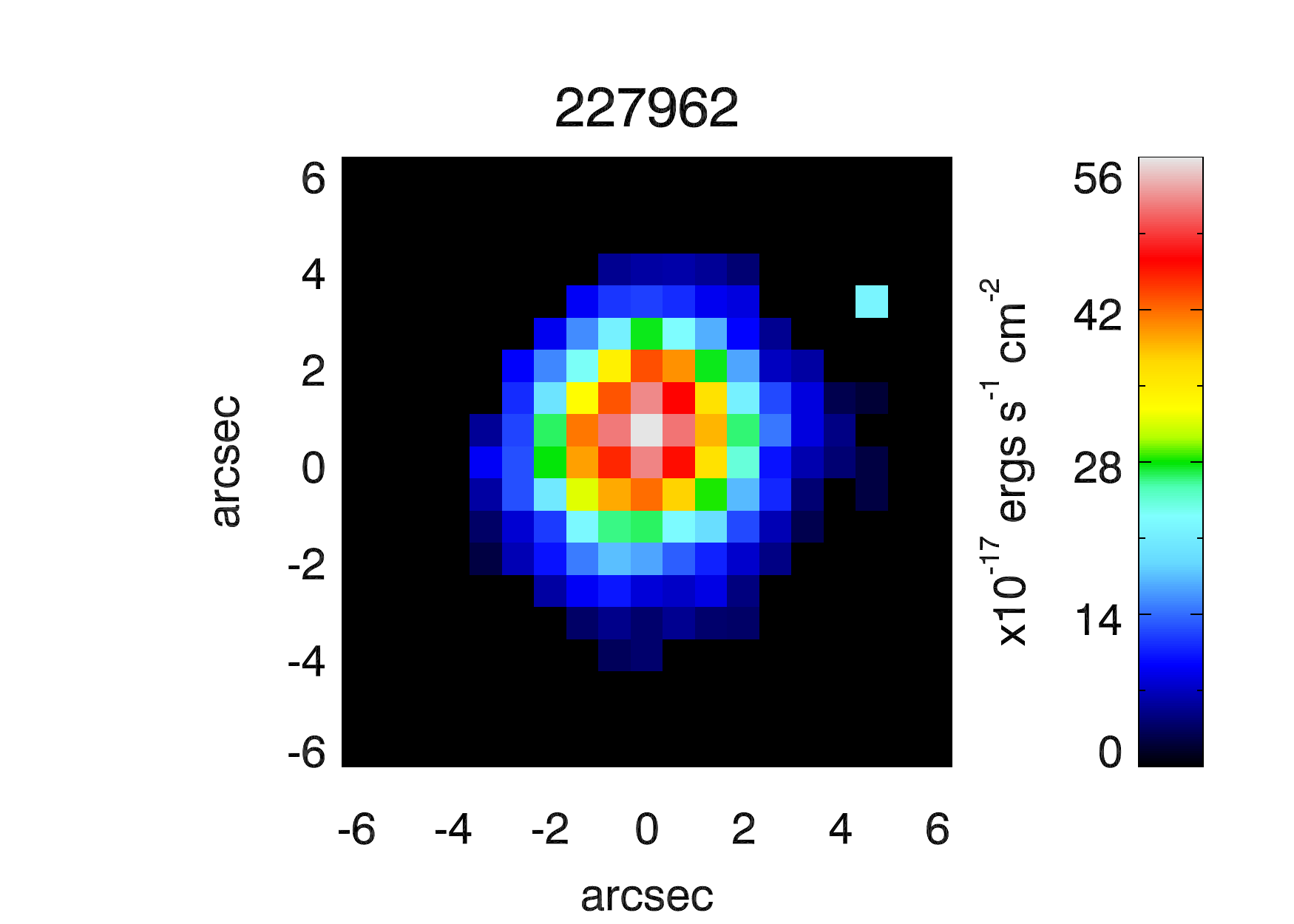} \includegraphics{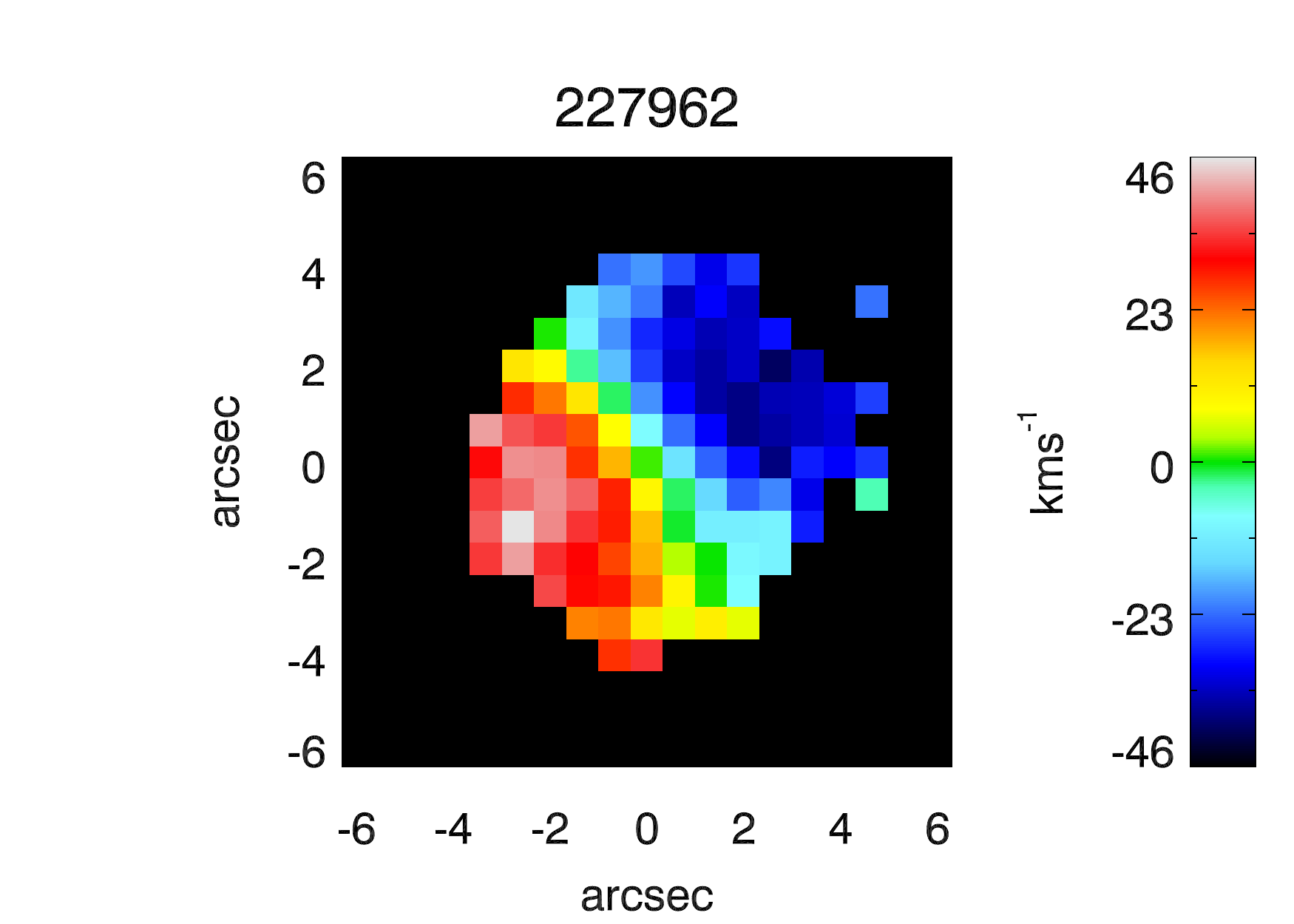}}
\resizebox{35pc}{!}{\includegraphics[scale=2.65,trim=0cm -0.8cm 0cm 0cm,clip]{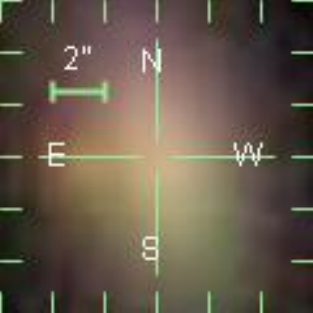}  \includegraphics{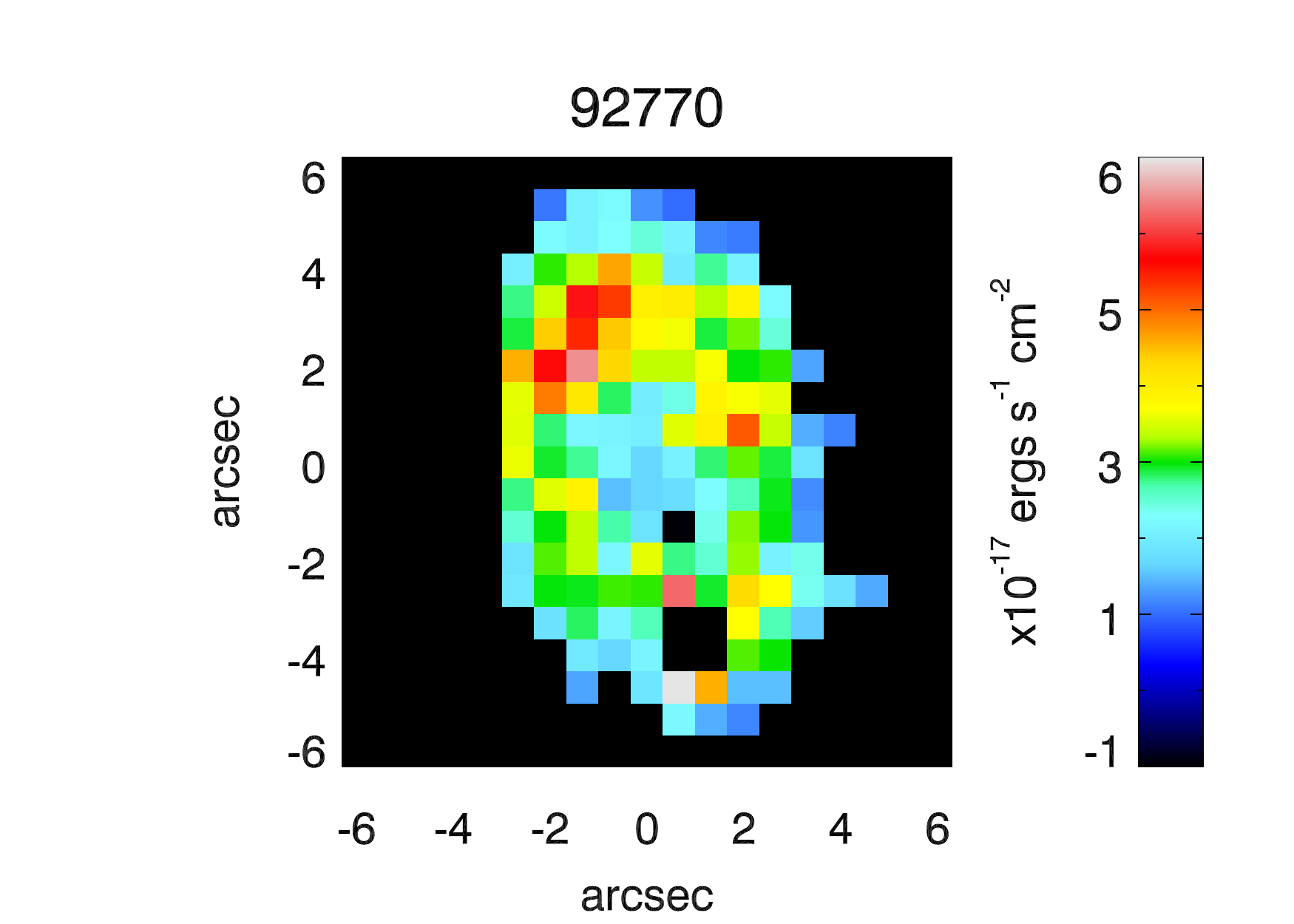} \includegraphics{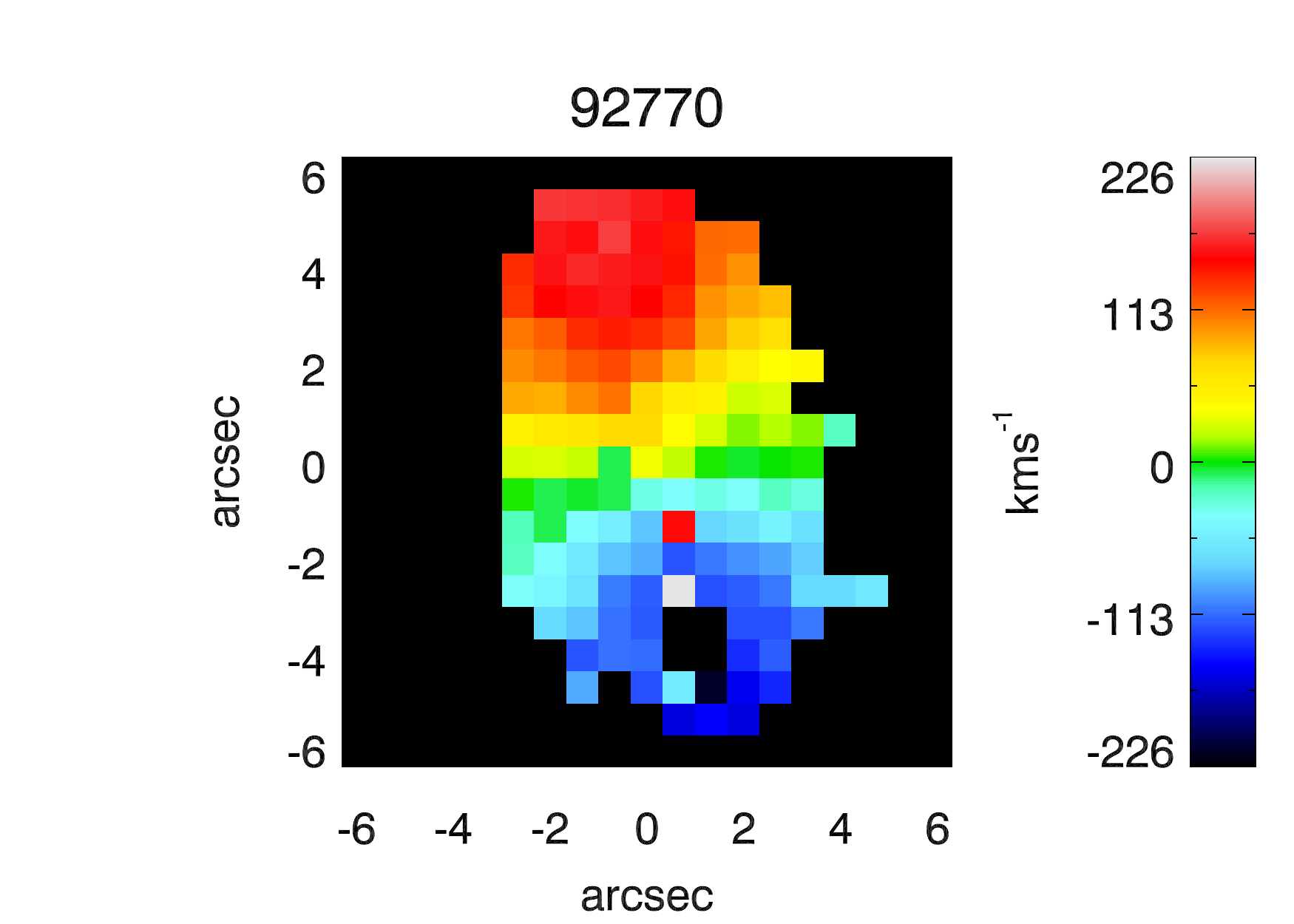} } 
\end{center}
\caption{Low-density environment galaxy sample. From left-to-right: SDSS thumbnail image of SPIRAL field-of-view, H$\alpha$ flux map of central region; H$\alpha$ velocity map of central region.  Only spaxels with signal-to-noise ratios $>3$ are shown.}
\label{piccies3}
\end{figure*}

\begin{figure*}
\begin{center}	
\resizebox{35pc}{!}{\includegraphics[angle=90,scale=2.65,trim=-0.8cm 0cm 0cm 0cm,clip]{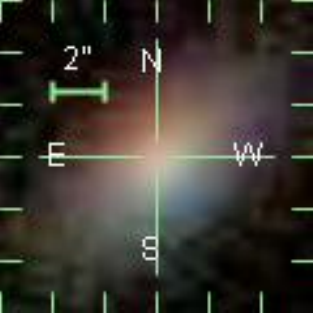} \includegraphics{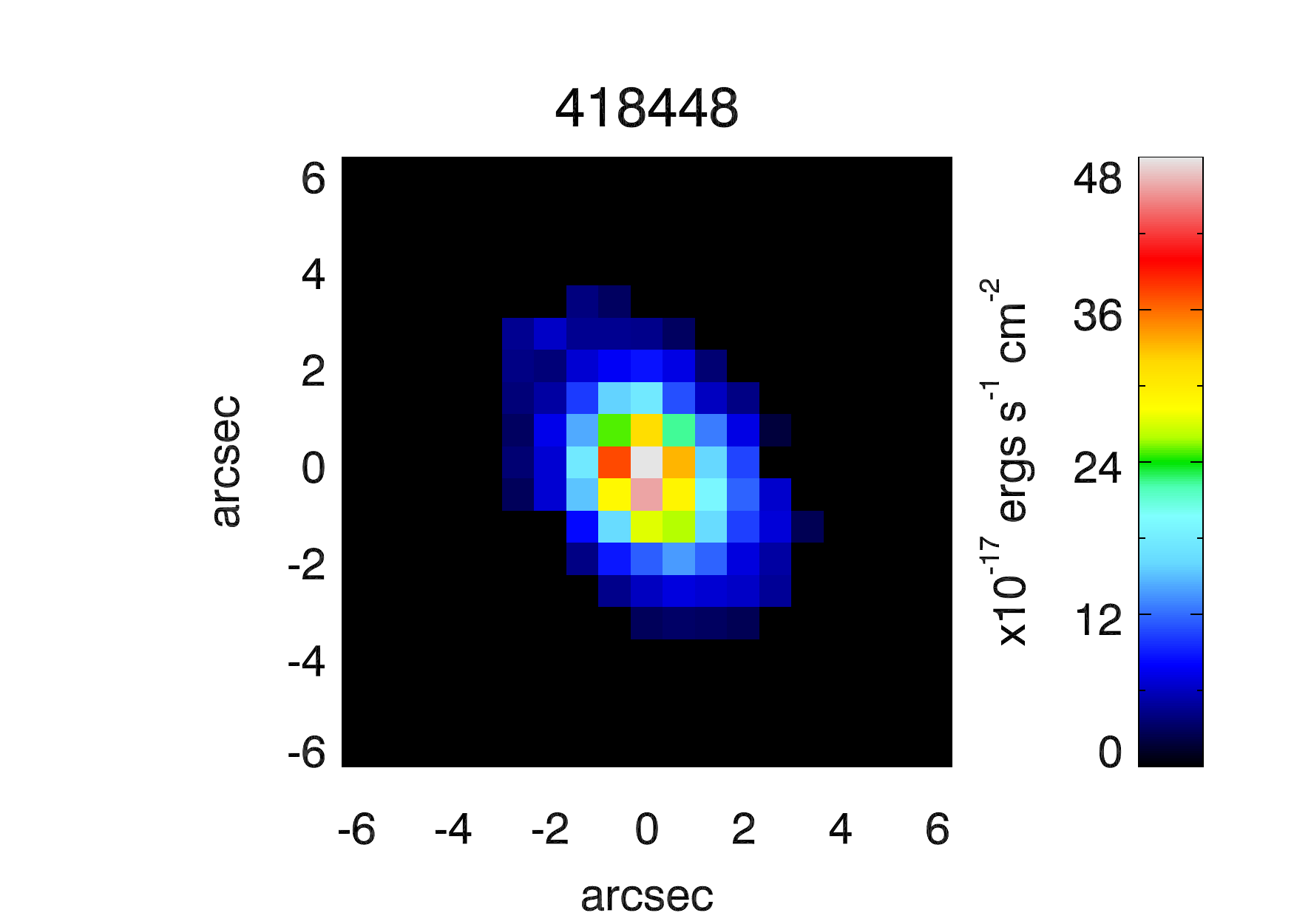} \includegraphics{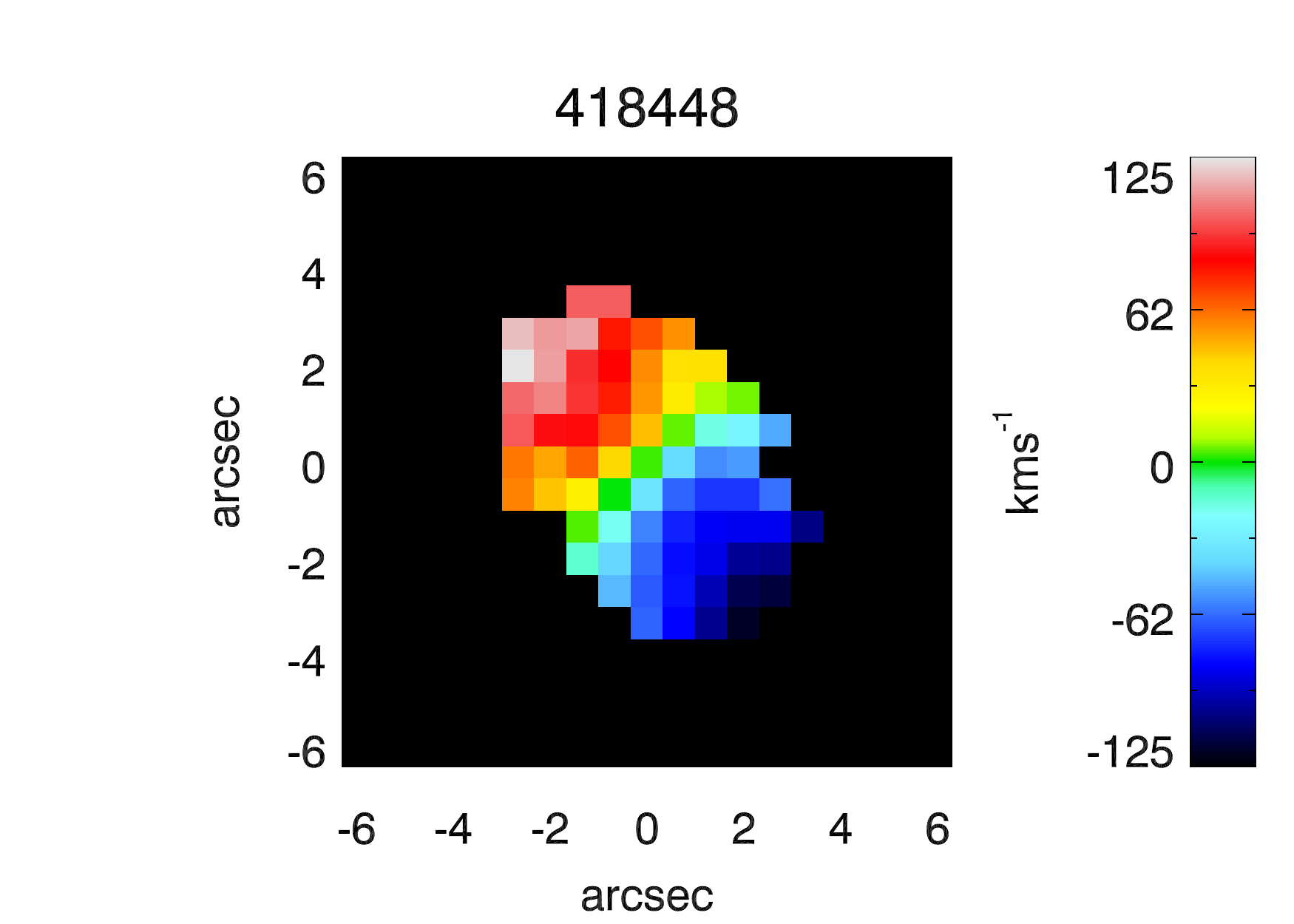}}
\resizebox{35pc}{!}{\includegraphics[angle=90,scale=2.65,trim=-0.8cm 0cm 0cm 0cm,clip]{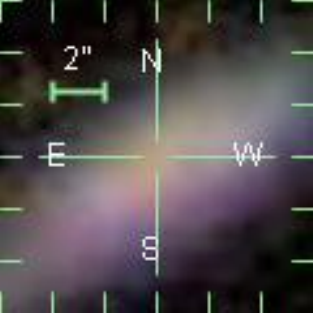} \includegraphics{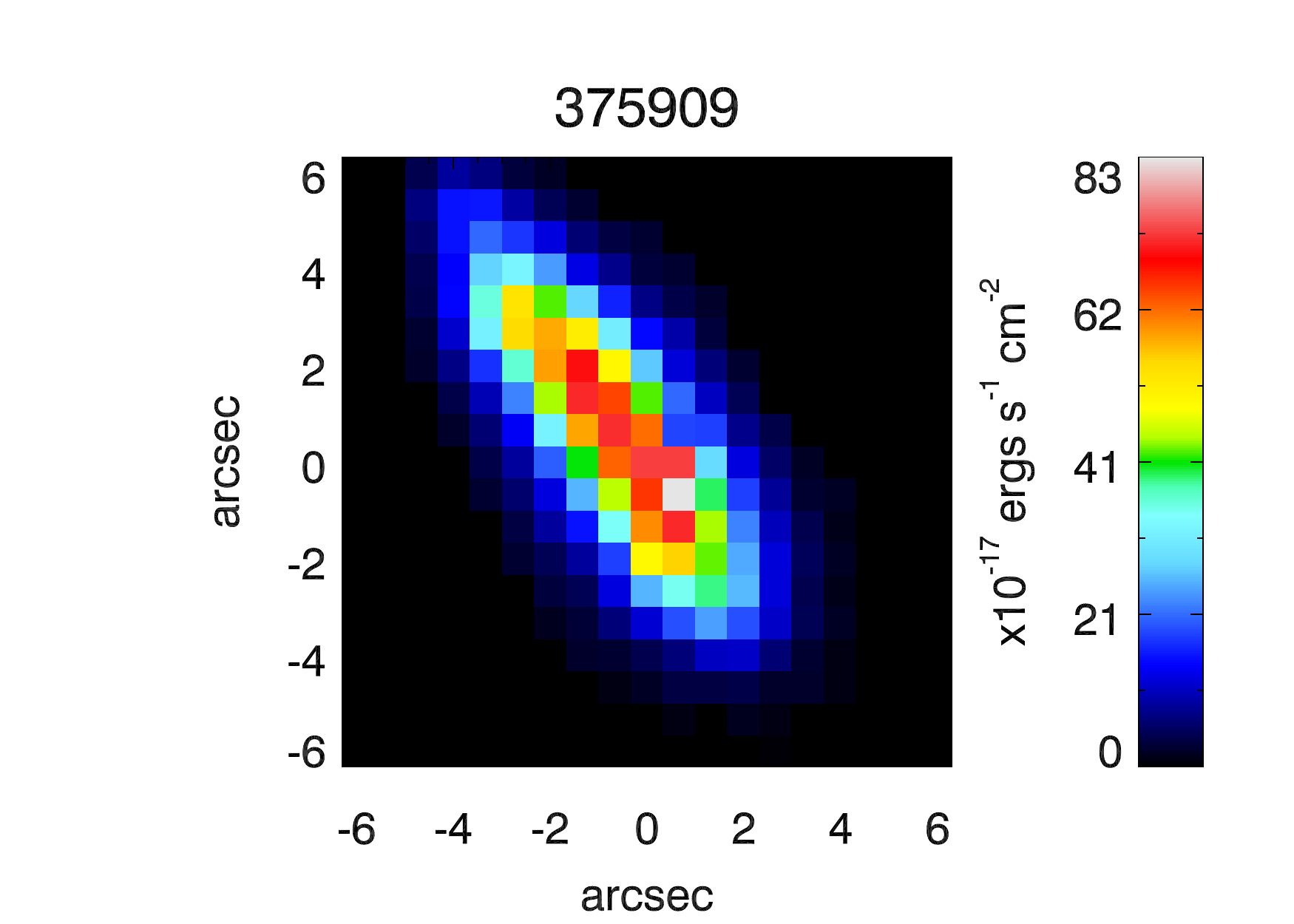} \includegraphics{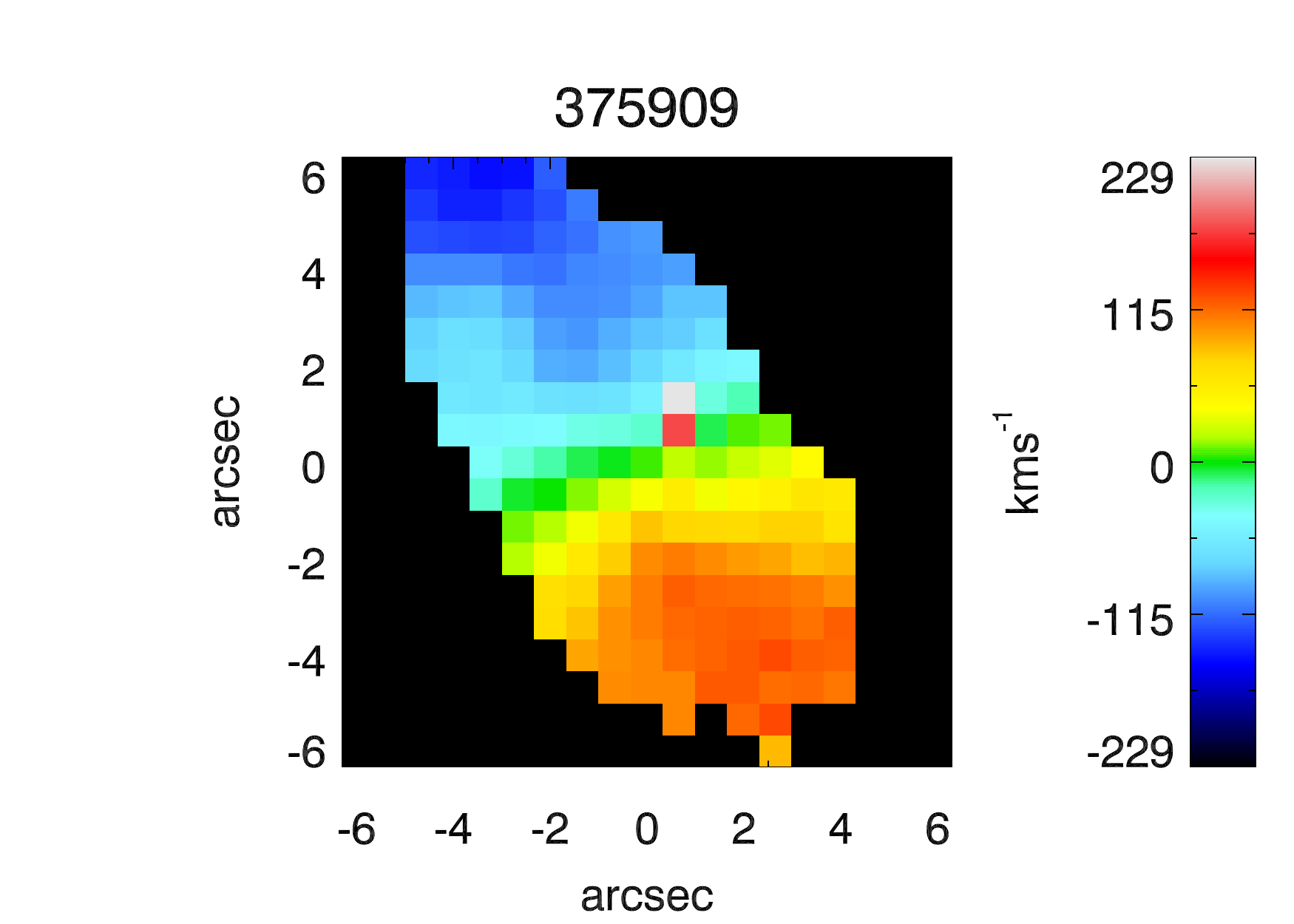}}
\resizebox{35pc}{!}{ \includegraphics[scale=2.65,trim=0cm -0.8cm 0cm 0cm,clip]{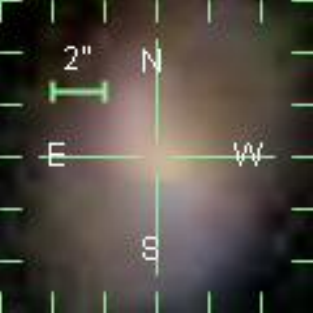} \includegraphics{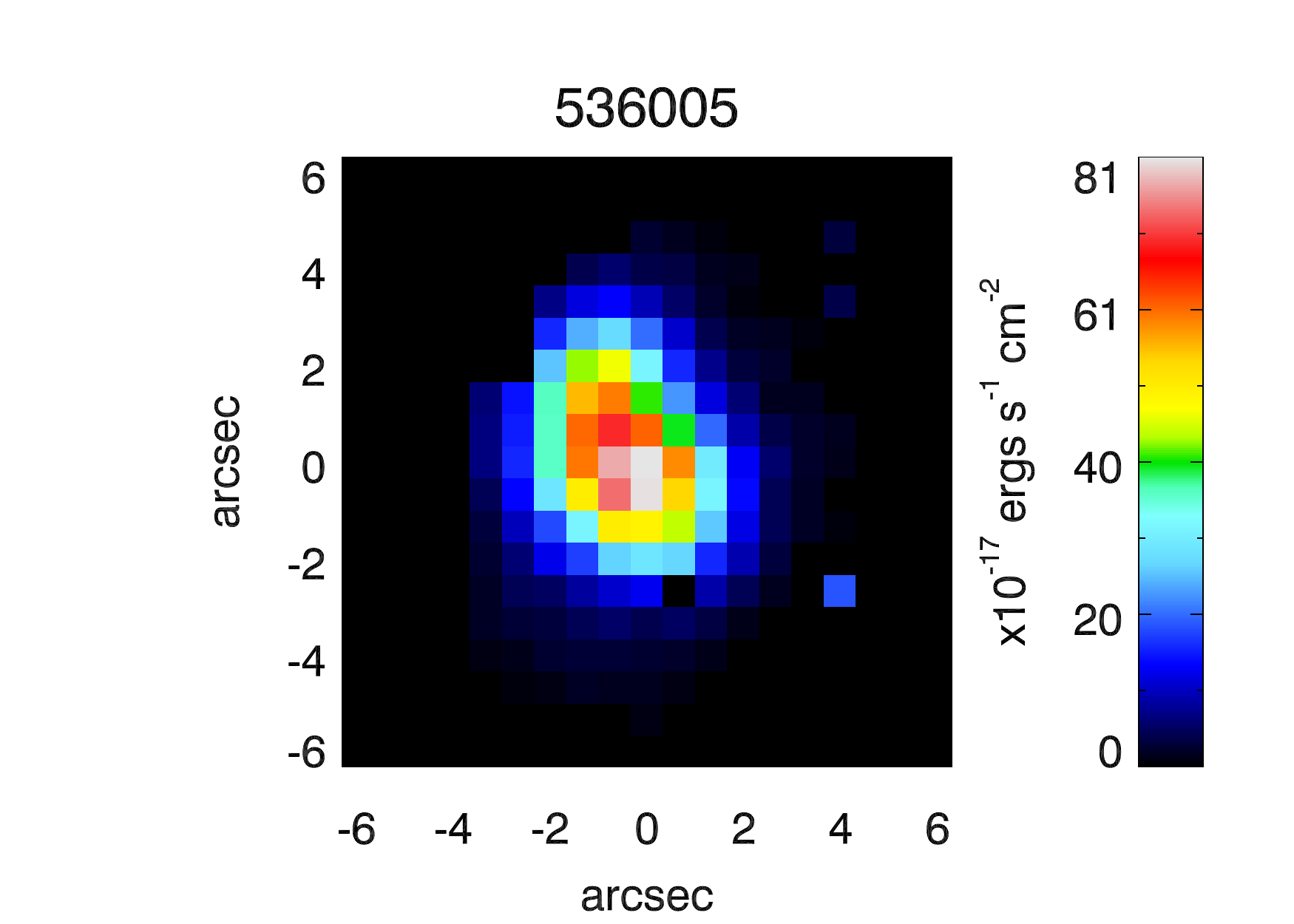} \includegraphics{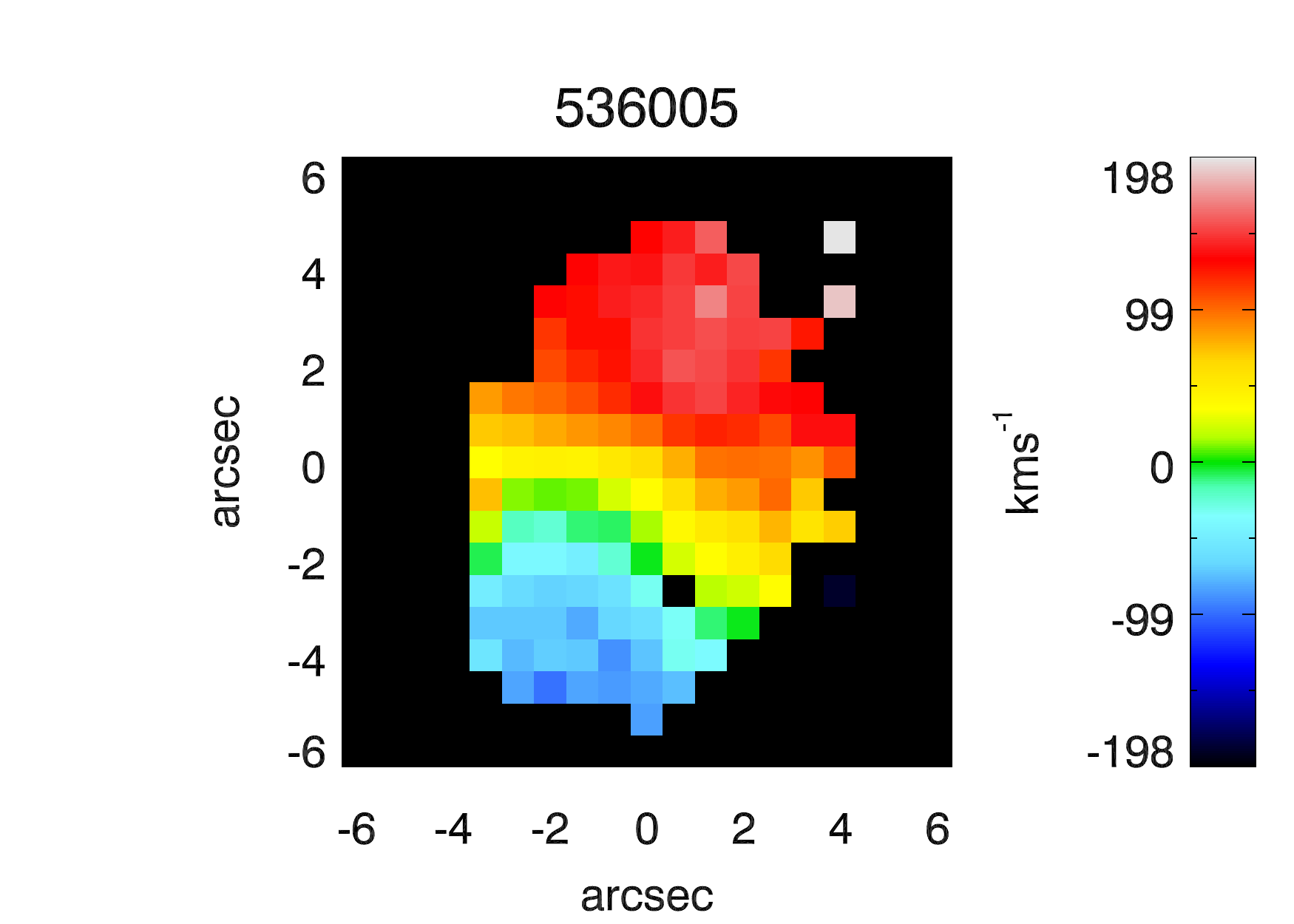}}
\resizebox{35pc}{!}{ \includegraphics[angle=90,scale=2.65,trim=-0.8cm 0cm 0cm 0cm,clip]{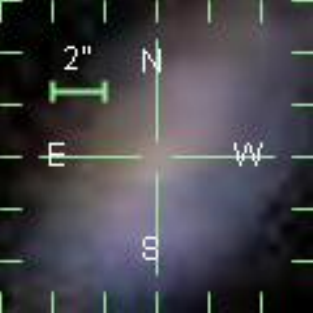} \includegraphics{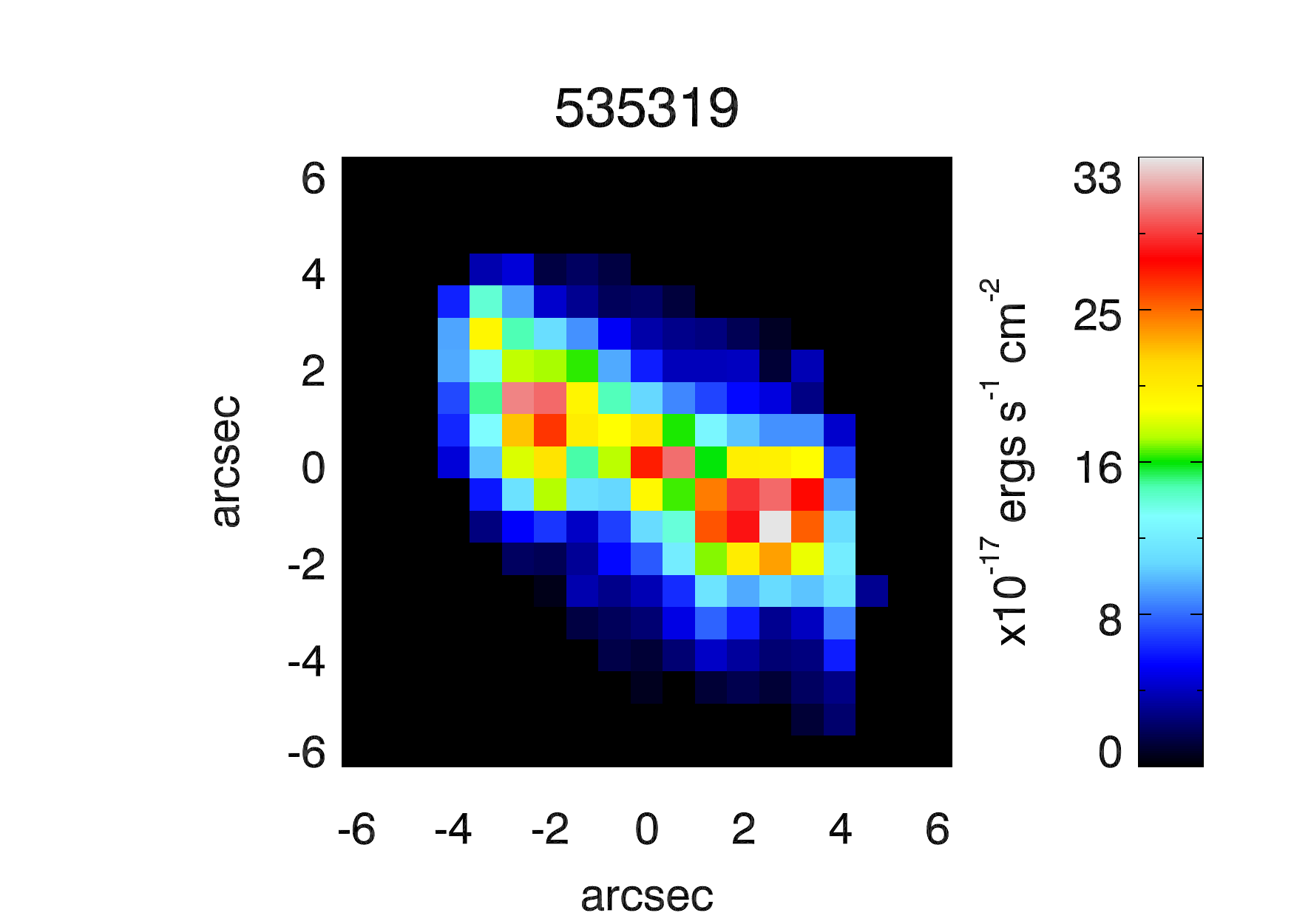} \includegraphics{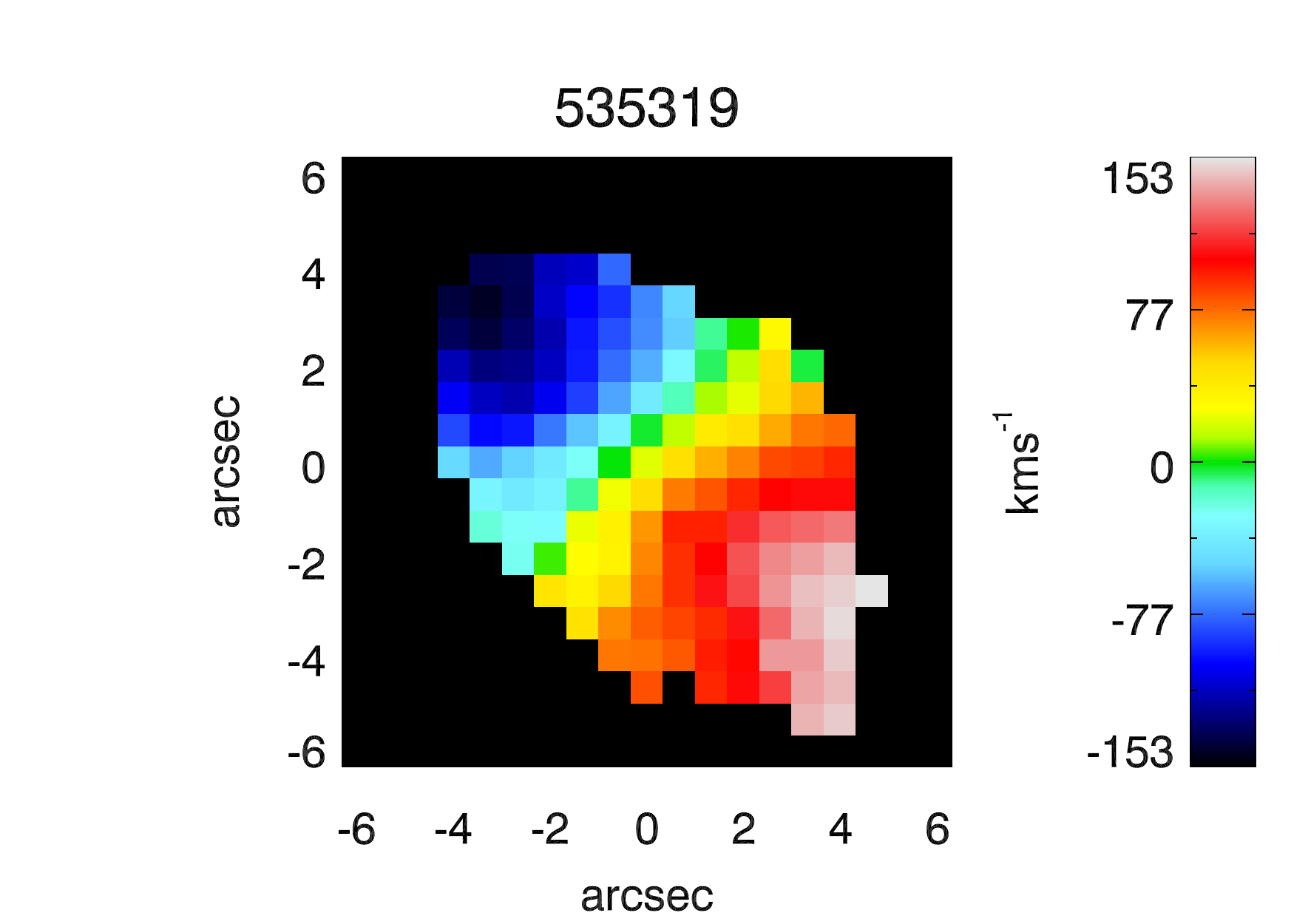}}

\end{center}
\caption{Low-density environment galaxy sample cont. From left-to-right: SDSS thumbnail image of SPIRAL field-of-view, H$\alpha$ flux map of central region; H$\alpha$ velocity map of central region.  Only spaxels with signal-to-noise ratios $>3$ are shown.}
\label{piccies4}
\end{figure*}

\begin{figure*}
\begin{center}
\resizebox{35pc}{!}{\includegraphics[angle=90,scale=2.65,trim=-0.8cm 0cm 0cm 0cm,clip]{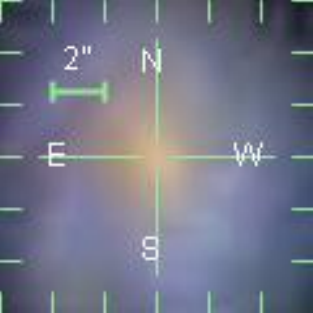} \includegraphics{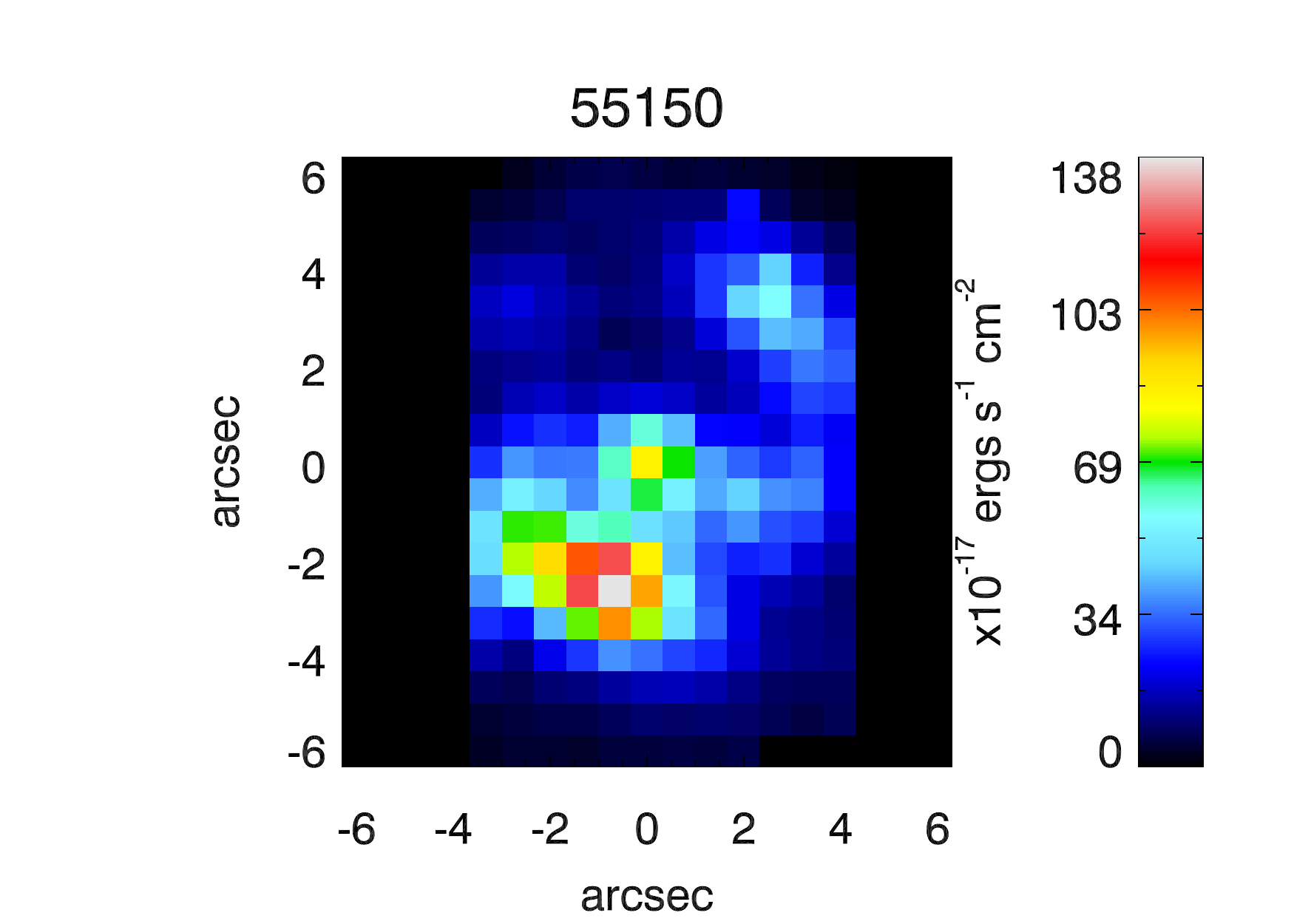} \includegraphics{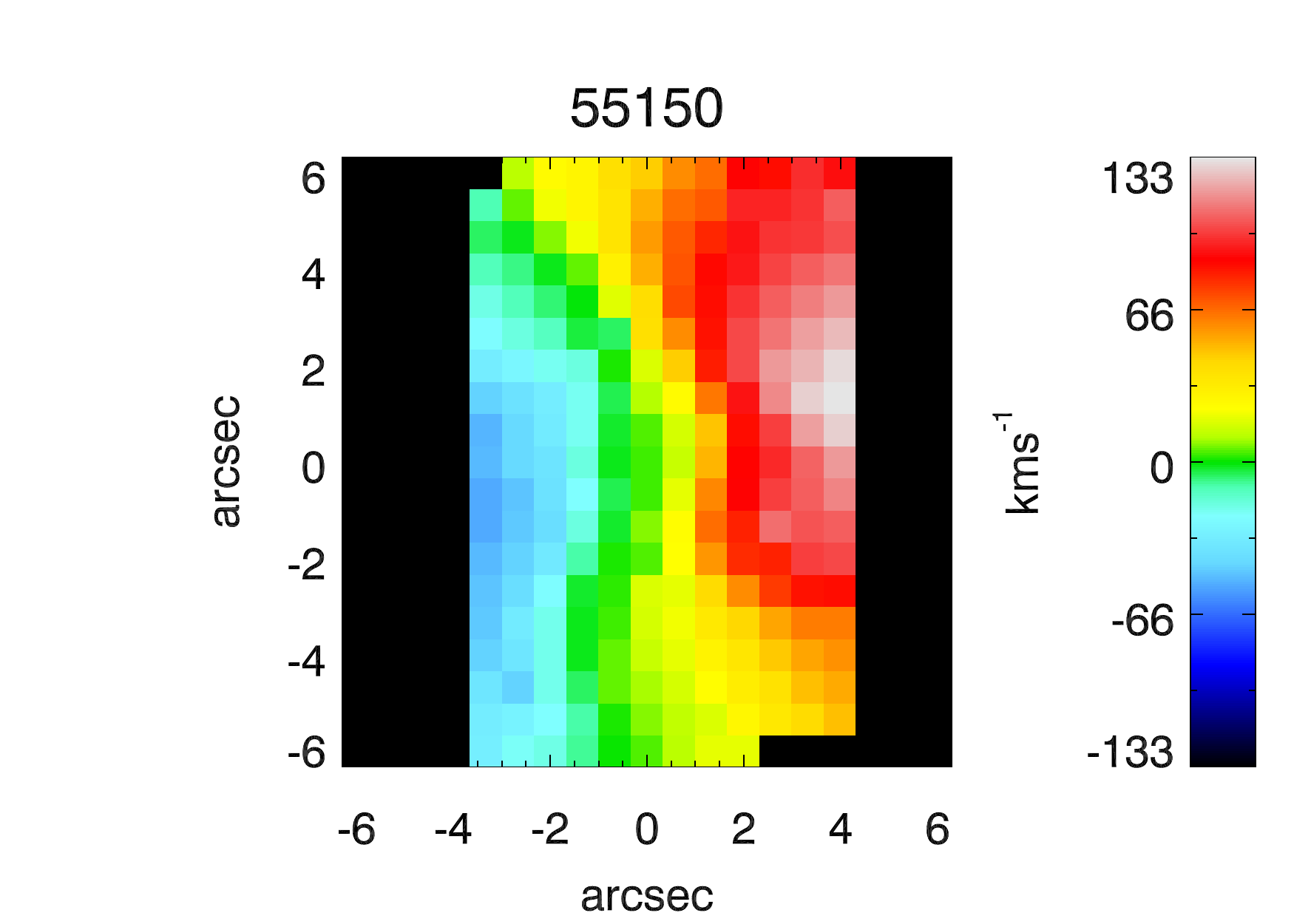}}
\resizebox{35pc}{!}{\includegraphics[scale=2.65,trim=0cm -0.8cm 0cm 0cm,clip]{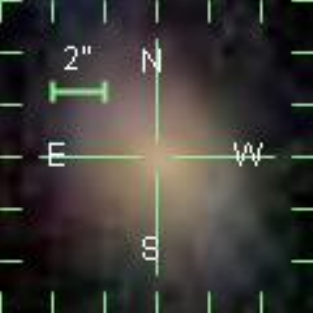} \includegraphics{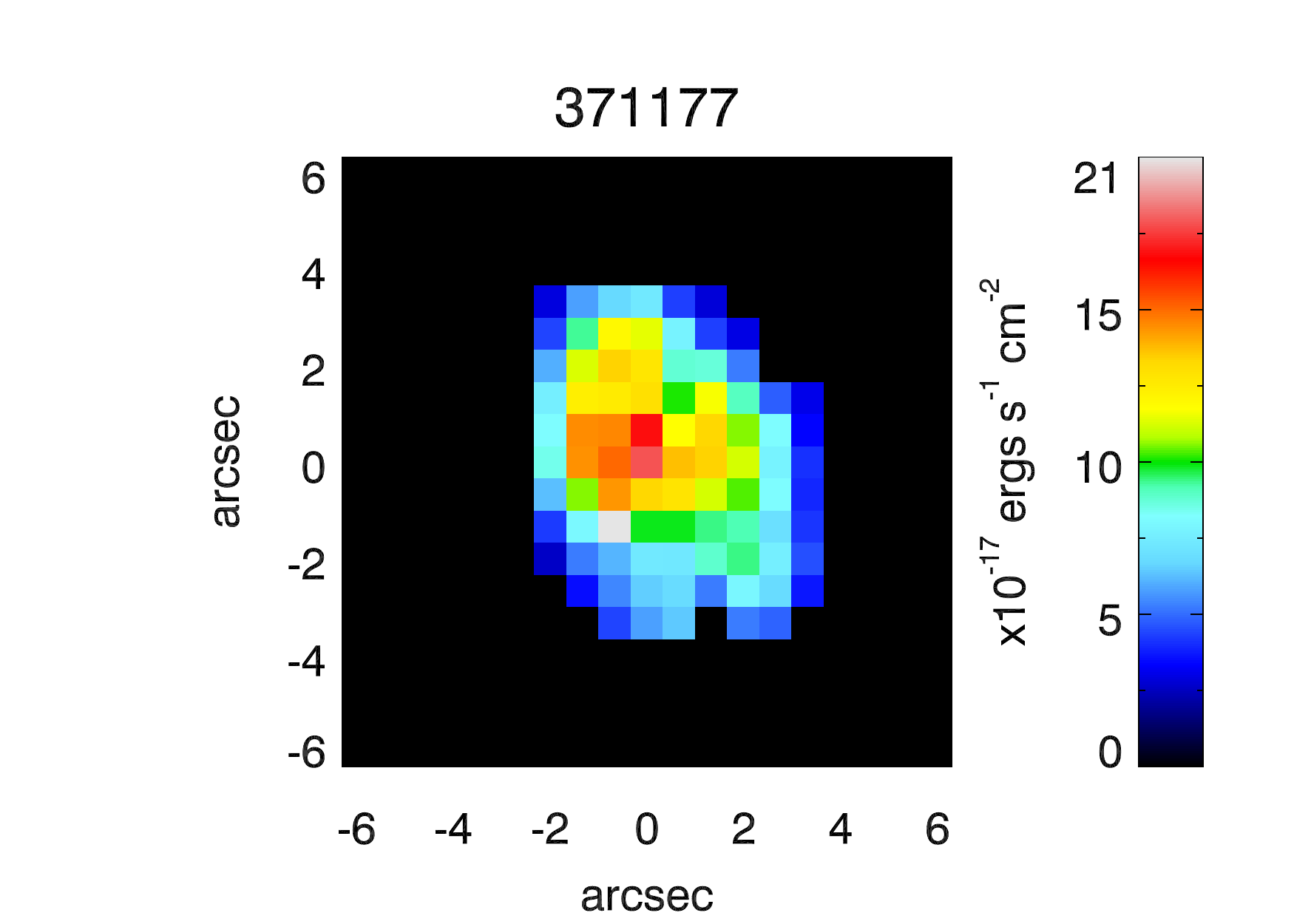} \includegraphics{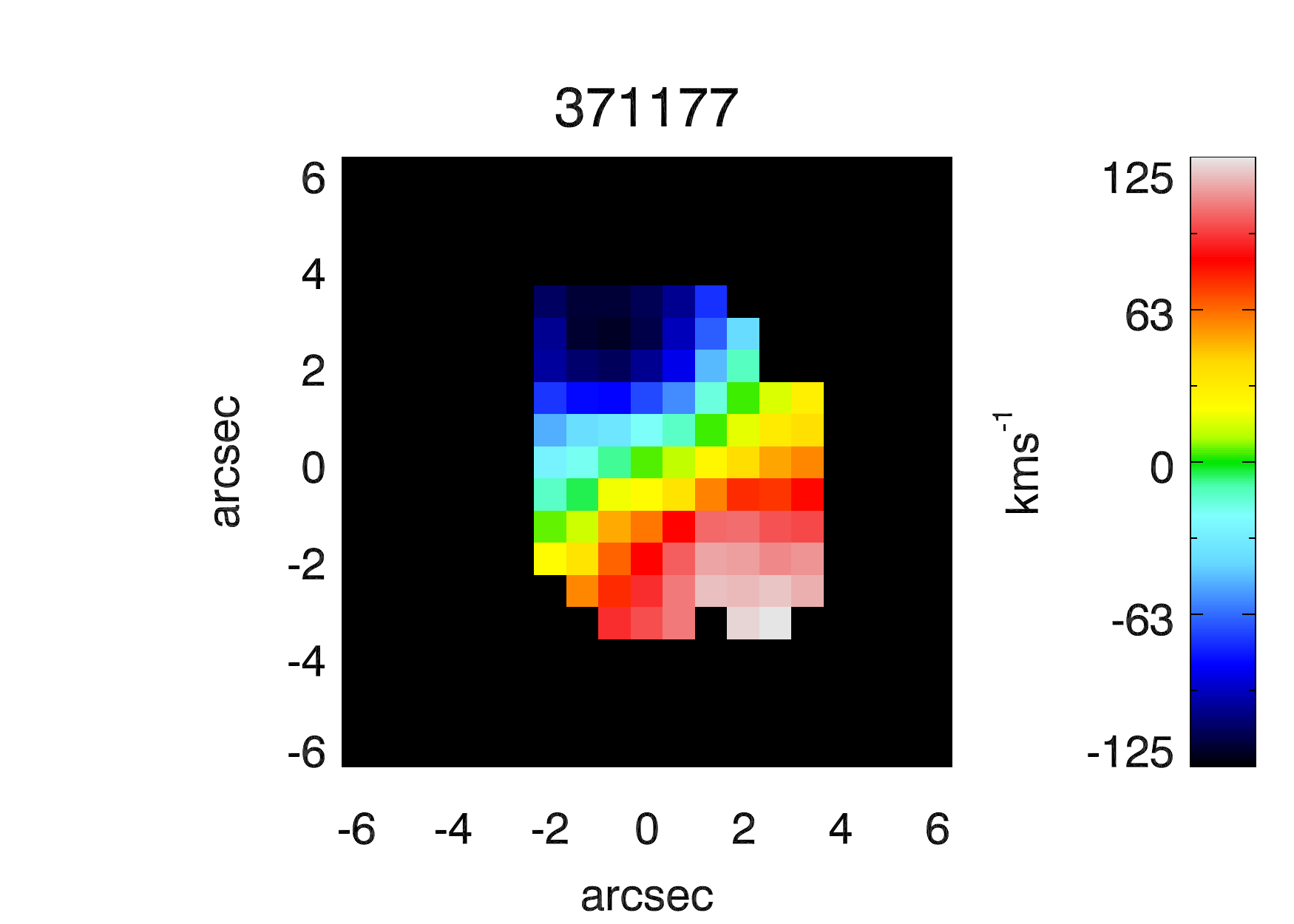}}
\resizebox{35pc}{!}{\includegraphics[angle=90,scale=2.65,trim=-0.8cm 0cm 0cm 0cm,clip]{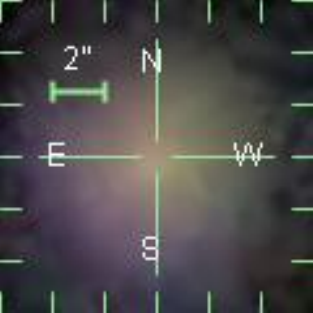} \includegraphics{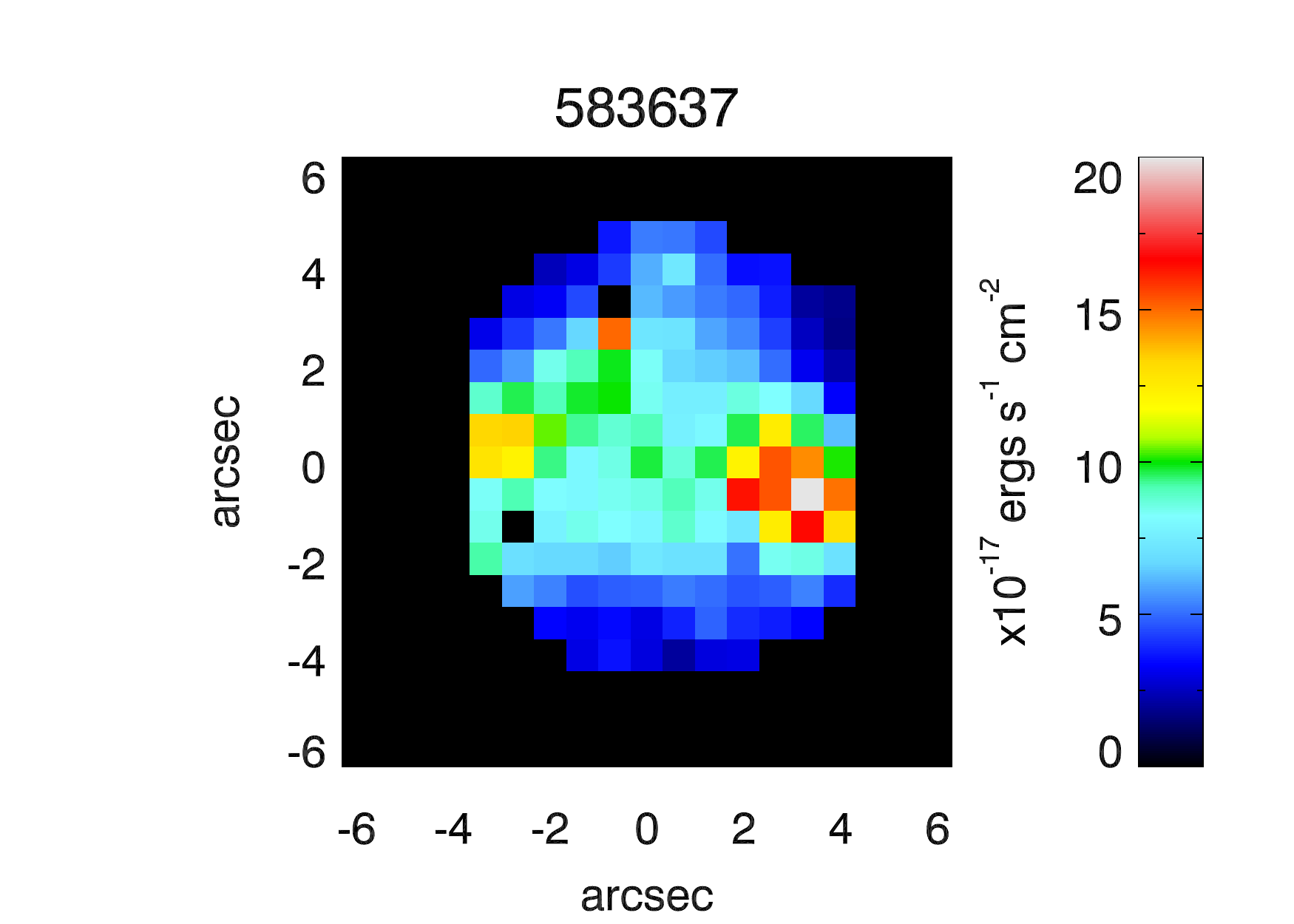} \includegraphics{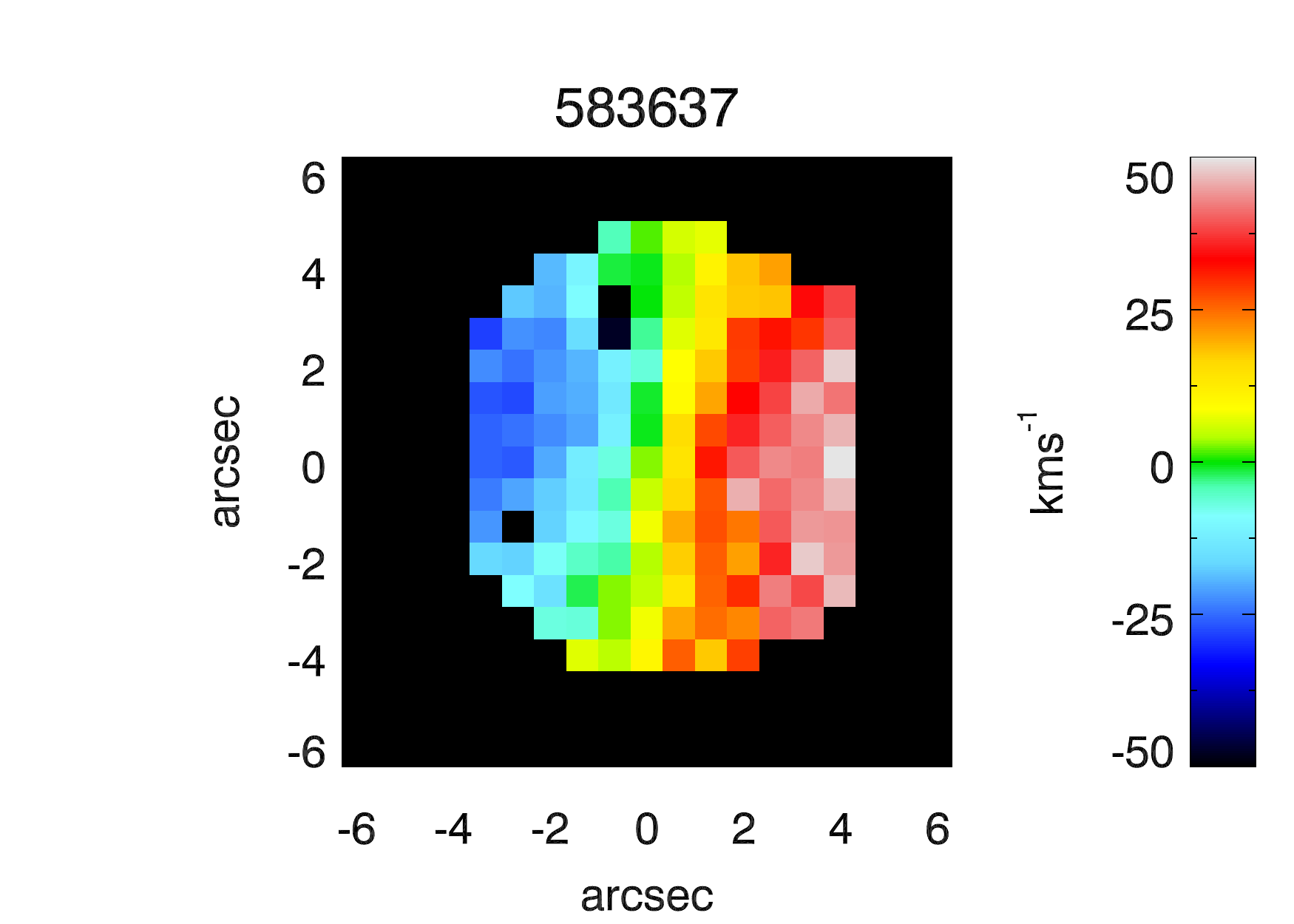}}
\end{center}
\caption{Low-density environment galaxy sample cont. From left-to-right: SDSS thumbnail image of SPIRAL field-of-view, H$\alpha$ flux map of central region; H$\alpha$ velocity map of central region.  Only spaxels with signal-to-noise ratios $>3$ are shown.}
\label{piccies5}
\end{figure*}

\bsp

\label{lastpage}

\end{document}